\documentclass[11pt]{article}
\pdfoutput=1
\usepackage[utf8x]{inputenc}	
\usepackage[english]{babel}
\usepackage{amssymb}
\usepackage{amsthm}
\usepackage{MnSymbol}
\usepackage{mathtools}		
\usepackage{dsfont}		
\usepackage{stmaryrd}		
\usepackage{cite}		
\usepackage[svgnames]{xcolor}
\usepackage{enumerate}
\usepackage{setspace}
\usepackage{booktabs}
\usepackage{tikz}
\usetikzlibrary{shapes}
\usetikzlibrary{positioning}
\usetikzlibrary{chains}
\usetikzlibrary{arrows,fit,decorations.pathreplacing, shapes.misc}
\tikzstyle{every picture}+=[remember picture]
\tikzstyle{na} = [baseline=-.5ex]
\usepackage{todonotes}
\usepackage{paralist}
\usepackage[font={small,it},labelfont=bf]{caption} 
\usepackage{subcaption}
\usepackage[pdftex,breaklinks,colorlinks=true,urlcolor=RoyalBlue,linkcolor=blue,
citecolor=blue]{hyperref}		
\usepackage[ddmmyyyy,hhmmss]{datetime}

\usepackage{multirow}

\catcode`@=11
\renewcommand{\section}{\@startsection{section}{1}{0pt}{\medskipamount}
{\medskipamount}{\Large\bf}}
\numberwithin{equation}{section}
\catcode`@=12

\def\dim{\mathrm{dim}}

\newcommand{\C}{\mathbb{C}}

\newcommand{\NN}{\mathbb{N}}  
\newcommand{\Z}{\mathbb{Z}}

\newcommand{\Coulomb}{\mathcal{C}}

\newcommand{\Higgs}{\mathcal{H}}

\newcommand{\clorbit}[1]{\overline{\mathcal{O}}_{#1}}
\newcommand{\height}[1]{\text{ht}(#1)}

\newcommand{\Ncal}{\mathcal{N}}

\newcommand{\su}{{{\rm SU}(2)}}

\newcommand{\uo}{{ \mathrm{U}(1)}}

\newcommand{\urm}{{{\rm U}}}

\newcommand{\surm}{{{\rm SU}}}
\newcommand{\surmL}{{{\mathfrak{su}}}}
\newcommand{\sorm}{{{\rm SO}}}
\newcommand{\orm}{{{\rm O}}}
\newcommand{\sormL}{{{\mathfrak{so}}}}

\newcommand{\usprm}{{{\rm USp}}}

\newcommand{\Rcal}{\mathcal{R}}
\newcommand{\Pcal}{\mathcal{P}}
\newcommand{\Ocal}{\mathcal{O}}

\newtheorem{myConj}{Conjecture}

\newcommand{\NS}{\text{NS5}}
\newcommand{\Done}{\text{D1}}
\newcommand{\Dtwo}{\text{D2}}
\newcommand{\Dthree}{\text{D3}}
\newcommand{\Df}{\text{D4}}
\newcommand{\Dfive}{\text{D5}}
\newcommand{\Ds}{\text{D6}}
\newcommand{\Dseven}{\text{D7}}
\newcommand{\De}{\text{D8}}
\newcommand{\Oe}{\text{O8}$^-$}
\newcommand{\Ost}{\text{O8}$^\ast$}
\newcommand{\Mf}{\text{M5}}
\newcommand{\Mn}{\text{M9}}
\newcommand{\magQuiv}{\mathsf{Q}}
\usepackage{xargs}

\newlength{\myline}
\newcommandx*{\doublearrow}[4][1=0, 2=1]{
  \draw[line width=\myline,double distance=3\myline,#3] #4;
}

\newcommandx*{\triplearrow}[4][1=0, 2=1]{
  \draw[line width=\myline,double distance=5\myline,#3] #4;
  \draw[line width=\myline,shorten <=#1\myline,shorten >=#2\myline,#3] #4;
}

\newcommandx*{\quadarrow}[4][1=0, 2=2.5]{
  \draw[line width=\myline,double distance=5\myline,#3] #4;
  \draw[line width=\myline,double distance=\myline,shorten <=#1\myline,shorten 
>=#2\myline,#3] #4;
}

\def\ns#1{
	\node[circle, draw, fill=white] at (#1){};
	\node[cross out, draw] at (#1){};
}

\def\DsixFree#1#2#3#4{
\draw(#1,#2)--(#3,#4);
}

\def\DsixSome#1#2{
	\foreach \i in {1,...,#1} {
		\draw (#2,{0.\i-#1/20-0.05})--(#2+1,{0.\i-#1/20-0.05});
	}
}

\def\DsixSomeMore#1#2#3#4{
	\foreach \i in {1,...,#1} {
	\draw (#2,{0.2*\i-#1/20-0.05+#4})--(#2+#3,{0.2*\i-#1/20-0.05+#4});
	}
}

\def\DsixMany#1#2{
		\draw(#2,0.09)--(#2+1,0.09);
		\draw(#2,0.03)--(#2+1,0.03);
		\draw(#2,-0.09)--(#2+1,-0.09);
	\draw[dotted](#2,-0.03)--(#2+1,-0.03);
	\node[label=above:{{\footnotesize{{#1}}}}] at (#2+0.6,-0.05) {};
}

\def\DsixEmpty#1{
		\draw[thick,dotted](#1+0.35,0)--(#1+1-0.35,0);
}

\def\Mnine#1{
 \foreach \i in {1,...,8}{
   \draw (#1+0.\i,-1)--(#1+0.\i,+1);
   }
   \draw[dashed,thick] (#1+1,-1)--(#1+1,+1);
}
\def\Oeight#1{
   \draw[dashed,thick] (#1,-1)--(#1,+1);
   }

\def\Ostar#1{
   \draw[dotted,thick] (#1,-1)--(#1,+1);
   }
   
\def\DeightMany#1#2#3{
   \foreach \i in {1,...,#1} {
	\draw({#2+#3*0.\i-0.1*#3},-1)--({#2+#3*0.\i-0.1*#3},+1);
      }
   }   
   
\def\Deight#1{
	  \draw(#1,-1)--(#1,1);
}

\def\DeightLong#1{
	  \draw(#1,-1)--(#1,2.3);
}

\def\ArrowMagnetic#1#2{
	\draw (#1,#2)--(#1+1,#2);
	\draw (#1+0.8,#2-.2)--(#1+1,#2);
	\draw (#1+0.8,#2+.2)--(#1+1,#2);
	\node (label) at (#1+0.5,#2+0.7)[]{magnetic quiver};
}

\allowdisplaybreaks

\begin{document}
\begin{titlepage}
\setcounter{page}{0}
%
%

\begin{flushright}
Imperial/TP/19/AH/01
\end{flushright}

\vskip 2cm

\begin{center}

{\Large\bf 
Magnetic Quivers, Higgs Branches, and $6$d $\Ncal=(1,0)$ Theories
}

\vspace{15mm}

{\large Santiago Cabrera${}^{1}$},\ {\large Amihay Hanany${}^{1}$}, and \ 
{\large 
Marcus Sperling${}^{2}$} 
\\[5mm]
\noindent ${}^1${\em Theoretical Physics Group, Imperial College London\\
Prince Consort Road, London, SW7 2AZ, UK}\\
{Email: {\tt santiago.cabrera13@imperial.ac.uk}, \\ {\tt 
a.hanany@imperial.ac.uk}}
\\[5mm]
\noindent ${}^{2}${\em Yau Mathematical Sciences Center, Tsinghua University}\\
{\em Haidian District, Beijing, 100084, China}\\
Email: {\tt marcus.sperling@univie.ac.at}
\\[5mm]

\vspace{15mm}

\begin{abstract}
The physics of \Mf\ branes placed near an \Mn\ plane on an $A$-type ALE 
singularity exhibits a variety of phenomena that introduce additional massless 
degrees of freedom. There are tensionless strings whenever two \Mf\ branes 
coincide or whenever an \Mf\ brane 
approaches the \Mn\ plane. 
These systems do not admit a low-energy Lagrangian 
description so new techniques are desirable to shed light on the physics 
of these phenomena.
The $6$-dimensional $\Ncal=(1,0)$ world-volume theory on the \Mf\ branes is 
composed of massless vector, tensor, and hyper multiplets, and has 
two branches of the vacuum moduli space 
where 
either the scalar fields in the tensor or hyper multiplets receive vacuum 
expectation values.
Focusing on the Higgs branch of the low-energy theory, 
previous works suggest the conjecture that a new Higgs branch arises whenever a 
BPS-string becomes tensionless. 
Consequently, a single theory admits a multitude of Higgs branches depending on 
the types of tensionless strings in the spectrum. 
The two main phenomena \emph{discrete gauging} and \emph{small $E_8$ instanton 
transition} can be treated in a concise and effective manner by means of
Coulomb branches of $3$-dimensional $\Ncal=4$ gauge theories.
In this paper, a formalism is introduced that allows to derive a novel object 
from a brane configuration, called the \emph{magnetic quiver}. The main 
features are as 
follows: (i) the 3d Coulomb branch of the magnetic quiver yields the Higgs 
branch of the 6d system, (ii) all discrete gauging and $E_8$ instanton 
transitions have an explicit brane realisation, 
and (iii) exceptional 
symmetries arise directly from brane configurations. The formalism 
facilitates 
the description of Higgs branches at finite and infinite gauge coupling as 
spaces of dressed monopole operators. 

\end{abstract}

\end{center}

\end{titlepage}

{\baselineskip=12pt
{\footnotesize
\tableofcontents
}
}
  \section{Introduction}
The world-volume theories of \Mf\ branes have led to interesting 
$6$-dimensional theories. A stack of coincident \Mf\ branes gives rise to 
world-volume theories with $\Ncal=(2,0)$ supersymmetry, 
which more generally admit an ADE classification 
\cite{Witten:1995zh,Strominger:1995ac}.
A larger class of $6$-dimensional supersymmetric 
theories has $\Ncal=(1,0)$ and their anomalies have been studied in works like 
\cite{Sagnotti:1992qw,Danielsson:1997kt,Bershadsky:1997sb}.
An example are the 
Type IIA constructions with 
\Ds-\De-\NS\ branes of \cite{Hanany:1997gh,Brunner:1997gk,Hanany:1997sa}. These 
brane constructions hinted at the existence of non-trivial conformal 
fixed-points at the origin of the tensor branch, i.e.\ when all \NS\ 
branes coincide.
Subsequently, a classification of $6$-dimensional superconformal field theories 
has been proposed in \cite{Heckman:2013pva,Heckman:2015bfa}.
Although these are local quantum field 
theories, no Lagrangian description is known and tensionless strings contribute 
to the low-energy degrees of freedom. 

The degrees of freedom of $6$-dimensional $\Ncal=(1,0)$ supersymmetric 
theory are given by vector multiplets, hypermultiplets, and tensor multiplets 
as well as other massless degrees of freedom which arise due to tensionless 
strings \cite{DelZotto:2014hpa}. 
Among others, the gravitational anomaly cancellation 
\cite{Green:1984bx} for such a theory requires 
\cite{RandjbarDaemi:1985wc,Dabholkar:1996zi} 
\begin{align}
 \#\{\text{hypers}\} + 29 
\#\{\text{tensors}\}-\#\{\text{vectors}\}=\text{constant} \; .
\end{align}
In $6$d theories the gauge coupling is a dynamical object as it is 
inversely related to a scalar field of a tensor multiplet which simultaneously 
serves as tension of a BPS string. 
At a generic point of the tensor branch, the $6$d $\Ncal=(1,0)$ theory may 
admit a low-energy effective description as all gauge couplings are finite. The 
corresponding Higgs branch at finite coupling is understood as a hyper-Kähler 
quotient due to the amount of supersymmetry \cite{Hitchin:1986ea}. At 
non-generic points of the tensor 
branch, the 6-dimensional theory is generically strongly coupled and a 
description of the 
corresponding Higgs branch is not straightforward. Since some BPS strings become 
tensionless whenever a gauge coupling is tuned to infinity, new massless degrees 
of freedom are expected to contribute to the Higgs branch such that it is 
still a hyper-Kähler space of larger dimension, but not a hyper-Kähler quotient 
any more.
Since these theories are non-Lagrangian, an alternative approach is desirable 
to investigate Higgs branches at infinite coupling.
Fortunately, another physical construction of hyper-Kähler singularities is 
known: the Coulomb branch of a $3$-dimensional $\Ncal=4$ gauge theory.
In fact, Coulomb branches have already been utilised to describe Higgs 
branches of 4-dimensional Argyres-Douglas theories 
\cite{DelZotto:2014kka}, 
$5$-dimensional gauge 
theories \cite{Cremonesi:2015lsa,Ferlito:2017xdq,Cabrera:2018jxt}, and
$6$-dimensional gauge theories 
\cite{Mekareeya:2017jgc,Hanany:2018uhm,Hanany:2018vph} at infinite 
coupling. 

The focus of this paper lies on a particular class of $6$d $\Ncal=(1,0)$ 
supersymmetric gauge theories obtained from the world-volume theories of 
multiple \Mf\ branes near an \Mn\ plane on an $A$-type ALE singularity.
This class has already been studied in some detail.
As discussed in \cite{DelZotto:2014hpa}, a system of multiple \Mf\ branes on 
an ALE singularity $\C^2 \slash \Gamma$, where $\Gamma \subset \surm(2)$ is a 
(discrete) ADE subgroup, undergoes a phase transition at the fixed point of 
the ALE space with new massless tensor multiplets appearing for $D$ and 
$E$-type, but not for $A$-type singularities.
The jumps in Higgs branch dimension at a generic point  
and the origin of the tensor branch for these theories has been computed in 
\cite{Mekareeya:2017sqh}.
The inclusion of the end-of-the-world \Mn\ plane with its global $E_8$ 
symmetry leads to the possibility of the small $E_8$ instanton 
transition\cite{Ganor:1996mu}, see also 
\cite{Seiberg:1996vs,Intriligator:1997kq,Blum:1997mm,Hanany:1997gh}. In 
particular, the Higgs branches exhibit an intimate relationship with the $E_8$ 
instanton moduli space on the $A$-type ALE space \cite{Mekareeya:2017jgc}, see 
also 
\cite{DelZotto:2014hpa,Heckman:2015bfa,Zafrir:2015rga,Ohmori:2015tka,
Hayashi:2015zka}.

The common feature of all the phenomena is the appearence of tensionless 
strings. Previous works indicate the following conjecture:
\begin{myConj}
Whenever a BPS-string becomes tensionless there is a singularity on the tensor 
branch and the associated massless degrees of freedom give rise to a new, 
finitely generated Higgs branch. 
\label{conj:Higgs}
\end{myConj}
The multitude of Higgs branches can be understood as \emph{phases} $\Pcal_i$ of 
the theory in the sense that the inverse gauge couplings and, hence, the scalar 
fields in the tensor multiplets serve as order parameters. Whenever at least one 
order 
parameter approaches zero, the Higgs branch changes discontinuously either
due to a gauging of a discrete group or due to a jump in dimension induced by
the small $E_8$ instanton transition\footnote{More phenomena show up in cases 
where the ALE singularity is not of $A$-type.}.

Here for a given phase $\Pcal_i$ of the $6$d $\Ncal=(1,0)$ theory, the emphasis 
lies on a systematic derivation of an associated \emph{magnetic 
quiver} $\magQuiv(\Pcal_i)$ such 
that their data considered as $3$d $\Ncal=4$ Coulomb branch does correctly 
describe the $6$d $\Ncal=(1,0)$ Higgs branch at the point $\Pcal_i$ 
of the tensor branch, i.e.\
\begin{align}
 \Higgs^{6d}\left( \text{phase } \Pcal_i \right) =  
\Coulomb^{3d}\left( 
\substack{\text{magnetic } \\ \text{quiver } } \magQuiv(\Pcal_i) \right) \;.
\label{eq:claim}
\end{align}
Since there are no gauge degrees of freedom on the \Mf\ brane there is the 
challenge to read off the low-energy gauge dynamics. It is useful to consider 
the dual Type IIA or Type I$^\prime$ 
description\cite{Hanany:1997gh,Brunner:1997gk} such that $6$d 
$\Ncal=(1,0)$ gauge dynamics can be deduced from the brane system involving \Ds 
, \De\ and \NS\ branes, possibly in the presence of orientifolds. The latter 
is known to be T-dual to the Type IIB construction 
\cite{Hanany:1996ie} of $3$d $\Ncal=4$ theories via \Dthree-\Dfive-\NS\ brane 
configurations.

The key tool to establish the objective \eqref{eq:claim} is to find a 
generalisation of the \emph{electric} 
and \emph{magnetic} theories within the Type IIB \Dthree-\Dfive-\NS\ brane 
configurations of $3$d $\Ncal=4$ theories. 
The derivation of the \emph{magnetic quiver} for the different phases of the 
$6$d theory can be summarised in 
the following two steps:
\begin{compactenum}[(i)]
 \item Change to the phase of the \Ds-\De-\NS\ brane system where all \Ds s are 
suspended 
between \De\ branes. This is analogous to the magnetic phase in 
\Dthree-\Dfive-\NS\ brane system, where the $\Dthree$ branes are in between 
$\Dfive$ branes and the \Done\ branes are the \emph{fundamental 
objects}.
\item Deduce the \emph{magnetic quiver} from this phase of the brane system by 
suspending \Df\ branes, which are the higher-dimensional analogous of the 
D-string.
\end{compactenum}
As as consequence, this procedure establishes a \emph{description of} $6$d 
$\Ncal=(1,0)$ \emph{Higgs 
branches as space of dressed monopole operators} as originally proposed in $3$d 
$\Ncal=4$ Coulomb branch set-up \cite{Cremonesi:2013lqa}.
The analysis of the phases as well as the transitions between the different 
phases requires many of the $3$d Coulomb branch techniques that have been 
developed 
recently. Starting from the realisation of the Coulomb branch as a space of 
dressed monopole operators and its description via the 
\emph{Hilbert series}\cite{Cremonesi:2013lqa}, useful techniques include: 
Kraft-Procesi transitions and transverse slices 
\cite{Cabrera:2016vvv,Cabrera:2017njm,Hanany:2018uhm}, quiver subtraction 
\cite{Cabrera:2018ann}, and discrete quotients 
\cite{Hanany:2018vph,Hanany:2018cgo,Hanany:2018dvd}.

It is worth pointing out that the Hilbert series is not an invariant quantity 
of the theory, in the sense that it varies between finite and infinite gauge 
coupling. In other words, the (Higgs branch) Hilbert series is not constant 
along the tensor branch. However, precisely this fact allows to utilise the 
Hilbert series as a tool to analyse the Higgs branch of vacuum moduli spaces as 
they vary along the tensor branch, see Conjecture 
\ref{conj:Higgs}. This has to be contrasted with quantities which are invariant 
under the choice of vacuum, i.e.\ constant along the tensor branch, because 
these would be insensitive to the different phases of the Higgs branch. 

The outline of the remainder is as follows: After introducing the set-up, the 
concept of \emph{electric and magnetic 
quiver} is explained in Section \ref{sec:Preliminaries} alongside with two 
paramount examples. Thereafter, in Section \ref{sec:body} the embedding of 
$\Z_k \hookrightarrow E_8$ is recalled and the cases of multiple \Mf\ branes 
near an \Mn\ plane on a $\C^2 \slash \Z_k$ singularity are elaborated on for 
$k=1,2,3,4$. The general case is presented in Section \ref{sec:general_case}.
An observation regarding the discrete 6d Theta-angle is discussed in Section 
\ref{sec:observation}. A conclusion and outlook is provided in Section 
\ref{sec:Conclusion}. Moreover, Appendix \ref{app:global_sym} provides details 
of background material.

  \section{Magnetic quivers}
\label{sec:Preliminaries}
\subsection{Set-up}
Consider \Mf\ branes and an \Mn\ plane as well as an $A_{k-1}$ ALE singularity 
stretching the space-time dimensions as indicated in Table \ref{tab:directions}. 
The singularity at the origin of $\C^2 \slash \Z_{k}$ is localized in directions 
$x^{7}$, $x^8$, $x^9$, and $x^{10}$, and spans directions $x^0,x^1, 
\ldots, x^6$. Therefore, it is represented as a horizontal line that ends on M9 
in the diagram below.
\begin{table}[t]
\centering
\begin{tabular}{c|ccccccccccc}
\toprule
 M-theory & $x^0$ &  $x^1$ & $x^2$ & $x^3$ & $x^4$ & $x^5$ & $x^6$ & $x^7$ & 
$x^8$ & $x^9$  & $x^{10}$ \\ \midrule 
\Mf\ & $\times$ & $\times$ & $\times$ & $\times$ & $\times$ & $\times$ & & & & 
& \\
\Mn\ & $\times$ & $\times$ & $\times$ & $\times$ & $\times$ & $\times$ & 
&$\times$ & $\times$ &$\times$ & $\times$  \\
$\C^2 \slash \Z_{k}$ & $\times$ & $\times$ & $\times$ & $\times$ & $\times$ & $\times$ & $\times$ & & & 
& \\ \midrule
Type IIA & $x^0$ &  $x^1$ & $x^2$ & $x^3$ & $x^4$ & $x^5$ & $x^6$ & $x^7$ & 
$x^8$ & $x^9$  & \\ \midrule 
\NS & $\times$ & $\times$ & $\times$ & $\times$ & $\times$ & $\times$ & & & & 
& \\
\Oe, \De & $\times$ & $\times$ & $\times$ & $\times$ & $\times$ & $\times$ & 
&$\times$ & $\times$ &$\times$ &   \\
\Ds & $\times$  & $\times$ & $\times$ & $\times$ & $\times$ & $\times$ & 
$\times$ &  &  & &   \\
\midrule
F1 & $\times$  &  &  & $\times$  &  &  &  &  &  &    \\
\Df & $\times$  &  &  & $\times$ & $\times$ & $\times$ &  & $\times$ &  &    
\\
\bottomrule
\end{tabular}
\caption{Upper part: Occupation of space-time directions by \Mf , \Mn , and 
$A_{k-1}$ 
singularity in M-theory. Lower part: Occupation of space-time directions by 
\NS, \Oe, \De, and \Ds\ in Type IIA. The fundamental string F1 and the \Df\ 
branes are virtual objects which are used to read off the electric and magnetic 
quivers.}
\label{tab:directions}
\end{table}
The M-theory picture can be presented as 
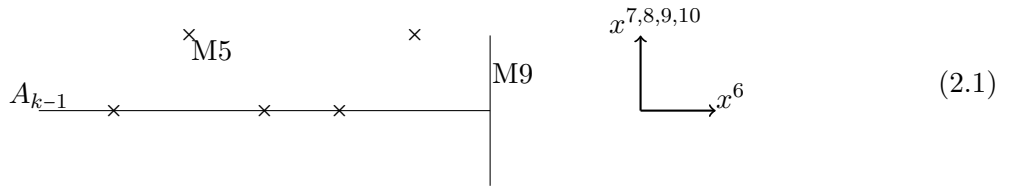
\begin{align}\label{branes:11d}
\raisebox{-.5\height}{
 \begin{tikzpicture}
  \draw(0,0)--(6,0);
  \draw (0,0.2) node {$A_{k-1}$};
  \draw (1,0) node {$\times$};
  \draw (2,1) node {$\times$};  
  \draw (3,0) node {$\times$};
  \draw (4,0) node {$\times$};
  \draw (5,1) node {$\times$};
  \draw (2.3,0.8) node {$\Mf$};
  \draw(6,-1)--(6,1);
  \draw (6.3,0.5) node {$\Mn$};
  \draw[thick,->] (8,0)--(9,0);
  \draw (9.2,0.2) node {$x^6$};
  \draw[thick,->] (8,0)--(8,1);
  \draw (8.2,1.2) node {$x^{7,8,9,10}$};
 \end{tikzpicture}
 }
\end{align}
The 
corresponding description in Type IIA is obtained by an identification as 
follows: the \NS\ originates from the \Mf\ which is point-like in the $x^{10}$ 
direction. The $E_8$ end-of-the world 9-plane \Mn\ gives 
rise to an \Oe\ orientifold together with $8$ \De s on top of it. Lastly, the 
$A_{k-1}$ ALE space $\C^2 \slash \Z_{k}$ in M-theory provides a local description 
of $k$ coincident \Ds\ branes in Type IIA on flat space. In particular, the 
directions $x^7, x^8, \ldots, x^{10}$ in which the singular origin of the
ALE singularity is localised become in the three directions transverse to the \Ds s and the 
direction of the M-theory circle. The corresponding Type IIA diagram is:
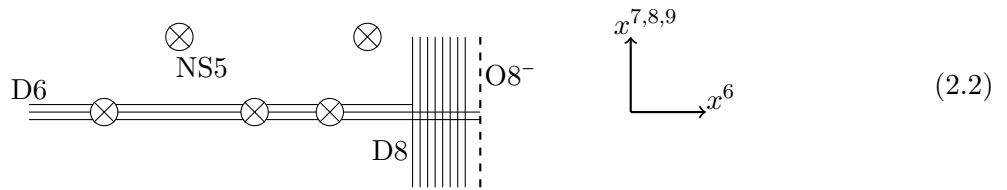
\begin{align}\label{branes:IIA}
\raisebox{-.5\height}{
 \begin{tikzpicture}
  \draw(0,0)--(6,0);
  \draw(0,-.1)--(6,-.1);
  \draw(0,.1)--(5.1,.1);
  \draw (0,0.3) node {\Ds};
  \ns{1,0};
  \ns{2,1}; 
  \ns{3,0}; 
  \ns{4,0}; 
  \ns{4.5,1}; 
  \draw (2.3,0.6) node {\NS};
  \Mnine{5};
  \draw (6.4,0.5) node {\Oe};
  \draw (4.8,-0.5) node {\De};
  \draw[thick,->] (8,0)--(9,0);
  \draw (9.2,0.2) node {$x^6$};
  \draw[thick,->] (8,0)--(8,1);
  \draw (8.2,1.2) node {$x^{7,8,9}$};
 \end{tikzpicture}
 }
\end{align}
Note that the \Ds\ branes have been assigned different boundary conditions along 
the $x^6$ direction. This is an essential part of our analysis and will be developed 
in full detail in Section \ref{sec:body}.

As an aside, some theories considered in later sections 
can admit 
\Ds, \De, and \NS\ branes  and additionally include \Ost planes, which occupy 
the same space-time dimensions as the \Oe.
%
%
\subsection{Electric and magnetic quiver}
As a detour, consider the \Dthree-\Dfive-\NS\ brane configurations of 
\cite{Hanany:1996ie} as summarised in Table \ref{tab:directions_3d}. A \Dthree\ 
brane suspended between two \NS s gives rise to an electric gauge group with a 
vector multiplet, whose 
gauge coupling is inversely proportional to the distance between the \NS s; while a 
\Dthree\ between \Dfive\ branes leads to an electric hypermultiplet. 
Consequently, the electric quiver gauge theory for the low-energy effective 
theory on the \Dthree\ world-volume is read off from the phase of the brane 
system in which all \Dthree s are suspended between \NS\ branes. In particular, 
the way fundamental strings can end on the branes gives rise the low-energy 
degrees of freedom.
On the other hand, the magnetic theory can be considered equally well: here, 
the 
\Dthree s between two \Dfive\ branes give rise to a magnetic gauge group with 
twisted vector multiplet \cite{Kapustin:1999ha}, whose 
gauge coupling inversely proportional to the distance between the \Dfive s. The 
magnetic hypermultiplet or twisted hypermultiplet \cite{Kapustin:1999ha} 
originates from \Dthree\ branes in 
between \NS\ branes. 
Taking this a step further, one can apply S-duality such that \Dfive\ and \NS\ 
branes are interchanged, while the \Dthree\ branes are invariant. Notably, the 
fundamental string is exchanged with the D-string, which is the fundamental 
object at large string coupling. Therefore, the degrees of freedom encoded on 
the magnetic quiver gauge theory are due to the way \Done\ branes stretch 
between \Dthree\ and \NS\ branes.
\begin{table}[t]
\centering
\begin{tabular}{c|cccccccccc}
\toprule
 Type IIB & $x^0$ &  $x^1$ & $x^2$ & $x^3$ & $x^4$ & $x^5$ & $x^6$ & $x^7$ & 
$x^8$ & $x^9$   \\ \midrule 
\NS & $\times$ & $\times$ & $\times$ & $\times$ & $\times$ & $\times$ & & & &\\
\Dfive & $\times$ & $\times$ & $\times$ & & & & &$\times$ & $\times$ &$\times$  
\\
\Dthree & $\times$  & $\times$ & $\times$ &  &  &  & $\times$ &  &  &    \\ 
\midrule
F1 & $\times$  &  &  &  $\times$ &  &  &  &  &  &    \\
\Done & $\times$  &  &  &  &  &  &  & $\times$ &  &    \\
\bottomrule
\end{tabular}
\caption{Occupation of space-time directions by \NS, \Dfive, and \Dthree\ in 
Type IIB. The fundamental string F1 induces the electric theory, while the 
D-string \Done\ induces the magnetic theory.}
\label{tab:directions_3d}
\end{table}

Returning to $6$-dimensional theories and \Ds-\De-\NS\ brane configurations 
of \cite{Hanany:1997gh}, one notices that these are obtained from the 
\Dthree-\Dfive-\NS\ configuration by three T-dualities along $x^3$, $x^4$, and 
$x^5$.
In the following, two phases of the Type IIA brane setting are important. 
Firstly, consider the phase in which all \Ds s are suspended between \NS s. Then 
the conventional $6$-dimensional low-energy effective field theory description 
is 
read off from fundamental strings stretching between \Ds s. The resulting theory 
can be expressed as a $6$-dimensional quiver gauge theory which is denoted 
as \emph{electric quiver} in what follows. Secondly, consider the phase in 
which all \Ds s are suspended between \De s and the \NS s are moved away from the 
\Ds s. In this phase one may suspend \Df\ branes between the \Ds s as well as 
between \Ds s and \NS s or \NS s and \NS s. The \Df\ suspension pattern is 
conveniently summarised in a quiver graph which one may call \emph{magnetic 
quiver}. The reason that \Df\ branes arise can be seen by following the 
T-dualities from the corresponding phase in the \Dthree-\Dfive-\NS\ 
configuration 
where \Done\ branes give rise to the magnetic quiver gauge theory. Thus, 
applying T-dualities along $x^3$, $x^4$, and $x^5$ to \Done\ branes naturally 
results in \Df\ branes. Nevertheless, a crucial difference arises in $6$d: 
since 
the \NS\ branes and the suspended \Ds\ branes share the same world-volume, the 
\NS s do contribute to the dynamics. Hence, lead to gauge nodes in the magnetic quiver as opposed to 
flavour nodes.

The main point of this paper is to argue that by taking the 
magnetic quiver data as input for the $3$-dimensional $\Ncal=4$ Coulomb branch 
description in terms of dressed monopole operators one can capture all phases 
of 
the $6$-dimensional Higgs branches in a systematic and concise fashion.
Therefore, it is imperative to distinguish the moduli spaces associated with 
the two kinds of quivers: the electric quiver data serves as definition of 
a low-energy effective $6$d $\Ncal=(1,0)$ theory and, in particular, its classical Higgs branch; 
whereas the magnetic quiver 
data defines a Coulomb branch of a $3$d $\Ncal=4$ gauge theory describing the Higgs branch of the 
strongly coupled $6$d $\Ncal=(1,0)$ theory. Both moduli 
spaces are hyper-Kähler singularities (symplectic singularities, see 
\cite{Beauville:2000}) with certain symmetries, as 
recalled in Appendix \ref{app:global_sym}.

Before considering the \Mf\ branes near the end-of-the world \Mn\ plane on an 
$A_{k-1}$ singularity, it is instructive to understand the two extreme cases: \Mf\ 
branes on a $A_{k-1}$ singularity and \Mf\ branes near an \Mn\ plane.

\paragraph{Notation.} 
In order to distinguish  the electric from the magnetic quiver, the following conventions 
are used: gauge nodes in the electric quiver are denoted explicitly by the 
gauge groups; whereas gauge nodes in the magnetic quiver are only labelled by the 
ranks $r_i$ of unitary gauge nodes $\urm(r_i)$. 
%
%
\subsection{Discrete gauging: M5 branes on A-type singularity}
\label{sec:discrete_gauging}
A system of multiple \Mf\ branes and an $A_{k-1}$ singularity can exhibit many 
phases 
\begin{align}
\raisebox{-.5\height}{
 \begin{tikzpicture}
   \draw(0,0)--(6,0);
  \draw (0,0.2) node {$A_{k-1}$};
  \draw (1,0) node {$\times$};
  \draw (2,1) node {$\times$};  
  \draw (3,0) node {$\times$};
  \draw (4,0) node {$\times$};
  \draw (4,0.5) node {$\times$};
  \draw (5,1) node {$\times$};
  \draw (5,0.3) node {$\times$};
  \draw (2.3,0.8) node {\Mf};
  \draw[thick,->] (8,0)--(9,0);
  \draw (9.2,0.2) node {$x^6$};
  \draw[thick,->] (8,0)--(8,1);
  \draw (8.2,1.2) node {$x^{7,8,9,10}$};
 \end{tikzpicture}
 }
\end{align}
depending on whether the \Mf s are on the singularity or away from it. 
Restricting to the phase where all \Mf s are on the singularity (i.e. they are at the origin of coordinates $x^7$, $x^8$,$x^9$ and $x^{10}$), there exist 
multiple phases describing the positions of the \Mf\ along the $x^6$ 
direction. In other words, if the \Mf s are separated or some of them coincide.

Consider $n$ \Mf\ branes on an $A_{k-1}$ singularity, in the phase where 
all \Mf\ are located at the singularity, but are separated along the $x^6$ direction:
\begin{align}
\raisebox{-.5\height}{
 \begin{tikzpicture}
   \draw(0,0)--(2.5,0) (3.5,0)--(6,0);
  \draw (0,0.2) node {$A_{k-1}$};
  \draw (1,0) node {$\times$};
  \draw (2,0) node {$\times$}; 
  \DsixEmpty{2.5};
  \draw (4,0) node {$\times$};
  \draw (5,0) node {$\times$};
  \draw[decoration={brace,mirror,raise=10pt},decorate,thick]
  (0.8,0) -- node[below=10pt] {$n$ \Mf} (5.2,0);
  \draw[thick,->] (8,0)--(9,0);
  \draw (9.2,0.2) node {$x^6$};
  \draw[thick,->] (8,0)--(8,1);
  \draw (8.2,1.2) node {$x^{7,8,9,10}$};
 \end{tikzpicture}
 }
\end{align}
The Type IIA 
description yields the following:
\begin{align}
\raisebox{-.5\height}{
 \begin{tikzpicture}
		\draw (0,-0.3) node {\Ds };
		\DsixMany{k}{0}
		\DsixMany{k}{1}
		\ns{1,0}
		\DsixMany{k}{2}
		\ns{2,0}
		\DsixEmpty{3}
		\ns{3,0}
		\DsixMany{k}{4}
		\ns{4,0}
		\DsixMany{k}{5}
		\ns{5,0}
  		\draw[decoration={brace,mirror,raise=10pt},decorate,thick]
  (0.8,0) -- node[below=10pt] {$n$ \NS } (5.2,0);
  \draw[thick,->] (8,0)--(9,0);
  \draw (9.2,0.2) node {$x^6$};
  \draw[thick,->] (8,0)--(8,1);
  \draw (8.2,1.2) node {$x^{7,8,9}$};
	\end{tikzpicture}
	} 
\end{align}
where the $k$ indicates that there are $k$ \Ds\ branes stacked together. The \emph{electric quiver} is read off from this brane system to be
\begin{align}
	\raisebox{-.5\height}{
 	\begin{tikzpicture}
	\tikzstyle{gauge} = [circle, draw,inner sep=3pt];
	\tikzstyle{flavour} = [regular polygon,regular polygon sides=4,inner 
sep=3pt, draw];
	\node (g1) [gauge,label={[rotate=-45]below right:{$\surm (k)$}}] {};
	\node (g2) [gauge, right of=g1,label={[rotate=-45]below right:{$\surm (k)$}}] {};
	\node (g3) [right of=g2] {$\ldots$};
	\node (g4) [gauge, right of=g3,label={[rotate=-45]below right:{$\surm (k)$}}] {};
	\node (f1) [flavour,left of=g1,label=below:{$k$}] {};
	\node (f4) [flavour,right of=g4,label=below:{$k$}] {};
	\draw (g1)--(g2) (g2)--(g3) (g3)--(g4) (g1)--(f1) (g4)--(f4);
	\draw[decoration={brace,raise=10pt},decorate,thick]
  (-0.2,0) -- node[above=10pt] {$n-1$} (3.2,0);
	\end{tikzpicture}
	}\,.
	\label{eq:6d_quiver_M5_on_A-type}
\end{align}
Note in particular, that the quiver describes a $6$d $\Ncal=(1,0)$ low energy 
effective field theory in which all gauge and flavour nodes denote $\surm(k)$ 
groups.

One may move to a different phase of the brane system by, firstly, pulling in 
\De\ branes from infinity and, secondly, rearranging the brane system such 
that all \Ds\ branes are suspended between \De\ branes.
As a result, one obtains
\begin{align}
\raisebox{-.5\height}{
 \begin{tikzpicture}
		\Deight{0}
		\Deight{0.2}
		\Deight{0.4}
		\draw (0.7,0) node {$\cdots$};
		\Deight{1}
		\Deight{1.2}
		\ns{2,1}
		\ns{2.4,1}
		\draw (2.9,1) node {$\cdots$};
		\ns{3.4,1}
		\Deight{4.2}
		\Deight{4.4}
		\draw (4.7,0) node {$\cdots$};
		\Deight{5}
		\Deight{5.2}
		\Deight{5.4}
		\DsixFree{0}{0}{0.2}{0}		
		\DsixFree{0.2}{0.1}{0.4}{0.1}	
		\DsixFree{0.2}{-0.1}{0.4}{-0.1}
		\DsixFree{1}{0.4}{1.2}{0.4}
		\DsixFree{1}{0.2}{1.2}{0.2}
		\draw (1.1,-0.1) node {$\vdots$};
		\DsixFree{1}{-0.4}{1.2}{-0.4}
		\DsixFree{1.2}{0.5}{4.2}{0.5}
		\DsixFree{1.2}{0.3}{4.2}{0.3}
		\DsixFree{1.2}{0.1}{4.2}{0.1}
		\draw (2.8,-0.1) node {$\vdots$};
		\DsixFree{1.2}{-0.3}{4.2}{-0.3}
		\DsixFree{1.2}{-0.5}{4.2}{-0.5}
		\DsixFree{4.2}{0.4}{4.4}{0.4}
		\DsixFree{4.2}{0.2}{4.4}{0.2}
		\draw (4.3,-0.1) node {$\vdots$};
		\DsixFree{4.2}{-0.4}{4.4}{-0.4}
		\DsixFree{5}{0.1}{5.2}{0.1}	
		\DsixFree{5}{-0.1}{5.2}{-0.1}
		\DsixFree{5.2}{0}{5.4}{0}	
		\draw[decoration={brace,mirror,raise=30pt},decorate,thick](-0.1,0) -- node[below=30pt] {$k$ \De } (1.3,0);
		\draw[decoration={brace,mirror,raise=30pt},decorate,thick](4.1,0) -- node[below=30pt] {$k$ \De } (5.5,0);
		\draw[decoration={brace,raise=37pt},decorate,thick](1.8,0) -- node[above=37pt] {$n$ \NS } (3.6,0);
  \draw[thick,->] (8,0)--(9,0);
  \draw (9.2,0.2) node {$x^6$};
  \draw[thick,->] (8,0)--(8,1);
  \draw (8.2,1.2) node {$x^{7,8,9}$};
		\end{tikzpicture}
		}
	\label{eq:multiple_M5_bouquet}
\end{align}
As aforementioned, one may now consider the possibility to suspend \Df\ 
branes between the \Ds\ and the \NS\ branes. In an interval between two \De s 
with $m$ \Ds s in between, the different ways to connect \Df\ branes between the \Ds s naturally 
furnishes the adjoint representation of $\urm(m)$.
This is analogous to 
Chan-Paton factors of an open string ending on D-branes. Hence, this is a 
$\frac{1}{4}$ BPS configuration which induces a magnetic vector multiplet. 
Likewise, the \Df\ branes 
stretching between two adjacent intervals with $m$ and $l$ \Ds\ branes furnish 
the bifundamental representation of $\urm(m)\times\urm(l)$ such that this 
$\tfrac{1}{4}$ BPS system contributes a bifundamental magnetic hypermultiplet.

For lower dimensional settings, like in Table \ref{tab:directions_3d}, this 
would be the end of the discussion, but here there are further 
possibilities. Recall that both the \NS\ and the suspended \Ds\ 
branes have $6$-dimensional world volumes; hence, the \NS s contribute to the 
dynamics too. More concretely, one may also stretch \Df\ branes between \NS\ 
branes. Since the \NS\ branes are not subject to boundary conditions on any 
other brane, they are $\frac{1}{2}$ BPS and as such contribute a magnetic 
vector multiple together with an adjoint magnetic hypermultiplet. If there are 
multiple \NS s in the same interval, but they are separated in $x^6$ direction 
then the \Df\ stretched between the \NS\ branes does not contribute massless 
degrees of freedom. In that case, each \NS\ 
contributes with a single $\uo$ gauge node and the corresponding adjoint 
hypermultiplet.
Furthermore, a \Df\ stretched between a \NS\ and a \Ds\  can only contribute 
massless degrees of freedom if 
they are in the same interval. consequently, a single \NS\ and stack of $m$ 
\Ds\ give rise to a bifundamental of $\urm(m)\times \uo$.  Following these 
observations results in the \emph{magnetic quiver} of the form 
\begin{align}
\raisebox{-.5\height}{
 	\begin{tikzpicture}
	\tikzstyle{gauge} = [circle, draw,inner sep=3pt];
	\tikzstyle{flavour} = [regular polygon,regular polygon sides=4,inner 
sep=3pt, draw];
	\node (g1) [gauge,label=below:{$1$}] {};
	\node (g2) [gauge, right of=g1,label=below:{$2$}] {};
	\node (g3) [right of=g2] {$\cdots$};
	\node (g4) [gauge, right of=g3,label=below:{$k{-}1$}] {};
	\node (g5) [gauge, right of=g4,label=below:{$k$}] {};
	\node (g6) [gauge, right of=g5,label=below:{$k{-}1$}] {};
	\node (g7) [right of=g6] {$\cdots$};
	\node (g8) [gauge, right of=g7,label=below:{$2$}] {};
	\node (g9) [gauge, right of=g8,label=below:{$1$}] {};
	\node (b1) [gauge,above left of=g5, label=left:{$1$}] {};
	\node (b0) [above of =g5,label=below:{$\ldots$},label=above:{$n$}] {} ;;
	\node (b2) [gauge,above right of=g5, label=right:{$1$}] {};
	\draw (g1)--(g2) (g2)--(g3) (g3)--(g4) (g4)--(g5) (g5)--(g6) 
(g6)--(g7) (g7)--(g8) (g8)--(g9) (g5)--(b1) (g5)--(b2);
	\end{tikzpicture}
	}
    \label{eq:3d_quiver_bouquet}
\end{align}
Note that there is a bouquet of $n$ separate $\uo$ nodes at the top, which results from the $n$ separated \NS\ 
branes\footnote{Since the attention is directed towards the Coulomb branch, 
the neutral adjoint hypermultiplet from any single \NS\ brane is neglected.}. 
If the data underlying the \emph{magnetic quiver} is understood as 
defining a $3$d $\Ncal=4$ quiver gauge theory, then the significance of this 
construction is
\begin{align}
 \Higgs^{6d}\left( \substack{ \text{electric} \\ \text{quiver}} 
\eqref{eq:6d_quiver_M5_on_A-type}\right) = 
 \Coulomb^{3d} \left( \substack{ \text{magnetic} \\ \text{quiver}} 
\eqref{eq:3d_quiver_bouquet}\right) \;,
\label{eq:equality_moduli-spaces}
\end{align}
as equality of moduli spaces. For consistency one may verify that the 
symmetries and dimensions agree, see Appendix \ref{app:global_sym}; indeed, one 
finds
\begin{align}
 G_F = \surm(k)^2 \times \uo^n = G_J  
 \qquad
 \text{and}
 \qquad
 \dim\, \Higgs^{6d} =k^2+n-1 = \dim\, \Coulomb^{3d} \,.
\end{align}
In fact, the equality \eqref{eq:equality_moduli-spaces} has been shown to 
arise in two steps in \cite{Hanany:2018vph}: firstly, S-duality or 3d mirror 
symmetry for the quiver \eqref{eq:6d_quiver_M5_on_A-type} where all special 
unitary nodes are replaced by unitary nodes. Secondly, employing the concept of 
\emph{implosion} \cite{Dancer:2012sv} on the $3$d mirror to arrive at the 
magnetic quiver \eqref{eq:3d_quiver_bouquet}.

As discussed in \cite{Hanany:2018vph,Hanany:2018cgo,Hanany:2018dvd}, the $6$d 
Higgs branch exhibits many more phases. The phases originate when some \Mf\ 
branes become coincident along the $x^6$ direction. Clearly, there exists no 6d low-energy effective 
quiver description as the distance between two neighbouring \NS\ branes 
determines the inverse gauge coupling of the corresponding gauge group. 
Therefore, there exists no \emph{electric quiver} for any of these strongly 
coupled phases.
In contrast, the magnetic phase can be readily applied to this setting.
Suppose that from the $n$ \Mf\ branes $n_i$ 
($i=1,\ldots, l$ such that $\sum_{i=1}^l n_i=n$) of these coincide at $x_i^6$, 
then the corresponding brane picture becomes

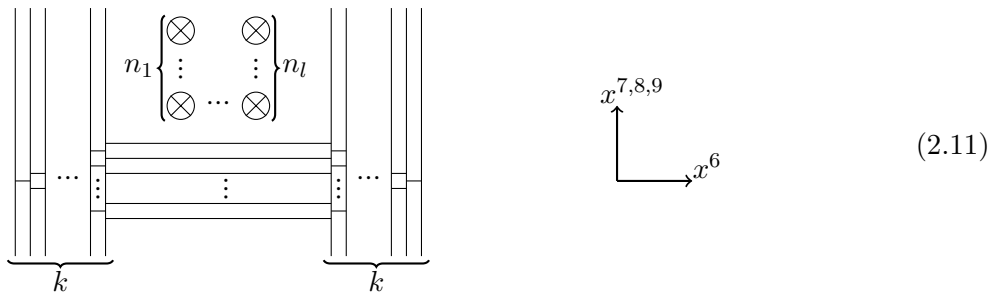
\begin{align}
\raisebox{-.5\height}{
 \begin{tikzpicture}
		\DeightLong{0}
		\DeightLong{0.2}
		\DeightLong{0.4}
		\draw (0.7,0) node {$\cdots$};
		\DeightLong{1}
		\DeightLong{1.2}
		\ns{2.2,1}
		\draw (2.2,1.5) node {$\vdots$};
		\ns{2.2,2}
		\draw[decoration={brace,raise=6pt},decorate,thick](2.2,0.8) -- node[left=6pt] {$n_1$} (2.2,2.2);
		\draw (2.7,1) node {$\cdots$};
		\ns{3.2,1}
		\draw (3.2,1.5) node {$\vdots$};
		\ns{3.2,2}
		\draw[decoration={brace,mirror,raise=6pt},decorate,thick](3.2,0.8) -- node[right=6pt] {$n_l$} (3.2,2.2);
		\DeightLong{4.2}
		\DeightLong{4.4}
		\draw (4.7,0) node {$\cdots$};
		\DeightLong{5}
		\DeightLong{5.2}
		\DeightLong{5.4}
		\DsixFree{0}{0}{0.2}{0}		
		\DsixFree{0.2}{0.1}{0.4}{0.1}	
		\DsixFree{0.2}{-0.1}{0.4}{-0.1}
		\DsixFree{1}{0.4}{1.2}{0.4}
		\DsixFree{1}{0.2}{1.2}{0.2}
		\draw (1.1,-0.1) node {$\vdots$};
		\DsixFree{1}{-0.4}{1.2}{-0.4}
		\DsixFree{1.2}{0.5}{4.2}{0.5}
		\DsixFree{1.2}{0.3}{4.2}{0.3}
		\DsixFree{1.2}{0.1}{4.2}{0.1}
		\draw (2.8,-0.1) node {$\vdots$};
		\DsixFree{1.2}{-0.3}{4.2}{-0.3}
		\DsixFree{1.2}{-0.5}{4.2}{-0.5}
		\DsixFree{4.2}{0.4}{4.4}{0.4}
		\DsixFree{4.2}{0.2}{4.4}{0.2}
		\draw (4.3,-0.1) node {$\vdots$};
		\DsixFree{4.2}{-0.4}{4.4}{-0.4}
		\DsixFree{5}{0.1}{5.2}{0.1}	
		\DsixFree{5}{-0.1}{5.2}{-0.1}
		\DsixFree{5.2}{0}{5.4}{0}	
		\draw[decoration={brace,mirror,raise=30pt},decorate,thick](-0.1,0) -- node[below=30pt] {$k$  } (1.3,0);
		\draw[decoration={brace,mirror,raise=30pt},decorate,thick](4.1,0) -- node[below=30pt] {$k$ } (5.5,0);
  \draw[thick,->] (8,0)--(9,0);
  \draw (9.2,0.2) node {$x^6$};
  \draw[thick,->] (8,0)--(8,1);
  \draw (8.2,1.2) node {$x^{7,8,9}$};
		\end{tikzpicture}
		}
		\label{eq:multiple_M5_discrete_gauging}
\end{align}

To determine the magnetic quiver for this brane system, one has to apply 
the above considerations to \Df\ branes stretched between coincident 
\NS\ branes. Since a stack of $m$ coincident \NS\ branes is a 
$\tfrac{1}{2}$ BPS configuration, there are two contributions: firstly, a 
magnetic vector multiplet for a $\urm(m)$ gauge node; 
secondly, a magnetic hypermultiplet in the adjoint representation of $\urm(m)$, 
which is denoted by a loop attached to the gauge node. 
From the brane 
configuration, the magnetic vector multiplet is associated with the motion in 
$x^7$, $x^8$, $x^9$ direction, while motions in $x^6$ direction give rise to 
the additional magnetic hypermultiplet. 
In other words, from the $3$d $\Ncal=4$ perspective each stack of $n_i$ \NS\ 
branes contributes an $\urm(n_i)$ together with an adjoint-valued hyper 
multiplet.
Equipped with these rules, the magnetic quiver associated to the brane 
configuration \eqref{eq:multiple_M5_discrete_gauging} is read off to be
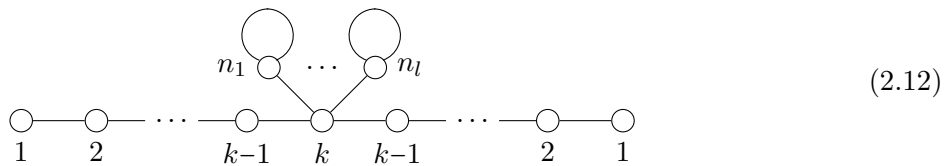
\begin{align}
\raisebox{-.5\height}{
 	\begin{tikzpicture}
	\tikzstyle{gauge} = [circle, draw,inner sep=3pt];
	\tikzstyle{flavour} = [regular polygon,regular polygon sides=4,inner 
sep=3pt, draw];
	\node (g1) [gauge,label=below:{$1$}] {};
	\node (g2) [gauge, right of=g1,label=below:{$2$}] {};
	\node (g3) [right of=g2] {$\ldots$};
	\node (g4) [gauge, right of=g3,label=below:{$k{-}1$}] {};
	\node (g5) [gauge, right of=g4,label=below:{$k$}] {};
	\node (g6) [gauge, right of=g5,label=below:{$k{-}1$}] {};
	\node (g7) [right of=g6] {$\ldots$};
	\node (g8) [gauge, right of=g7,label=below:{$2$}] {};
	\node (g9) [gauge, right of=g8,label=below:{$1$}] {};
	\node (b1) [gauge,above left of=g5, label=left:{$n_1$}] {};
	\node (b0) [above of =g5,label=below:{$\ldots$}] {} ;
	\node (b2) [gauge,above right of=g5, label=right:{$n_l$}] {};
	\draw [-] (3.275+0.11,0.805) arc (-70:90:10pt);
	\draw [-] (3.275+-0.11,0.805) arc (250:90:10pt);
	\draw [-] (4.7+0.11,0.805) arc (-70:90:10pt);
	\draw [-] (4.7-0.11,0.805) arc (250:90:10pt);
	\draw (g1)--(g2) (g2)--(g3) (g3)--(g4) (g4)--(g5) (g5)--(g6) 
(g6)--(g7) (g7)--(g8) (g8)--(g9) (g5)--(b1) (g5)--(b2);
	\end{tikzpicture}
	}
	\label{eq:3d_quiver_discrete_gauged}
\end{align}
Some comments are in order. Firstly, the magnetic quiver prescription provides 
a systematic description of all the (weakly and strongly coupled) phases of the 
6d Higgs branches. The underlying 3d Coulomb branch quiver has already been 
discussed in \cite{Hanany:2018vph,Hanany:2018cgo,Hanany:2018dvd}. Secondly, the 
novel perspective in the present paper is the brane realisation of these magnetic quivers via 
suspended \Df s. Thirdly, the use of branes makes the \emph{discrete gauging} 
relation between the various Higgs branches manifest. In more 
detail, the conjecture of \cite{Hanany:2018vph} asserts that the 6d Higgs 
branches corresponding to \eqref{eq:3d_quiver_bouquet} and 
\eqref{eq:3d_quiver_discrete_gauged} are related via gauging of discrete 
permutation groups. The Type IIA picture in phase \eqref{eq:multiple_M5_bouquet} 
exhibits an $S_n$ symmetry due to the indistinguishable nature of the \NS s. 
When the \NS\ branes are coincident as in 
\eqref{eq:multiple_M5_discrete_gauging} the discrete $\prod_i S_{n_i}$ group is 
gauged.
%
%
\subsection{Small \texorpdfstring{$E_8$}{E8} instanton transition: \Mf\ branes 
near \Mn\ plane}
\label{sec:small_E8}
The other extreme is a system of \Mf\ branes near an \Mn\ plane which do also 
exhibit various phases
\begin{align}
\raisebox{-.5\height}{
 \begin{tikzpicture}
  \draw(0,0)--(6,0);
  \draw (0,0.2) node {$A_0$};
  \draw (1,0) node {$\times$};
  \draw (2,0) node {$\times$};  
  \draw (3,0) node {$\times$};
  \draw (4,0) node {$\times$};
  \draw (5,0) node {$\times$};
  \draw (2.3,0.3) node {\Mf};
  \draw(6,-1)--(6,1);
  \draw (6.3,0.5) node {\Mn};
  \draw[thick,->] (8,0)--(9,0);
  \draw (9.2,0.2) node {$x^6$};
  \draw[thick,->] (8,0)--(8,1);
  \draw (8.2,1.2) node {$x^{7,8,9,10}$};
 \end{tikzpicture}
 }
\end{align}
depending on whether the \Mf\ branes are outside the \Mn\ or inside. Here, 
$\C^2$ 
is conveniently treated as $\C^2 \slash \Z_1$, i.e.\ the $A_0$ singularity. 
Correspondingly, in the Type IIA picture there is a single \Ds\ brane.
\paragraph{Single \Mf.}
Consider a single \Mf\ near an \Mn . The phase where the \Mf\ is outside the \Mn\ has the following brane system:
\begin{align}
\raisebox{-.5\height}{
 \begin{tikzpicture}
  \draw(3,0)--(6,0);
  \draw (3,0.2) node {$A_0$};
  \draw (4.5,0) node {$\times$};
  \draw (4.5,0.3) node {\Mf};
  \draw(6,-1)--(6,1);
  \draw (6.3,0.5) node {\Mn};
  \draw[thick,->] (8,0)--(9,0);
  \draw (9.2,0.2) node {$x^6$};
  \draw[thick,->] (8,0)--(8,1);
  \draw (8.2,1.2) node {$x^{7,8,9,10}$};
 \end{tikzpicture}
 }
\end{align}

It can be described in Type IIA as follows\footnote{In the remaining brane diagrams we will omit the labels for the different branes. The brane diagrams are either M-theory or Type IIA diagrams and follow the conventions established in diagrams \eqref{branes:11d} and \eqref{branes:IIA} respectively.}:
\begin{align}
\raisebox{-.5\height}{
 \begin{tikzpicture}
		\DsixFree{0}{0}{1}{0}
		\DsixFree{1}{0}{2.1}{0}
		\ns{1,0}
		\Mnine{2}
 \end{tikzpicture} 
 }
\qquad \Leftrightarrow \qquad \text{electric quiver:} \quad
	\raisebox{-.5\height}{
 	\begin{tikzpicture}
	\tikzstyle{gauge} = [circle, draw,inner sep=3pt];
	\tikzstyle{flavour} = [regular polygon,regular polygon sides=4,inner 
sep=3pt, draw];
	\node (f1) [flavour,label=below:{$1$}] {};
	\node (f2) [flavour,right of=f1,label=below:{$1$}] {};
	\draw (f1)--(f2);
	\end{tikzpicture}
	}
	\label{eq:6d_quiver_1M5_out_M9}
\end{align}
Note that there is no choice of boundary condition involved. Since there is 
only one \Ds\ and all eight \De\ are strictly speaking on top of the \Oe\ 
orientifold, one may connect the \Ds\ to any of the \De s. In addition, the brane 
system in \eqref{eq:6d_quiver_1M5_out_M9} only displays one side of the entire 
brane content as all the mirror objects outside the \Oe\ behave identical to 
their counterparts. That being said, note that the depicted \NS s are 
technically half \NS\ branes.

Similar to above, one can move to the phase of 
the brane system where all \Ds s are suspended between 
\De s by pulling one \De\ from infinity, one obtains
\begin{align}
\raisebox{-.5\height}{
 \begin{tikzpicture}
		\Deight{0}
		\DsixFree{0}{0}{1}{0}
		\DsixFree{1}{0}{2.1}{0}
		\ns{1,1}
		\Mnine{2}
 \end{tikzpicture} 
 }
\qquad \Leftrightarrow \qquad \text{magnetic quiver:} \quad
	\raisebox{-.5\height}{
 	\begin{tikzpicture}
	\tikzstyle{gauge} = [circle, draw,inner sep=3pt];
	\tikzstyle{flavour} = [regular polygon,regular polygon sides=4,inner 
sep=3pt, draw];
	\node (g1) [gauge,label=below:{$1$}] {};
	\node (g2) [gauge,above of=g1,label=above:{$1$}] {};
	\draw (g1)--(g2);
	\end{tikzpicture}
	}
	\label{eq:quiver_1M5_out_M9}
\end{align}
Note again that one $\uo$ gauge node originates from the \Ds\ suspended between 
two \De s, while the other $\uo$ stems from the \NS\ (once again the neutral hypermultiplet 
that also corresponds to the \NS\ has been omitted in the depiction of the 
magnetic quiver, 
since it does not contribute to the 3d Coulomb branch). The relation between the 
electric and magnetic quiver is given in terms of their associated moduli 
spaces
\begin{align}
 \Higgs^{6d}\left(\substack{\text{electric} \\ \text{quiver}} 
\eqref{eq:6d_quiver_1M5_out_M9} \right)
=
\Coulomb^{3d}\left(\substack{\text{magnetic} \\ \text{quiver}} 
\eqref{eq:quiver_1M5_out_M9} \right)  = \mathbb{C}^2 = \mathbb{H}\;.
\end{align}

However, there is another phase of the $6$d system which is reached  when the 
\Mf\ 
approaches the \Mn\ plane.
In Type IIA the half \NS\ can be moved towards the \Oe\ through the \De s via 
a transition with brane creation \cite[Sec.\ 3.2]{Hanany:1997sa}. As first 
step, one moves 
the half \NS\ behind the last \De\ and takes care of brane creation as follows:
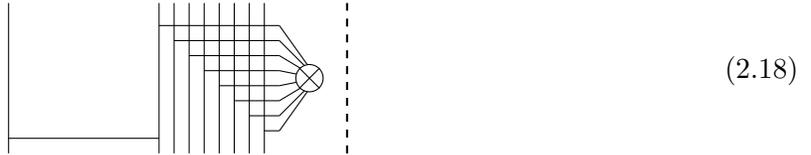
\begin{align}
\raisebox{-.5\height}{
 \begin{tikzpicture}
		\Deight{0}
		\Deight{2}
		\Deight{2.2}
		\Deight{2.4}
		\Deight{2.6}
		\Deight{2.8}
		\Deight{3}
		\Deight{3.2}
		\Deight{3.4}
		\DsixFree{0}{-0.8}{2}{-0.8}
		\DsixFree{2}{0.7}{3.6}{0.7}
		\DsixFree{2.2}{0.5}{3.6}{0.5}
		\DsixFree{2.4}{0.3}{3.6}{0.3}
		\DsixFree{2.6}{0.1}{3.6}{0.1}
		\DsixFree{2.8}{-0.1}{3.6}{-0.1}
		\DsixFree{3}{-0.3}{3.6}{-0.3}		
		\DsixFree{3.2}{-0.5}{3.6}{-0.5}
		\DsixFree{3.4}{-0.7}{3.6}{-0.7}
		\DsixFree{3.6}{0.7}{4.1}{0}
		\DsixFree{3.6}{0.5}{4.1}{0}
		\DsixFree{3.6}{0.3}{4.1}{0}
		\DsixFree{3.6}{0.1}{4.1}{0}
		\DsixFree{3.6}{-0.1}{4.1}{0}
		\DsixFree{3.6}{-0.3}{4.1}{0}		
		\DsixFree{3.6}{-0.5}{4.1}{0}
		\DsixFree{3.6}{-0.7}{4.1}{0}
		\ns{4,0}
		\Oeight{4.5}
\end{tikzpicture} 
 }
\end{align}

Next, one merges the half \NS\ on the orientifold with its mirror image, then 
splits them along the \Oe\ such that these are free to move vertically. All the newly 
created \Ds s become unfrozen and are now free to move along the vertical directions as well. 
Recalling that a \Ds\ stretched between a \De\ and its mirror image does not 
lead to a massless BPS state, the \Ds s in the last two segments closest to the 
\Oe\ need to be rearranged as follows:
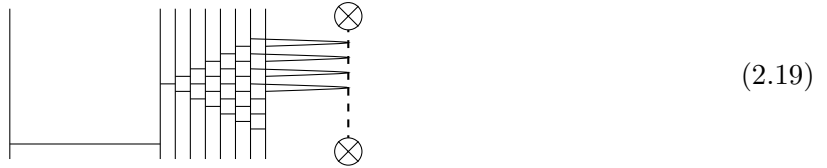
\begin{align}
\raisebox{-.5\height}{
 \begin{tikzpicture}
		\Deight{0}
		\DeightMany{8}{2}{2}
		\DsixFree{0}{-0.8}{2}{-0.8}
		\DsixFree{2}{0}{2.2}{0}
		\DsixFree{2.2}{0.1}{2.4}{0.1}
		\DsixFree{2.2}{-0.1}{2.4}{-0.1}
		\DsixFree{2.4}{0.2}{2.6}{0.2}
		\DsixFree{2.4}{0}{2.6}{0}		
		\DsixFree{2.4}{-0.2}{2.6}{-0.2}
		\DsixFree{2.6}{0.3}{2.8}{0.3}
		\DsixFree{2.6}{0.1}{2.8}{0.1}
		\DsixFree{2.6}{-0.1}{2.8}{-0.1}
		\DsixFree{2.6}{-0.3}{2.8}{-0.3}
		\DsixFree{2.8}{0.4}{3}{0.4}
		\DsixFree{2.8}{0.2}{3}{0.2}
		\DsixFree{2.8}{0}{3}{0}
		\DsixFree{2.8}{-0.2}{3}{-0.2}
		\DsixFree{2.8}{-0.4}{3}{-0.4}
		\DsixFree{3}{0.5}{3.2}{0.5}
		\DsixFree{3}{0.3}{3.2}{0.3}
		\DsixFree{3}{0.1}{3.2}{0.1}
		\DsixFree{3}{-0.1}{3.2}{-0.1}		
		\DsixFree{3}{-0.3}{3.2}{-0.3}
		\DsixFree{3}{-0.5}{3.2}{-0.5}
		\DsixFree{3.2}{-0.2}{3.4}{-0.2}		
		\DsixFree{3.2}{-0.4}{3.4}{-0.4}
		\DsixFree{3.2}{-0.6}{3.4}{-0.6}
		\draw (3.2,0.6) .. controls (4.9,0.55) .. (3.4,0.5);
		\draw (3.2,0.4) .. controls (4.9,0.35) .. (3.4,0.3);
		\draw (3.2,0.2) .. controls (4.9,0.15) .. (3.4,0.1);
		\draw (3.2,0.0) .. controls (4.9,-0.05) .. (3.4,-0.1);
		\Oeight{4.5}
		\ns{4.5,0.9}
		\ns{4.5,-0.9}		
\end{tikzpicture} 
 }
 \label{eq:1M5_in_M9}
\end{align}
In the last brane system the $8$ \Ds s in the interval between the rightmost \De\ 
and 
the \Oe\ have been connected with their mirror images. From this, one can 
read off the magnetic quiver using the rules established before
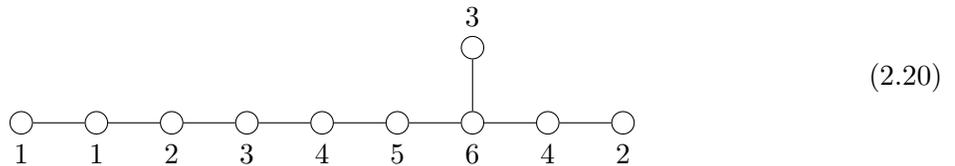
\begin{align}
\raisebox{-.5\height}{
 	\begin{tikzpicture}
	\tikzstyle{gauge} = [circle, draw,inner sep=3pt];
	\tikzstyle{flavour} = [regular polygon,regular polygon sides=4,inner 
sep=3pt, draw];
	\node (g1) [gauge,label=below:{$1$}] {};
	\node (g2) [gauge,right of=g1,label=below:{$1$}] {};
	\node (g3) [gauge,right of=g2,label=below:{$2$}] {};
	\node (g4) [gauge,right of=g3,label=below:{$3$}] {};
	\node (g5) [gauge,right of=g4,label=below:{$4$}] {};
	\node (g6) [gauge,right of=g5,label=below:{$5$}] {};
	\node (g7) [gauge,right of=g6,label=below:{$6$}] {};
	\node (g8) [gauge,right of=g7,label=below:{$4$}] {};
	\node (g9) [gauge,right of=g8,label=below:{$2$}] {};
	\node (g10) [gauge,above of=g7,label=above:{$3$}] {};
	\draw (g1)--(g2) (g2)--(g3) (g3)--(g4) (g4)--(g5) (g5)--(g6) (g6)--(g7) 
(g7)--(g8) (g8)--(g9) (g7)--(g10);
	\end{tikzpicture}
	}
	 \label{eq:quiver_1M5_in_M9}
\end{align}
This result deserves some comments. Firstly, the bifurcation in the 
magnetic quiver is a direct consequence of the brane picture 
\eqref{eq:1M5_in_M9}. 
In more detail, there is a stack of three \Ds s between the 7th and 8th \De s 
starting from the left, as 
well as a stack of four \Ds s between the 8th and the 7th \De s, but these \Ds s 
go all the way through the \Oe. By the previous arguments, the stack of three 
and four \Ds\ give rise to an $\urm(3)$ and an $\urm(4)$ magnetic vector multiplets, 
respectively, which are both connected via magnetic bifundamental 
hypermultiplets to the $\urm(6)$ gauge node from the stack of six \Ds s in 
between the 6th and 7th \De s.
Secondly, the $\urm(2)$ node at the very right of 
the quiver results from the two  half \NS\ branes that can move freely along the \Oe . 
The setting is similar to the discrete gauging argument of 
\eqref{eq:multiple_M5_discrete_gauging}: the two half \NS\ branes on the \Oe\ are 
coincident with the difference that the magnetic adjoint hypermultiplet is frozen due to the 
orientifold projection; we would like to relate this effect also to the fact that the \NS s on the \Oe\
cannot move in the $x^6$ 
direction.
The resulting $\urm(2)$ magnetic gauge node is connected via a magnetic 
bifundamental hypermultiplet due to \Df\ branes stretching between the stack of 
four \Ds s and the stuck \NS\ branes.
Thirdly, treating the magnetic quiver from 
\eqref{eq:quiver_1M5_out_M9} and 
\eqref{eq:quiver_1M5_in_M9} as 3d $\Ncal=4$ Coulomb branch quiver, one 
observes that the difference in dimension is 
$29$ and the symmetry of \eqref{eq:quiver_1M5_in_M9} is enhanced to 
$E_8$ in contrast to \eqref{eq:quiver_1M5_out_M9}.
This effect is known as small $E_8$ instanton transition, as discussed in 
\cite{Hanany:2018uhm}. 

It is important for later discussion that the same quiver can be read off from 
a different (but also maximal) subdivision of the \Ds s of the brane system
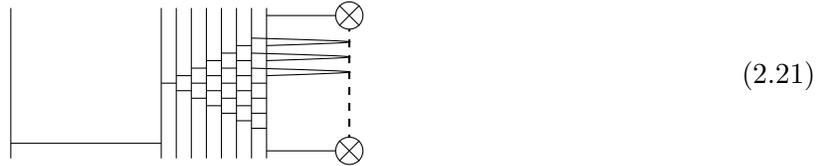
\begin{align}
\raisebox{-.5\height}{
 \begin{tikzpicture}
		\Deight{0}
		\DeightMany{8}{2}{2}
		\DsixFree{0}{-0.8}{2}{-0.8}
		\DsixFree{2}{0}{2.2}{0}
		\DsixFree{2.2}{0.1}{2.4}{0.1}
		\DsixFree{2.2}{-0.1}{2.4}{-0.1}
		\DsixFree{2.4}{0.2}{2.6}{0.2}
		\DsixFree{2.4}{0}{2.6}{0}		
		\DsixFree{2.4}{-0.2}{2.6}{-0.2}
		\DsixFree{2.6}{0.3}{2.8}{0.3}
		\DsixFree{2.6}{0.1}{2.8}{0.1}
		\DsixFree{2.6}{-0.1}{2.8}{-0.1}
		\DsixFree{2.6}{-0.3}{2.8}{-0.3}
		\DsixFree{2.8}{0.4}{3}{0.4}
		\DsixFree{2.8}{0.2}{3}{0.2}
		\DsixFree{2.8}{0}{3}{0}
		\DsixFree{2.8}{-0.2}{3}{-0.2}
		\DsixFree{2.8}{-0.4}{3}{-0.4}
		\DsixFree{3}{0.5}{3.2}{0.5}
		\DsixFree{3}{0.3}{3.2}{0.3}
		\DsixFree{3}{0.1}{3.2}{0.1}
		\DsixFree{3}{-0.1}{3.2}{-0.1}		
		\DsixFree{3}{-0.3}{3.2}{-0.3}
		\DsixFree{3}{-0.5}{3.2}{-0.5}
		\DsixFree{3.2}{-0.2}{3.4}{-0.2}		
		\DsixFree{3.2}{-0.4}{3.4}{-0.4}
		\DsixFree{3.2}{-0.6}{3.4}{-0.6}
		\DsixFree{3.2}{0}{3.4}{0}
		\draw (3.2,0.6) .. controls (4.9,0.55) .. (3.4,0.5);
		\draw (3.2,0.4) .. controls (4.9,0.35) .. (3.4,0.3);
		\draw (3.2,0.2) .. controls (4.9,0.15) .. (3.4,0.1);
		\Oeight{4.5}
		\DsixFree{3.4}{0.9}{4.5}{0.9}
		\DsixFree{3.4}{-0.9}{4.5}{-0.9}
		\ns{4.5,0.9}
		\ns{4.5,-0.9}		
\end{tikzpicture} 
 }
 \label{eq:1M5_in_M9_alternative}
\end{align}
from which one would read off
\begin{align}
\raisebox{-.5\height}{
 	\begin{tikzpicture}
	\tikzstyle{gauge} = [circle, draw,inner sep=3pt];
	\tikzstyle{flavour} = [regular polygon,regular polygon sides=4,inner 
sep=3pt, draw];
	\node (g1) [gauge,label=below:{$1$}] {};
	\node (g2) [gauge,right of=g1,label=below:{$1$}] {};
	\node (g3) [gauge,right of=g2,label=below:{$2$}] {};
	\node (g4) [gauge,right of=g3,label=below:{$3$}] {};
	\node (g5) [gauge,right of=g4,label=below:{$4$}] {};
	\node (g6) [gauge,right of=g5,label=below:{$5$}] {};
	\node (g7) [gauge,right of=g6,label=below:{$6$}] {};
	\node (g8) [gauge,above of=g7,label=left:{$4$}] {};
	\node (g9) [gauge,above of=g8,label=left:{$2$}] {};
	\node (g10) [gauge,right of=g7,label=below:{$3$}] {};
	\draw (g1)--(g2) (g2)--(g3) (g3)--(g4) (g4)--(g5) (g5)--(g6) (g6)--(g7) 
(g7)--(g8) (g8)--(g9) (g7)--(g10);
	\end{tikzpicture}
	}
	 \label{eq:quiver_1M5_in_M9_alternative}
\end{align}
i.e.\ the difference between \eqref{eq:1M5_in_M9} and 
\eqref{eq:1M5_in_M9_alternative} is that the gauge nodes after the bifurcation 
are interchanged. Moreover, the two half \NS\ branes on the \Oe\ have each a 
\Ds\ ending on them. These \Ds s do not contribute any degrees of freedom as 
they are frozen between a \De\ and a \NS . However, the \NS s still 
contribute the gauge degrees as these are free to move along the \Oe.

The point to appreciate here is that the prescription of \emph{magnetic quiver} 
is capable to produce a quiver that contains an affine $E_8$ Dynkin diagram in 
its balanced set of nodes. Therefore, the moduli naturally has an $E_8$ 
symmetry. Again, the relevant 3d $\Ncal=4$ Coulomb branch quiver has been 
proposed before \cite{Hanany:2018uhm}, but the proposal of this paper provides 
an 
explicit brane realisation. 
\paragraph{Remark.}
The two brane configurations \eqref{eq:1M5_in_M9} and 
\eqref{eq:1M5_in_M9_alternative} deserve to be commented on. At first glance, 
there are two different brane systems in which the numbers of freely moving 
\Ds\ branes are identical. The corresponding magnetic quivers differ only by an 
exchange of $(4')$---$(2')$ and $(3')$ legs, using the Dynkin labels of the 
affine $E_8$ Dynkin diagram. Therefore, the Coulomb branches of 
\eqref{eq:1M5_in_M9} and \eqref{eq:1M5_in_M9_alternative} are isomorphic.

It is not clear whether these two brane systems do hint on a geometric 
phenomenon. One possibility might be \emph{moduli spaces which are the union 
of two 
cones}, as observed in $4$d $\Ncal=2$ $\su$ gauge theory with $2$ flavours 
\cite{Seiberg:1994aj}, in 
$3$d $\Ncal=4$ $\usprm(2n)$ gauge theory with $2n$ flavours 
\cite{Ferlito:2016grh}, or in $5$d SQCD \cite{Cabrera:2018jxt}. However, the 
remainder of this paper consists of a detailed study of the brane configurations 
of type \eqref{eq:1M5_in_M9} since they provide novel insights on many of the 
physical features already presented in 
\cite{Mekareeya:2017jgc,Hanany:2018uhm,Hanany:2018vph}. 
 The possibility of a geometric significance of the two configurations 
\eqref{eq:1M5_in_M9} and 
\eqref{eq:1M5_in_M9_alternative} is interesting, but 
further analysis is required and postponed to future work.
\paragraph{Multiple \Mf\ branes on $A_0$.}
One can readily repeat the analysis for multiple \Mf s near an \Mn . There are 
multiple phases that can be realised: \Mf s outside can either be coincident 
or 
separated along the $x^6$ direction, while one may also move \Mf s into the \Mn.  Suppose there are $n$ 
\Mf\ in total from which $n_0$ moved inside the \Mn\ and from the remaining 
\Mf s there are $n_i$ coincident at position $x_i^6$, for $i=1,\ldots,l$. Of 
course $\sum_{i=0}^l n_i =n$. The relevant \emph{magnetic quiver} can be extracted from 
the 
previous arguments: each \Mf\ that moves inside the \Mn\ creates branes in the 
pattern of \eqref{eq:1M5_in_M9}.
Moreover, the coincident branes outside the \Mn\ affect 
the brane picture as in \eqref{eq:multiple_M5_discrete_gauging} for $k=1$. Hence, 
the magnetic quiver reads
\begin{align}
\raisebox{-.5\height}{
 	\begin{tikzpicture}
	\tikzstyle{gauge} = [circle, draw,inner sep=3pt];
	\tikzstyle{flavour} = [regular polygon,regular polygon sides=4,inner 
sep=3pt, draw];
	\node (g1) [gauge,label=below:{$1$}] {};
	\node (g2) [gauge,right of=g1,label=below:{$n_0$}] {};
	\node (g3) [gauge,right of=g2,label=below:{$2n_0$}] {};
	\node (g4) [gauge,right of=g3,label=below:{$3n_0$}] {};
	\node (g5) [gauge,right of=g4,label=below:{$4n_0$}] {};
	\node (g6) [gauge,right of=g5,label=below:{$5n_0$}] {};
	\node (g7) [gauge,right of=g6,label=below:{$6n_0$}] {};
	\node (g8) [gauge,right of=g7,label=below:{$4n_0$}] {};
	\node (g9) [gauge,right of=g8,label=below:{$2n_0$}] {};
	\node (g10) [gauge,above of=g7,label=above:{$3n_0$}] {};
	\node (b1) [gauge,above left of=g1, label=left:{$n_1$}] {};
	\node (b0) [above of =g1,label=below:{$\ldots$}] {} ;
	\node (b2) [gauge,above right of=g1, label=right:{$n_l$}] {};
	\draw [-] (-0.7+0.11,0.85) arc (-70:90:10pt);
	\draw [-] (-0.7-0.11,0.85) arc (250:90:10pt);
	\draw [-] (0.7+0.11,0.85) arc (-70:90:10pt);
	\draw [-] (0.7-0.11,0.85) arc (250:90:10pt);
	\draw (g1)--(g2) (g2)--(g3) (g3)--(g4) (g4)--(g5) (g5)--(g6) (g6)--(g7) 
(g7)--(g8) (g8)--(g9) (g7)--(g10) (g1)--(b1) (g1)--(b2);
	\end{tikzpicture}
	}
	 \label{eq:quiver_multiple_M5_in_M9}
\end{align}
Again, the symmetry contains and $E_8$ factor which is recognised by the 
pattern of balanced nodes. Consequently, the magnetic quivers provide a 
systematic description for all 
(weakly and strongly coupled) phases of the 6d Higgs branch.
%
%
%
\subsection{Derivation rules}
\label{sec:rules}
Following the discussion of Sections 
\ref{sec:discrete_gauging}--\ref{sec:small_E8}, the procedure for deriving the 
magnetic quiver can be formalised by a few rules.

\begin{myConj}[Magnetic quiver]
For a \Ds-\De-\NS\ brane system, cf.\ Table \ref{tab:directions}, in which all 
\Ds\ branes are suspended between \De\ branes, the massless BPS states, deduced 
from stretching virtual \Df\ branes, arise from the following configurations:
\begin{compactenum}[(i)]
 \item Stack of $m$ \Ds\ branes suspended between two \De s in a finite $x^6$ 
interval:  the vertical motion along the $x^7$, $x^8$, $x^9$ directions gives rise to a 
$\urm(m)$ magnetic vector multiplet due to \Df s stretched between them. 
\begin{align}
\raisebox{-.5\height}{
	\begin{tikzpicture}
	\tikzstyle{gauge} = [circle, draw,inner sep=3pt];
		\Deight{0}
		\Deight{1}
		\DsixFree{0}{0.4}{1}{0.4}
		\DsixFree{0}{0.2}{1}{0.2}
		\draw (.5,-0.1) node {$\vdots$};
		\DsixFree{0}{-0.4}{1}{-0.4}
		\draw[decoration={brace,raise=3pt},decorate,thick](0,-0.5) -- 
node[left=6pt] {$m$ \Ds } (0,0.5);
		\ArrowMagnetic{3}{0}
		\node (g) at (6,0) [gauge,label=below:{$m$}] {};
  		\draw (0,-1.3) node {\De };
	\end{tikzpicture}
	}
	\label{eq:rule_magQuiv_1}
\end{align}
%
\item Stacks of $m$ \Ds\ and $l$ \Ds\ branes in adjacent \De\ intervals along the 
$x^6$ direction: the \Df\ branes suspended between \Ds s of different intervals induce a 
magnetic bifundamental hypermultiplet of $\urm(m)\times \urm(l)$.
\begin{align}
\raisebox{-.5\height}{
	\begin{tikzpicture}
	\tikzstyle{gauge} = [circle, draw,inner sep=3pt];
		\Deight{0}
		\Deight{1}
		\Deight{2}
  		\draw (0,-1.3) node {\De };
		\DsixFree{0}{0.4}{1}{0.4}
		\DsixFree{0}{0.2}{1}{0.2}
		\draw (.5,-0.1) node {$\vdots$};
		\DsixFree{0}{-0.4}{1}{-0.4}
		\draw[decoration={brace,raise=3pt},decorate,thick](0,-0.5) -- 
node[left=6pt] {$m$ \Ds } (0,0.5);
		\DsixFree{1}{0.5}{2}{0.5}
		\DsixFree{1}{0.3}{2}{0.3}
		\draw (1.5,-0.1) node {$\vdots$};
		\DsixFree{1}{-0.5}{2}{-0.5}	
\draw[decoration={brace,raise=3pt,mirror},decorate,thick](2,-0.6) -- 
node[right=6pt] {$l$ \Ds } (2,0.6);
		\ArrowMagnetic{5}{0}
		\node (g1) at (8,0) [gauge,label=below:{$m$}] {};
		\node (g2) at (9,0) [gauge,label=below:{$l$}] {};
		\draw (g1)--(g2);
	\end{tikzpicture}
		}
	\label{eq:rule_magQuiv_2}
\end{align}
%
\item Stack of $m$ \NS\ branes at coincident $x^6$ position: the vertical 
motion along the $x^7$, $x^8$, $x^9$ 
directions gives rise to a 
$\urm(m)$ magnetic vector multiplet due to \Df s stretched between.
If the \NS s are free to move along the $x^6$ direction, there is an additional 
 hypermultiplet in 
the adjoint representation of $\urm(m)$ (this is in contrast to the \NS s being stacked at the \Oe\ plane, where there is no adjoint hypermultiplet in the magnetic quiver).
\begin{align}
\raisebox{-.5\height}{
	\begin{tikzpicture}
	\tikzstyle{gauge} = [circle, draw,inner sep=3pt];
		\Deight{-1}
		\Deight{2}
		\ns{1,0.7}
		\ns{1,0.3}
		\ns{1,-0.7}
		\draw (1,-0.2) node {$\vdots$};
		\draw[decoration={brace,raise=3pt},decorate,thick](.9,-.9) -- 
node[left=6pt] {$m$ \NS } (.9,.9);
		\ArrowMagnetic{4}{0}
		\node (g) at (7,0) [gauge,label=below:{$m$}] {};
	\draw [-] (7+0.11,0.105) arc (-70:90:10pt);
	\draw [-] (7+-0.11,0.105) arc (250:90:10pt);
  		\draw (-1,-1.3) node {\De };
	\end{tikzpicture}
	}
	\label{eq:rule_magQuiv_3}
\end{align}
%
\item Stacks of $l$ \Ds\ and $m$ \NS\ branes between two \De\ in a finite $x^6$ 
interval: the vertical distance in the $x^7$, $x^8$, $x^9$ directions leads to a 
magnetic bifundamental hypermultiplet of $\urm(l)\times \urm(m)$. 
\begin{align}
\raisebox{-.5\height}{
	\begin{tikzpicture}
	\tikzstyle{gauge} = [circle, draw,inner sep=3pt];
		\DeightLong{-1}
		\DeightLong{2}
		\ns{1,1.3+0.7}
		\ns{1,1.3+0.3}
		\ns{1,1.3-0.7}
		\draw (1,1.3-0.2) node {$\vdots$};
		\draw[decoration={brace,raise=3pt},decorate,thick](.9,1.3-.9) -- 
node[left=6pt] {$m$ \NS } (.9,1.3+.9);
		\DsixFree{-1}{-.5+0.4}{2}{-.5+0.4}
		\DsixFree{-1}{-.5+0.2}{2}{-.5+0.2}
		\draw (.5,-.5-0.1) node {$\vdots$};
		\DsixFree{-1}{-.5-0.4}{2}{-.5-0.4}
		\draw[decoration={brace,raise=3pt},decorate,thick](-1,-1) -- 
node[left=6pt] {$l$ \Ds } (-1,0);
		\ArrowMagnetic{4}{0}
		\node (g1) at (7,0) [gauge,label=below:{$l$}] {};
		\node (g2) at (7,1) [gauge,label=right:{$m$}] {};
	\draw [-] (7+0.11,1.105) arc (-70:90:10pt);
	\draw [-] (7+-0.11,1.105) arc (250:90:10pt);
		\draw (g1)--(g2);
  		\draw (2,-1.3) node {\De };
	\end{tikzpicture}
		}
	\label{eq:rule_magQuiv_4}
\end{align}
%
\end{compactenum}
The massless degrees of freedom can be encoded in a quiver diagram in the 
familiar way.
\label{conj:magnetic_quiver}
\end{myConj}
%
%
\subsection{Phases and their geometry}
\label{sec:geometry}
The two fundamental cases of Section \ref{sec:discrete_gauging} and 
\ref{sec:small_E8} are sufficient to treat all cases of $n$ \Mf\ branes near 
an \Mn\ plane on $\C^2 \slash \Z_k$, provided the embeddings $\Z_k 
\hookrightarrow 
E_8$ are known, see Section \ref{sec:embedding_E8}.
Before proceeding to the general case, some remarks are in order.

Firstly, two Higgs branches $\Higgs_{1,2}^{6d}$ which are related via discrete 
gauging of a discrete $S_l$ permutation group satisfy
\begin{align}
 \Higgs_{1}^{6d}= \Coulomb_1^{3d} = \Coulomb_2^{3d}\slash S_l = 
\Higgs_2^{6d}\slash S_l \qquad \Rightarrow \qquad \Rcal(\Higgs_{1}^{6d}) 
\subset 
\Rcal(\Higgs_{2}^{6d}) \,,
\end{align}
where $\Rcal$ denotes the associated chiral rings. Note both moduli spaces 
have the same dimension as only a discrete group has been 
gauged in the electric theory or quotient by in the magnetic theory 
\cite{Hanany:2018vph,Hanany:2018cgo,Hanany:2018dvd}. However, the inclusion 
holds only on the space of functions or, equivalently, the space of protected 
operators.

Secondly, two Higgs branches $\Higgs_{1,2}^{6d}$ which are related via a small 
$E_8$ instanton transition satisfy the following \emph{quiver subtraction} 
relation on their magnetic quivers
\begin{align}
 \Higgs^{6d}_i &= \Coulomb^{3d}\left(\substack{\text{magnetic} \\ \text{quiver} 
}\, \magQuiv_i\right)  \,, i=1,2\, , \qquad 
\overline{\Ocal}_{\text{min}}^{E_8} = \Coulomb^{3d} \left( \magQuiv_{E_8} 
\right) \\
&\Rightarrow \qquad \magQuiv_1 - \magQuiv_2 \ = \magQuiv_{E_8} ,.
\end{align}
Note in particular that $ \Higgs^{6d}_2 \subset  \Higgs^{6d}_1$ and that the 
transverse slice of $ \Higgs^{6d}_2$ inside $ \Higgs^{6d}_1$ is given by the 
closure of the minimal nilpotent orbit $\overline{\Ocal}_{\text{min}}^{E_8}$ of 
$E_8$. As simplest example, consider the quivers \eqref{eq:quiver_1M5_out_M9}, 
\eqref{eq:quiver_1M5_in_M9} and perform the quiver subtraction 
\cite{Cabrera:2018ann,Hanany:2018uhm} as follows:
\begin{equation}
\begin{aligned}
 &\raisebox{-.5\height}{
 	\begin{tikzpicture}
	\tikzstyle{gauge} = [circle, draw,inner sep=3pt];
	\tikzstyle{flavour} = [regular polygon,regular polygon sides=4,inner 
sep=3pt, draw];
	\node (g1) [gauge,label=below:{$1$}] {};
	\node (g2) [gauge,right of=g1,label=below:{$1$}] {};
	\node (g3) [gauge,right of=g2,label=below:{$2$}] {};
	\node (g4) [gauge,right of=g3,label=below:{$3$}] {};
	\node (g5) [gauge,right of=g4,label=below:{$4$}] {};
	\node (g6) [gauge,right of=g5,label=below:{$5$}] {};
	\node (g7) [gauge,right of=g6,label=below:{$6$}] {};
	\node (g8) [gauge,right of=g7,label=below:{$4$}] {};
	\node (g9) [gauge,right of=g8,label=below:{$2$}] {};
	\node (g10) [gauge,above of=g7,label=above:{$3$}] {};
	\draw (g1)--(g2) (g2)--(g3) (g3)--(g4) (g4)--(g5) (g5)--(g6) (g6)--(g7) 
(g7)--(g8) (g8)--(g9) (g7)--(g10);
	\end{tikzpicture}
	} 
	\\
	-& 
	\raisebox{-.5\height}{
 	\begin{tikzpicture}
	\tikzstyle{gauge} = [circle, draw,inner sep=3pt];
	\tikzstyle{flavour} = [regular polygon,regular polygon sides=4,inner 
sep=3pt, draw];
	\node (g1) [gauge,label=below:{$1$}] {};
	\node (g2) [gauge,above of=g1,label=above:{$1$}] {};
	\draw (g1)--(g2);
	\end{tikzpicture}
	} \\
	=&\quad \qquad
	\raisebox{-.5\height}{
 	\begin{tikzpicture}
	\tikzstyle{gauge} = [circle, draw,inner sep=3pt];
	\tikzstyle{flavour} = [regular polygon,regular polygon sides=4,inner 
sep=3pt, draw];
	\node (g2) [gauge,label=below:{$1$}] {};
	\node (g3) [gauge,right of=g2,label=below:{$2$}] {};
	\node (g4) [gauge,right of=g3,label=below:{$3$}] {};
	\node (g5) [gauge,right of=g4,label=below:{$4$}] {};
	\node (g6) [gauge,right of=g5,label=below:{$5$}] {};
	\node (g7) [gauge,right of=g6,label=below:{$6$}] {};
	\node (g8) [gauge,right of=g7,label=below:{$4$}] {};
	\node (g9) [gauge,right of=g8,label=below:{$2$}] {};
	\node (g10) [gauge,above of=g7,label=above:{$3$}] {};
	\draw (g2)--(g3) (g3)--(g4) (g4)--(g5) (g5)--(g6) (g6)--(g7) 
(g7)--(g8) (g8)--(g9) (g7)--(g10);
	\end{tikzpicture}
	}
\end{aligned}
\end{equation}
such that
\begin{align}
 \Higgs^{6d}_1 = \Coulomb^{3d}\left(\substack{\text{magnetic} \\ \text{quiver} 
}\, \eqref{eq:quiver_1M5_out_M9}\right) = \mathbb{H}
\; \subset \;
 \Higgs^{6d}_2 = \Coulomb^{3d}\left(\substack{\text{magnetic} \\ \text{quiver} 
}\, \eqref{eq:quiver_1M5_in_M9}\right) = \mathbb{H} \times 
\clorbit{\mathrm{min}}^{E_8}
\end{align}
and the transverse slice $\clorbit{\mathrm{min}}^{E_8}$ becomes apparent in 
this example.

This highlights and clarifies the interplay between the size of the 
$\uo$-bouquet and the $E_8$ transition, because it originates from motions of 
\NS\ branes in the brane configuration. Moreover, all phase transitions that 
were originally derived from the brane setting can be equally well understood 
from operations on the magnetic quivers.

  \section{Multiple \Mf\ branes near an \Mn\ plane on 
\texorpdfstring{$A_{k-1}$}{A(k-1)} singularity}
\label{sec:body}
After establishing the usefulness of magnetic quivers and the phase of the Type 
IIA brane setup in which all \Ds s are suspended between \De\ branes, the 
generic case of \Mf\ branes near an \Mn\ plane on a $\C^2 \slash \Z_k$ 
singularity can be approached. The arising difficulty is the need to specify 
the embedding of $\Z_k$ into the $E_8$ symmetry of the end-of-the-world \Mn\ 
or, put differently, to assign boundary conditions of the \Ds\ on the \De\ 
branes in the Type IIA or Type I${}^\prime$ set-up.
\subsection{Embedding of \texorpdfstring{$\Z_{k}$}{Zk} into 
\texorpdfstring{$E_8$}{E8} }
\label{sec:embedding_E8}
Following \cite{Kac:1994,Heckman:2015bfa}, the embedding of $\Z_k 
\hookrightarrow 
E_8$ can be labelled by non-negative integer fluxes
\begin{align}
 \begin{pmatrix}
   &  &  &  &  & m'_3 &  &   \\
  m_1 & m_2 & m_3 & m_4 & m_5 & m_6 & m'_4 & m'_2
 \end{pmatrix} \; .
 \label{eq:fluxes}
\end{align}
Using the Dynkin labels $a_i$ of affine $E_8$
\begin{align}
 \raisebox{-.5\height}{
 	\begin{tikzpicture}
	\tikzstyle{gauge} = [circle, draw,inner sep=3pt];
	\tikzstyle{flavour} = [regular polygon,regular polygon sides=4,inner 
sep=3pt, draw];
	\node (g1) [gauge] {$1$};
	\node (g2) [gauge,right of=g1] {$2$};
	\node (g3) [gauge,right of=g2] {$3$};
	\node (g4) [gauge,right of=g3] {$4$};
	\node (g5) [gauge,right of=g4] {$5$};
	\node (g6) [gauge,right of=g5] {$6$};
	\node (g7) [gauge,right of=g6] {$4$};
	\node (g8) [gauge,right of=g7] {$2$};
	\node (g9) [gauge,above of=g6] {$3$};
	\draw (g1)--(g2) (g2)--(g3) (g3)--(g4) (g4)--(g5) (g5)--(g6) (g6)--(g7) 
(g7)--(g8) (g6)--(g9);
	\end{tikzpicture}
	} \; ,
\end{align}
the fluxes determine the order $k$ of the $A_{k-1}$ singularity via 
\begin{align}
 k=\sum_{i=1}^6 a_i m_i + \sum_{i=2,3,4} a'_i m'_i \; ,
 \end{align}
such that $a_i = i$ and $a'_i = i$.
The particular choice of embedding has an immediate physical 
consequence on the $6$d theory: the commutant of the image of 
$\Z_k$ inside $E_8$ is isomorphic to the global symmetry. In fact, the 
commutant can be read off from the affine $E_8$ Dynkin diagram by deleting the 
nodes that take non-trivial flux \eqref{eq:fluxes}.

Most of the considerations will be within the Type IIA framework; hence, it is 
useful to reformulate the embeddings $\Z_k \hookrightarrow E_8$ via 
partitions $\vec{\lambda} = (\lambda_1,\ldots,\lambda_9)$ which determine the 
boundary conditions of the $k$ \Ds\ ending on the $8$ \De\ branes on top of the 
\Oe\ orientifold. The following choice is useful:
\begin{equation}
\begin{aligned}
 m_i &= \lambda_i -\lambda_{i+1} \; , \qquad \text{for } i=1,\ldots, 6 \qquad 
\text{and} \\
 m'_3 &= \lambda_7 + \lambda_8 \; , \qquad 
 m'_4= \lambda_7 - \lambda_8 \; , \qquad 
 m'_2= \lambda_8 - \lambda_9 \; .
\end{aligned} 
\label{eq:fluxes_via_lambda}
\end{equation}
Note that this parametrisation suggests that $m_i$ together with $m'_3$ and 
$m'_4$ form the simple roots of $\sorm(16)$, which is natural since there are 
$8$ \De\ present.
Similarly, the $m_i$ together with $m'_4$ and $m'_2$ furnish the simple roots 
of $\surm(9)$.
Alternatively, one can think in terms of linking numbers $l_i$ for the $i$-th 
\De\ branes, which are defined as 
\begin{align}
 l_i \coloneqq \#\{\text{D6 ending from the left}\} 
 -  \#\{\text{D6 ending from the right}\}
 +\#\{\text{NS5 to the right}\}
\end{align}
such that $l_i = \lambda_i$ for all $i$.
As a comment, if the $\lambda_i$ are such that only the first eight linking 
numbers are non-negative, then Type IIA with 8 \De\ branes is the useful 
setting. If, 
however, all nine linking numbers are non-negative then
Type I${}^\prime$ with 9 \De\ branes becomes convenient.

Given the embedding $\Z_k \hookrightarrow E_8$, one can now consider the first 
few cases and finally present the general result.
%
%
\subsection{Case \texorpdfstring{$k=1$}{k=1}}
There is only one possibility: $m_1=1$ and all other fluxes \eqref{eq:fluxes} 
vanish. The linking numbers are $(1,0^7)$ and the Type IIA setting has already 
been discussed in Section \ref{sec:small_E8}.
%
%
\subsection{Case \texorpdfstring{$k=2$}{k=2}}
There exist three possibilities, cf.\ \cite[Sec.\ 5.1]{Mekareeya:2017jgc}:
\begin{align}
m_1=2\; ,  \qquad 
m_2=1 \;, \quad \text{or} \quad
m'_2=1 \;,
\end{align}
which will be discussed in turn below. The details of the discrete gauging phase 
transitions or small $E_8$ instanton transition will not be spelled out, as 
these are straight forward operations on 
the brane picture that manifest themselves either as local operations on the 
associated bouquet or as inverse operations of quiver subtraction.
\subsubsection{Symmetry \texorpdfstring{$\surm(2)\times E_8$}{SU(2)xE8} --- 
case \texorpdfstring{$m_1=2$}{m1=2}}
\label{sec:k=2_m1=2}
The linking numbers read $l= (2,0^7)$ and the Type IIA brane system is given by
\begin{align}
 \raisebox{-.5\height}{
 \begin{tikzpicture}
      \DsixSome{2}{0}
      \DsixSome{2}{1}
      \ns{1,0}
      \DsixEmpty{2}
      \ns{2,0}
      \DsixSome{2}{3}
      \ns{3,0}
      \DsixSome{2}{4}
      \ns{4,0}
      \DsixSome{2}{5}
      \ns{5,0}
      \DeightMany{8}{6}{1}
      \Oeight{7}
 \end{tikzpicture} 
 }
 \\
  \raisebox{-.5\height}{
  \begin{tikzpicture}
      \DsixSome{2}{0}
      \DsixSome{2}{1}
      \ns{1,0}
      \DsixEmpty{2}
      \ns{2,0}
      \DsixSome{2}{3}
      \ns{3,0}
      \Deight{3.5}
      \DsixSome{1}{4}
      \ns{4,0}
      \ns{5,0}
      \DeightMany{7}{6.1}{1}
      \Oeight{7}
 \end{tikzpicture}
 }
\end{align}
which gives rise to the electric quiver
\begin{align}
 \raisebox{-.5\height}{
 	\begin{tikzpicture}
	\tikzstyle{gauge} = [circle, draw,inner sep=3pt];
	\tikzstyle{flavour} = [regular polygon,regular polygon sides=4,inner 
sep=3pt, draw];
	\node (g1) [gauge,label={[rotate=-45]below right:{$\su$}}] {};
	\node (g2) [right of=g1] {$\ldots$};
	\node (g3) [gauge,right of=g2,label={[rotate=-45]below right:{$\su$}}] 
{};
	\node (g4) [gauge,right of=g3,label={[rotate=-45]below right:{$\urm(1)$}}] 
{};
	\node (f1) [flavour,left of=g1,label=below:{$2$}] {};
	\node (f3) [flavour,above of=g3,label=above:{$1$}] {};
	\draw (g1)--(g2) (g2)--(g3) (g3)--(g4) (g1)--(f1) (g3)--(f3);
	\draw[decoration={brace,mirror,raise=20pt},decorate,thick]
  (-0.25,-0.6) -- node[below=20pt] {$n-2$} (2.75,-0.6);
	\end{tikzpicture}
	} 
	\qquad \Rightarrow \qquad
\raisebox{-.5\height}{
 	\begin{tikzpicture}
	\tikzstyle{gauge} = [circle, draw,inner sep=3pt];
	\tikzstyle{flavour} = [regular polygon,regular polygon sides=4,inner 
sep=3pt, draw];
	\node (g1) [gauge,label={[rotate=-45]below right:{$\su$}}] {};
	\node (g2) [right of=g1] {$\ldots$};
	\node (g3) [gauge,right of=g2,label={[rotate=-45]below right:{$\su$}}] {};
	\node (f1) [flavour,left of=g1,label=below:{$2$}] {};
	\node (f3) [flavour,above of=g3,label=above:{$2$}] {};
	\draw (g1)--(g2) (g2)--(g3) (g1)--(f1) (g3)--(f3);
	\draw[decoration={brace,mirror,raise=20pt},decorate,thick]
  (-0.25,-0.6) -- node[below=20pt] {$n-2$} (2.75,-0.6);
	\end{tikzpicture}
	} 
	\; .
\end{align}
The last step is needed as there is no $\uo$ gauge symmetry in $6$ dimensions.
The 
resulting $6$d quiver gauge theory has flavour symmetry and Higgs 
branch dimension given by
\begin{align}
 G_F= (\sorm(4))^2 \times \usprm(2)^{n-3} \cong \surm(2)^{n+1} 
 \qquad 
 \text{and}
 \qquad
  \dim\,\Higgs^{6d} &=n+2 \,.
\end{align}
Likewise, one may change to the brane system in which all \Ds s are suspended 
between \De\ branes and reads off the magnetic quiver
\begin{align}
   \raisebox{-.5\height}{
  \begin{tikzpicture}
  \DeightMany{2}{-0.2}{2}
      \DsixFree{-0.2}{0}{0}{0}
      \DsixSome{2}{0}
      \DsixSome{2}{1}
      \DsixSome{2}{2}
      \ns{1,1}
      \ns{1.5,1}
      \draw (2,1) node {$\ldots$};
      \ns{2.5,1}
    \DeightMany{8}{3}{1}
      \Oeight{4}
 \end{tikzpicture}
 }
 \qquad \Leftrightarrow \qquad
  \raisebox{-.5\height}{
 	\begin{tikzpicture}
	\tikzstyle{gauge} = [circle, draw,inner sep=3pt];
	\tikzstyle{flavour} = [regular polygon,regular polygon sides=4,inner 
sep=3pt, draw];
	\node (g1) [gauge,label=below:{$1$}] {};
	\node (g2) [gauge,right of=g1,label=below:{$2$}] {};
	\node (b1) [gauge,above left of=g2,label=left:{$1$}] {};
	\node (b0) [above of=g2,label=below:{$\ldots$},label=above:{$n$}] {} ;
	\node (b2) [gauge,above right of=g2,label=right:{$1$}] {};
	\draw (g1)--(g2) (g2)--(b1) (g2)--(b2);
	\end{tikzpicture}
	} 
\end{align}
where the  topological symmetry and Coulomb branch dimension of the magnetic 
quiver are
\begin{align}
 G_J = \surm(2)^{n+1}
 \qquad 
 \text{and}
 \qquad 
 \dim \,\Coulomb^{3d} =n+2 \,.
\end{align}
The electric and magnetic quiver for the weakly coupled phase are related via 
their associated moduli spaces: $\Higgs^{6d}(\text{electric 
quiver}) = 
\Coulomb^{3d}(\text{magnetic quiver})$. 
\subsubsection{Symmetry \texorpdfstring{$\surm(2)\times (E_7\times 
\uo)$}{SU(2)xE7xU(1)} 
--- 
case \texorpdfstring{$m_2=1$}{m2=1}}
\label{sec:k=2_m2=1}
The linking numbers read $l= (1^2,0^6)$ and the Type IIA brane system is given 
by
\begin{subequations}
\begin{align}
 \raisebox{-.5\height}{
 \begin{tikzpicture}
      \DsixSome{2}{0}
      \DsixSome{2}{1}
      \ns{1,0}
      \DsixEmpty{2}
      \ns{2,0}
      \DsixSome{2}{3}
      \ns{3,0}
      \DsixSome{2}{4}
      \ns{4,0}
      \DsixSome{2}{5}
      \ns{5,0}
      \DsixFree{6}{-0.05}{6.2}{-0.05}
      \DeightMany{8}{6}{2}
      \Oeight{7.5}
 \end{tikzpicture} 
 }
 \\
  \raisebox{-.5\height}{
  \begin{tikzpicture}
      \DsixSome{2}{0}
      \DsixSome{2}{1}
      \ns{1,0}
      \DsixEmpty{2}
      \ns{2,0}
      \DsixSome{2}{3}
      \ns{3,0}
      \DeightMany{2}{4.4}{2}
      \DsixSome{2}{4}
      \ns{4,0}
      \ns{5,0}
      \DeightMany{6}{6.4}{2}
      \Oeight{7.5}
 \end{tikzpicture}
 }
\end{align}
\end{subequations}
from which one can read off the electric quiver
\begin{align}
 \raisebox{-.5\height}{
 	\begin{tikzpicture}
	\tikzstyle{gauge} = [circle, draw,inner sep=3pt];
	\tikzstyle{flavour} = [regular polygon,regular polygon sides=4,inner 
sep=3pt, draw];
	\node (g1) [gauge,label={[rotate=-45]below right:{$\su$}}] {};
	\node (g2) [right of=g1] {$\ldots$};
	\node (g3) [gauge,right of=g2,label={[rotate=-45]below right:{$\su$}}] {};
	\node (g4) [gauge,right of=g3,label={[rotate=-45]below right:{$\su$}}] {};
	\node (f1) [flavour,left of=g1,label=below:{$2$}] {};
	\node (f4) [flavour,above of=g4,label=above:{$2$}] {};
	\draw (g1)--(g2) (g2)--(g3) (g3)--(g4) (g1)--(f1) (g4)--(f4);
	\draw[decoration={brace,mirror,raise=20pt},decorate,thick]
  (-0.25,-0.6) -- node[below=20pt] {$n-2$} (2.75,-0.6);
	\end{tikzpicture}
	}  \; ,
\end{align}
with flavour symmetry and Higgs branch dimension
\begin{align}
 G_F = \sorm(4)^2 \times \usprm(2)^{n-2} \cong \surm(2)^{n+2}
 \qquad 
 \text{and}
 \qquad
 \dim \, \Higgs^{6d} = n+3
 \,.
\end{align}
Likewise, one may change to the phase of the brane system which yields the 
magnetic quiver
\begin{align}
   \raisebox{-.5\height}{
  \begin{tikzpicture}
  \DeightMany{2}{-0.2}{2}
      \DsixFree{-0.2}{0}{0}{0}
      \DsixSome{2}{0}
      \DsixSome{2}{1}
      \DsixSome{2}{2}
      \ns{1,1}
      \ns{1.5,1}
      \draw (2,1) node {$\ldots$};
      \ns{2.5,1}
      \DeightMany{8}{3}{2}
      \DsixFree{3}{0}{3.2}{0}
      \Oeight{4.5}
 \end{tikzpicture}
 }
 \qquad \Leftrightarrow \qquad
  \raisebox{-.5\height}{
 	\begin{tikzpicture}
	\tikzstyle{gauge} = [circle, draw,inner sep=3pt];
	\tikzstyle{flavour} = [regular polygon,regular polygon sides=4,inner 
sep=3pt, draw];
	\node (g1) [gauge,label=below:{$1$}] {};
	\node (g2) [gauge,right of=g1,label=below:{$2$}] {};
	\node (g3) [gauge,right of=g2,label=below:{$1$}] {};
	\node (b1) [gauge,above left of=g2,label=left:{$1$}] {};
	\node (b0) [above of=g2,label=below:{$\ldots$},label=above:{$n$}] {} ;
	\node (b2) [gauge,above right of=g2,label=right:{$1$}] {};
	\draw (g1)--(g2) (g2)--(g3) (g2)--(b1) (g2)--(b2);
	\end{tikzpicture}
	} 
\end{align}
where the topological symmetry and Coulomb branch dimension read
\begin{align}
 G_J = \surm(2)^{n+2}
 \qquad 
 \text{and}
 \qquad
 \dim \, \Coulomb^{3d} = n+3
 \,.
\end{align}
Again, the electric and magnetic quiver for the weakly coupled phase are 
related via $\Higgs^{6d}(\text{electric 
quiver}) = 
\Coulomb^{3d}(\text{magnetic quiver})$.
\subsubsection{Symmetry \texorpdfstring{$\surm(2)\times 
\sorm(16)$}{SU(2)xSO(16)} --- 
case \texorpdfstring{$m'_2=1$}{mp2=1}}
\label{sec:k=2_mp2=1}
%
The linking numbers read $l= (0^8)$ and the Type 
IIA brane system is given by
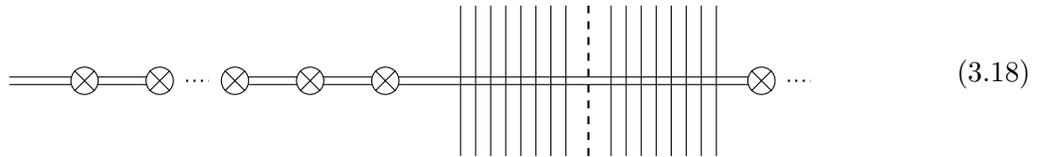
\begin{align}
 \raisebox{-.5\height}{
 \begin{tikzpicture}
      \DsixSome{2}{0}
      \DsixSome{2}{1}
      \ns{1,0}
      \DsixEmpty{2}
      \ns{2,0}
      \DsixSome{2}{3}
      \ns{3,0}
      \DsixSome{2}{4}
      \ns{4,0}
      \DsixSome{2}{5}
      \ns{5,0}
      \DeightMany{8}{6}{2}
      \Oeight{7.7}
      \DeightMany{8}{8}{2}
      \DsixSome{2}{6}
      \DsixSome{2}{7}
      \DsixSome{2}{8}
      \DsixSome{2}{9}
      \DsixEmpty{10}
      \ns{10,0}
 \end{tikzpicture} 
 }
\end{align}
which gives rise to the electric quiver
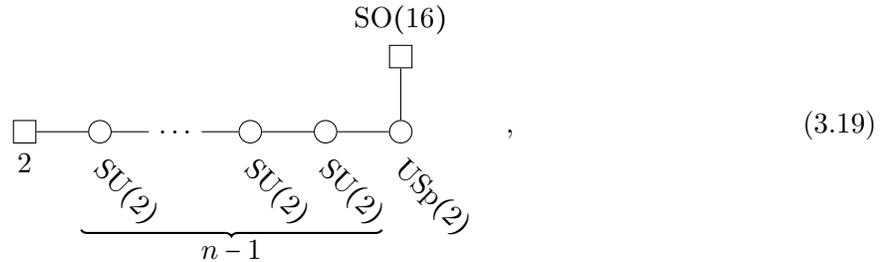
\begin{align}
 \raisebox{-.5\height}{
 	\begin{tikzpicture}
	\tikzstyle{gauge} = [circle, draw,inner sep=3pt];
	\tikzstyle{flavour} = [regular polygon,regular polygon sides=4,inner 
sep=3pt, draw];
	\node (g1) [gauge,label={[rotate=-45]below right:{$\su$}}] {};
	\node (g2) [right of=g1] {$\ldots$};
	\node (g3) [gauge,right of=g2,label={[rotate=-45]below right:{$\su$}}] {};
	\node (g4) [gauge,right of=g3,label={[rotate=-45]below right:{$\su$}}] {};
	\node (g5) [gauge,right of=g4,label={[rotate=-45]below right:{$\usprm(2)$}}] 
{};
	\node (f1) [flavour,left of=g1,label=below:{$2$}] {};
	\node (f5) [flavour,above of=g5,label=above:{$\sorm(16)$}] {};
	\draw (g1)--(g2) (g2)--(g3) (g3)--(g4) (g1)--(f1) (g4)--(g5) (g5)--(f5);
  \draw[decoration={brace,mirror,raise=20pt},decorate,thick]
  (-0.25,-0.6) -- node[below=20pt] {$n-1$} (3.75,-0.6);
	\end{tikzpicture}
	} \; ,
\end{align}
with flavour symmetry and Higgs branch dimension
\begin{align}
  G_F = \sorm(4) \times \usprm(2)^{n-1} \times \sorm(16)\cong \surm(2)^{n+1} 
\times \sorm(16)
  \qquad 
  \text{and}
  \qquad
  \dim\, \Higgs^{6d} =n+16
  \; .
\end{align}
Note that the theory is anomaly free as $\usprm(N_c)$ is equipped with $N_f 
=N_c +8$ flavours.
As before, one may change to the following phase of the brane system 
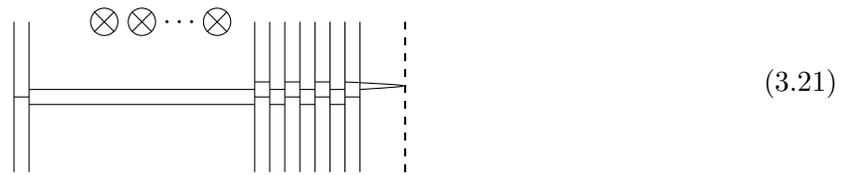
\begin{align}
   \raisebox{-.5\height}{
  \begin{tikzpicture}
  \DeightMany{2}{-0.2}{2}
      \DsixFree{-0.2}{0}{0}{0}
      \DsixSomeMore{2}{0}{3}{-0.15}
      \ns{1,1}
      \ns{1.5,1}
      \draw (2,1) node {$\ldots$};
      \ns{2.5,1}
      \DeightMany{8}{3}{2}
      \Oeight{5}
      \DsixSomeMore{2}{3}{0.2}{-0.05}
      \DsixSomeMore{2}{3.2}{0.2}{-0.15}
      \DsixSomeMore{2}{3.4}{0.2}{-0.05}
      \DsixSomeMore{2}{3.6}{0.2}{-0.15}
      \DsixSomeMore{2}{3.8}{0.2}{-0.05}
      \DsixSomeMore{2}{4.0}{0.2}{-0.15}     
      \draw (4.2,0.2) .. controls (5.2,0.15) .. (4.4,0.1);
      \DsixFree{4.2}{0}{4.4}{0}
 \end{tikzpicture}
 } 
 \end{align}
which yields the magnetic quiver
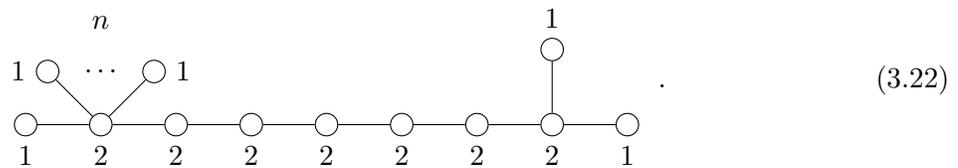
\begin{align} 
  \raisebox{-.5\height}{
 	\begin{tikzpicture}
	\tikzstyle{gauge} = [circle, draw,inner sep=3pt];
	\tikzstyle{flavour} = [regular polygon,regular polygon sides=4,inner 
sep=3pt, draw];
	\node (g1) [gauge,label=below:{$1$}] {};
	\node (g2) [gauge,right of=g1,label=below:{$2$}] {};
	\node (g3) [gauge,right of=g2,label=below:{$2$}] {};
	\node (g4) [gauge,right of=g3,label=below:{$2$}] {};
	\node (g5) [gauge,right of=g4,label=below:{$2$}] {};
	\node (g6) [gauge,right of=g5,label=below:{$2$}] {};
	\node (g7) [gauge,right of=g6,label=below:{$2$}] {};
	\node (g8) [gauge,right of=g7,label=below:{$2$}] {};
	\node (g9) [gauge,right of=g8,label=below:{$1$}] {};
	\node (g10) [gauge,above of=g8,label=above:{$1$}] {};
	\node (b1) [gauge,above left of=g2,label=left:{$1$}] {};
	\node (b0) [above of=g2,label=below:{$\ldots$},label=above:{$n$}] {} ;
	\node (b2) [gauge,above right of=g2,label=right:{$1$}] {};
	\draw (g1)--(g2) (g2)--(g3) (g3)--(g4) (g4)--(g5) (g5)--(g6) (g6)--(g7) 
(g7)--(g8) (g8)--(g9) (g8)--(g10) (g2)--(b1) (g2)--(b2);
	\end{tikzpicture}
	} \,.
\end{align}
The dimensions and symmetries of the magnetic quiver are 
\begin{align}
G_J= \surm(2)^{n+1} \times \sorm(16)
  \qquad 
  \text{and}
  \qquad
  \dim\, \Coulomb^{3d} =n+16
  \; .
\end{align}
\subsection{Case \texorpdfstring{$k=3$}{k=3}}
There exist five possibilities to embed $\Z_3$ into $E_8$:
\begin{align}
 m_1 = 3 \; , \qquad 
 m_1=1, \; m_2 =1 \; , \qquad 
 m_1=1, \; m'_2 =1 \; , \qquad
m_3=1 \; , \quad \text{or} \quad   m'_3=1 \; .
\end{align}
This will be discussed in detail below. Again, discrete 
gauging or small $E_8$ instanton transitions will not be elaborated on. The 
focus is put on deriving the associated magnetic quiver for the electric quiver 
using the rules of Conjecture \ref{conj:magnetic_quiver}.
%
%
\subsubsection{Symmetry \texorpdfstring{$\surm(3)\times E_8$}{SU(3)xE8} --- 
case \texorpdfstring{$m_1=3$}{m1=3}}
\label{sec:k=3_m1=3}
%
The linking numbers read $l= (3,0^7)$ and the Type IIA brane system is given by
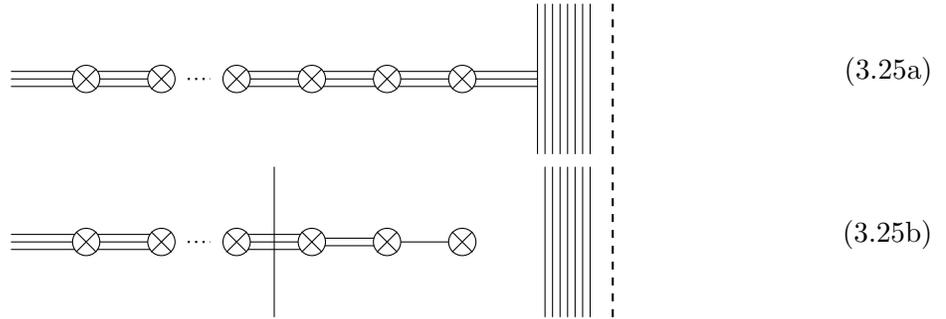
\begin{subequations}
\begin{align}
 \raisebox{-.5\height}{

 }
\end{align}
\end{subequations}
which gives rise to the electric quiver
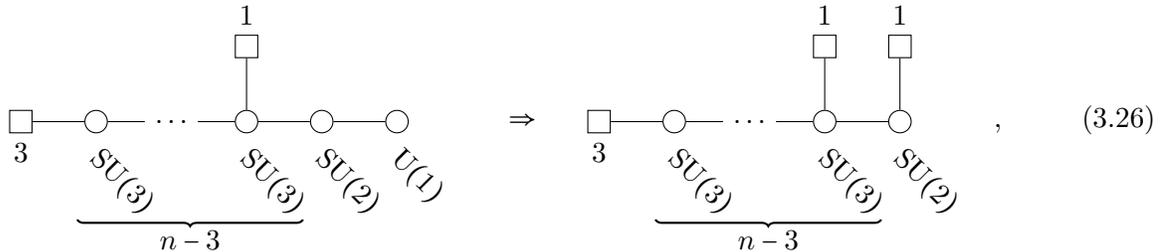
\begin{align}
 \raisebox{-.5\height}{
 	%
	}
	\; ,
\end{align}
wherein the last step is necessary as there are no $\uo$ gauge nodes in 6d. The 
flavour symmetry and Higgs branch dimension of the electric quiver are
\begin{align}
 G_F = \surm(3) \times\uo^n
 \qquad 
 \text{and}
 \qquad
 \dim\,\Higgs^{6d} = n+5 \,.
\end{align}
Changing the brane system to the phase were \Ds s are suspended between 
\De\ branes leads to the magnetic quiver
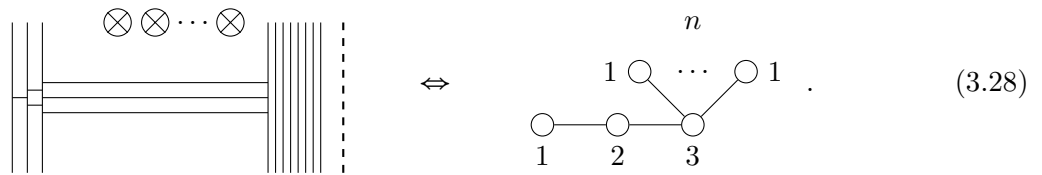
\begin{align}
   \raisebox{-.5\height}{
  %
	} \,.
\end{align}
The Coulomb branch dimension and symmetries are 
\begin{align}
 G_J=  \surm(3) \times\uo^n
 \qquad 
 \text{and}
 \qquad
 \dim\,\Coulomb^{3d} = n+5 \,.
\end{align}
\subsubsection{Symmetry \texorpdfstring{$\surm(3)\times(\uo\times 
E_7) $}{SU(3)xU(1)xE7} --- 
case \texorpdfstring{$m_1=m_2=1$}{m1=m2=1}}
\label{sec:k=3_m1=m2=1}
%
The linking numbers read $l= (2,1,0^6)$ and the Type IIA brane system is given 
by
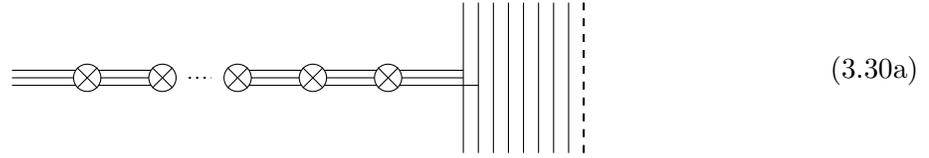
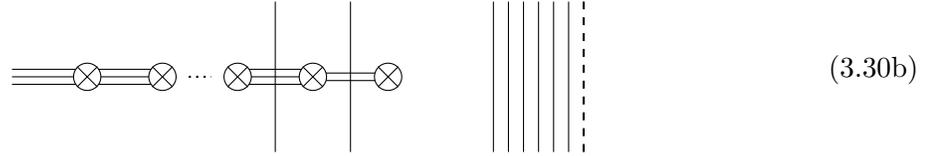
\begin{subequations}
\begin{align}
 \raisebox{-.5\height}{
 %
 }
\end{align}
\end{subequations}
which gives rise to the electric quiver
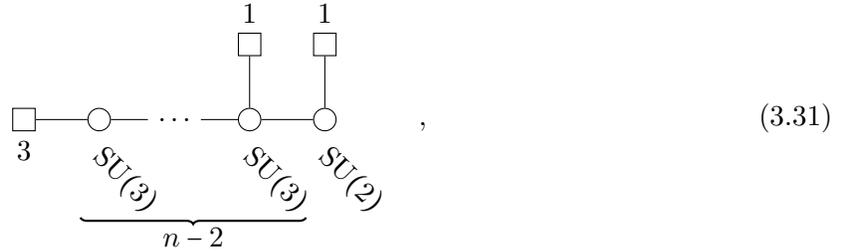
\begin{align}
 \raisebox{-.5\height}{
 	%
	} \; ,
\end{align}
with flavour symmetry and Higgs branch dimension
\begin{align}
 G_F = \surm(3) \times \uo^{n+1}
 \qquad
 \text{and}
 \qquad
 \dim\, \Higgs^{6d} = n+6
 \, .
\end{align}
Passing to the brane system for the magnetic quiver, one obtains
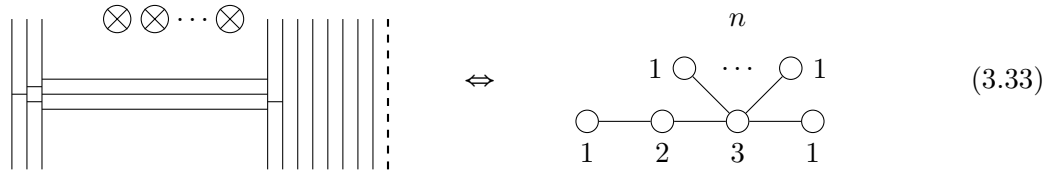
\begin{align}
   \raisebox{-.5\height}{
  %
	} 
\end{align}
and the dimensions and symmetries of the magnetic quiver  are
\begin{align}
 G_J = \surm(3)  \times \uo^{n+1}
 \qquad
 \text{and}
 \qquad
 \dim\, \Coulomb^{3d} = n+6
 \, .
\end{align}
\subsubsection{Symmetry \texorpdfstring{$\surm(3)\times(\surm(3)\times 
E_6)$}{SU(3)xSU(3)xE6} --- 
case \texorpdfstring{$m_3=1$}{m3=1}}
\label{sec:k=3_m3=1}
%
The linking numbers read $l= (1^3,0^5)$ and the Type IIA brane system is 
given by
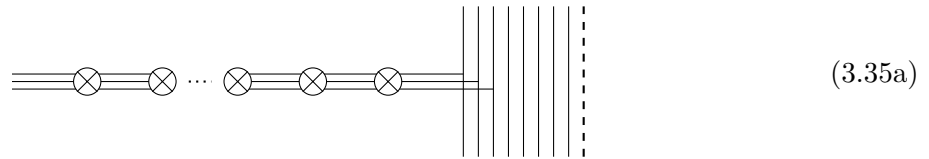
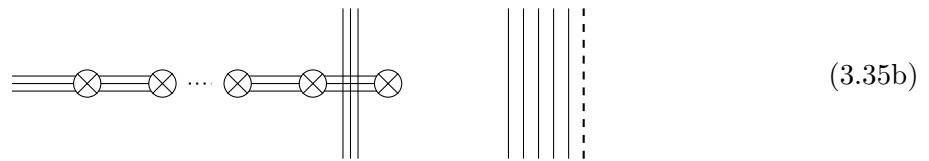
\begin{subequations}
\begin{align}
 \raisebox{-.5\height}{
 %
 }
\end{align}
\end{subequations}
which gives rise to the electric quiver
\begin{align}
 \raisebox{-.5\height}{
 	%
	} \; ,
\end{align}
with flavour symmetry and Higgs branch dimension
\begin{align}
  G_F=\surm(3)^2\times \uo^{n}
  \qquad
  \text{and}
    \qquad
    \dim\,\Higgs^{6d}=n+8
  \; .
\end{align}
To derive the magnetic quiver, one passes to the following phase of the brane 
system:
\begin{align}
   \raisebox{-.5\height}{
  %
	} 
\end{align}
where the dimensions and symmetries are 
\begin{align}
  G_J =\surm(3)^2\times \uo^{n}
  \qquad
  \text{and}
    \qquad
    \dim\,\Coulomb^{3d}=n+8
  \; .
\end{align}
\subsubsection{Symmetry \texorpdfstring{$\surm(3)\times (\uo \times
\sorm(14))$}{SU(3)xU(1)xSO(14)} --- 
case \texorpdfstring{$m_1=m'_2=1$}{m1=mp2=1}}
\label{sec:k=3_m1=mp2=1}
%
The linking numbers read $l= (1,0^7)$ and the Type IIA brane system is given 
by
\begin{subequations}
\begin{align}
 \raisebox{-.5\height}{
 %
 
 }
\end{align}
\end{subequations}
which gives rise to the electric quiver
\begin{align}
 \raisebox{-.5\height}{
 	%
	} \; ,
\end{align}
with flavour symmetry and Higgs branch dimension
\begin{align}
 G_F  =\surm(3)\times \uo^{n+1} \times \sorm(14)
 \qquad
 \text{and}
 \qquad
 \dim\, \Higgs^{6d}=n+19
 \; .
\end{align}
Note that the theory is anomaly free as $\usprm(N_c)$ is equipped with $N_f 
=N_c +8$ flavours.
Changing to the brane system for the magnetic quiver results in
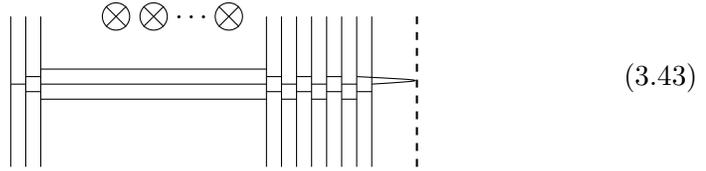
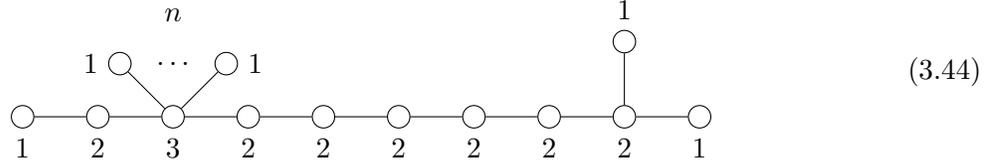
\begin{align}
   \raisebox{-.5\height}{
  %
	} 
\end{align}
and the symmetry and Coulomb branch dimension are
\begin{align}
 G_J  =\surm(3)\times \uo^{n+1} \times \sorm(14)
 \qquad
 \text{and}
 \qquad
 \dim\, \Coulomb^{3d}=n+19
 \; .
\end{align}
\subsubsection{Symmetry \texorpdfstring{$\surm(3)\times \surm(9)$}{SU(3)xSU(9)} 
--- 
case \texorpdfstring{$m'_3=1$}{mp3=1}}
\label{sec:k=3_mp3=1}
%
The linking numbers read $l= (\frac{1}{2}^8)$ and the Type IIA brane system is 
given by
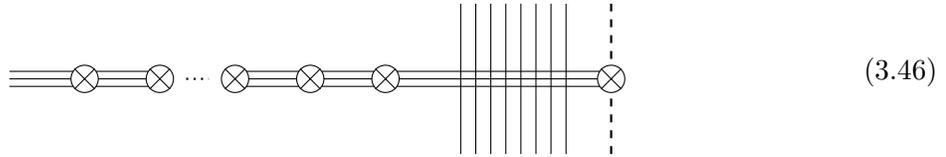
\begin{align}
 \raisebox{-.5\height}{
 %
 
 } 
\end{align}
Here, the linking number of $\frac{1}{2}$ has been realised by a stuck half 
\NS\ on the orientifold. This gives rise to the following electric quiver:
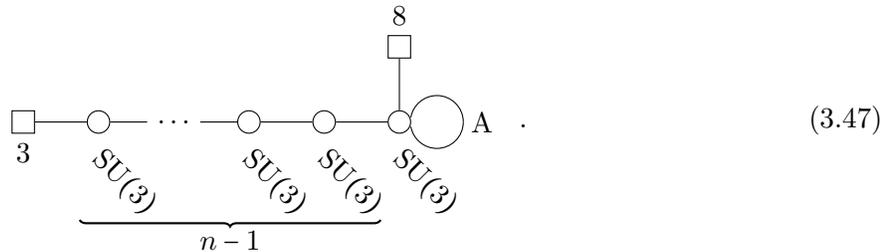
\begin{align}
 \raisebox{-.5\height}{
 	%
	} \; .
\end{align}
Here, the anti-symmetric loop at the last gauge node is necessary for anomaly 
cancellation; this is a clear consequence of the brane picture, cf.\ 
\cite[Sec.\ 2.1]{Hanany:1997gh}. Nonetheless, 
for $\surm(3)$ one has $ \Lambda^2[1,0]=[0,1]\cong [1,0]$. Consequently, the 
quiver is equivalent to
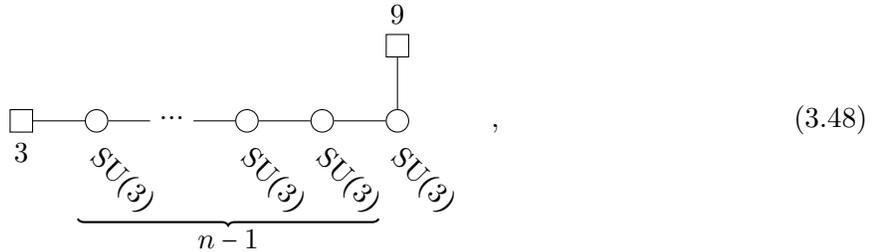
\begin{align}
 \raisebox{-.5\height}{
 	%
	} \; ,
\end{align}
with flavour symmetry and Higgs branch dimension
\begin{align}
 G_F=  \surm(3)\times \uo^{n+1} \times \surm(9)
 \qquad
\text{and}
\qquad 
 \dim\, \Higgs^{6d}= n+27 \,. 
\end{align}
As usual, one may change to the phase of the brane system for the magnetic 
quiver
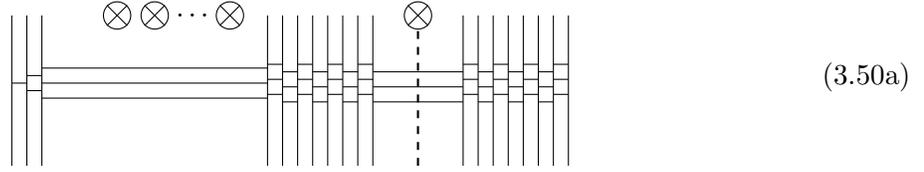
\begin{subequations}
\begin{align}
   \raisebox{-.5\height}{
  %
 } 
 \end{align}
which can be rearranged by connecting the \Ds s in the interval between the 
last 
\De\ and the \Oe\ with their mirror images. In addition, one \Ds\ will be 
stretched between the stuck half \NS\ and the last \De\ such that the brane 
system becomes 
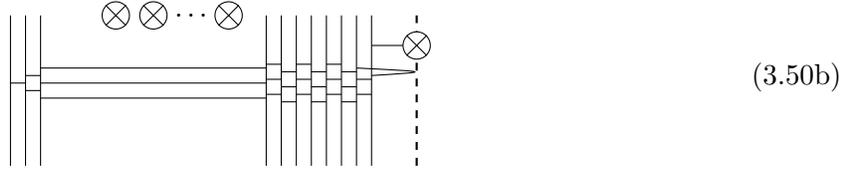
\begin{align}
    \raisebox{-.5\height}{
  %
 } 
\end{align}
and the magnetic quiver reads
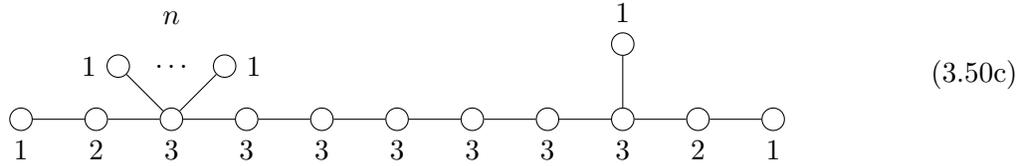
\begin{align}
  \raisebox{-.5\height}{
 	%
	}  
\end{align}
\end{subequations}
with Coulomb branch dimension and symmetry given by
\begin{align}
 G_J=  \surm(3)\times \uo^{n+1} \times \surm(9)
 \qquad
\text{and}
\qquad 
 \dim\, \Coulomb^{3d}= n+27 \,. 
\end{align}
\subsection{Case \texorpdfstring{$k=4$}{k=4}}
For the $A_3$ singularity $\C^2 \slash \Z_4$, there exist ten possibilities 
for the embedding $\Z_4 \hookrightarrow E_8$, cf.\ \cite[Sec.\ 
5.3]{Mekareeya:2017jgc}:
\begin{alignat}{5}
\begin{aligned}
 m_1 &= 4 \; , &\qquad
 m_1 &= 2, \; m_2=1 \; , &\qquad 
 m_1&=2, \; m'_2 =1 \; , &\qquad 
 m_2&=1, \; m'_2 =1 \; , &\qquad 
 m_2&=2 \,, \\
 m'_2&=2,  \; &\qquad
 m_1&=1, \; m_3 =1 \; , &\qquad 
 m_1&=1, \; m'_3 =1 \; , &\qquad 
m_4&=1 \; , \quad \text{or} &\quad   m'_4&=1\, .
\end{aligned}
\end{alignat}
As before, the brane systems and the electric as well as magnetic quivers will 
be provided. Since discrete gauging and $E_8$ transitions are 
straightforwardly derived from the discussion above, these transitions will not 
be detailed any further. It is sufficient to specify the derivation of the 
magnetic quiver in the weakly coupled phase by means of Conjecture 
\ref{conj:magnetic_quiver}.
%
%
\subsubsection{Symmetry \texorpdfstring{$\surm(4)\times E_8$}{SU(4)xE8} 
--- 
case \texorpdfstring{$m_1=4$}{m1=4}}
The linking numbers are $l= (4,0^7)$ such that the brane system reads
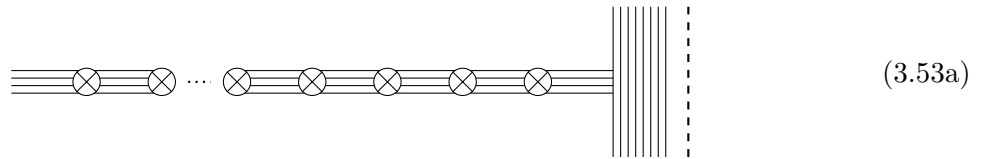
\begin{subequations}
\begin{align}
 \raisebox{-.5\height}{

 }
\end{align}
\end{subequations}
with associated electric quiver
\begin{align}
 \raisebox{-.5\height}{
 	%
	}
	\; ,
\end{align}
as there is no $\uo$ gauge symmetry in 6d. The flavour symmetry and Higgs 
branch dimension become
\begin{align}
 G_F = \surm(4)\times \uo^{n}
 \qquad
 \text{and}
 \qquad
 \dim\,\Higgs^{6d} = n+9 \,. 
\end{align}
The phase of the brane system for the magnetic quiver is derived as
\begin{align}
   \raisebox{-.5\height}{
  %
	} 
\end{align}
and the symmetry and Coulomb branch dimension are 
\begin{align}
 G_J = \surm(4)\times \uo^{n}
 \qquad
 \text{and}
 \qquad
 \dim\,\Coulomb^{3d} = n+9 \,. 
\end{align}
%
%
%
\subsubsection{Symmetry \texorpdfstring{$\surm(4)\times(\uo\times 
E_7)$}{SU(4)xU(1)xE7} 
--- 
case \texorpdfstring{$m_1=2, m_2=1$}{m1=2,m2=1}}
The linking numbers are $l= (3,1,0^6)$ such that the Type IIA brane system 
becomes
\begin{subequations}
\begin{align}
 \raisebox{-.5\height}{
 %
 }
\end{align}
\end{subequations}
and the electric quiver is read off to be
\begin{align}
 \raisebox{-.5\height}{
 	%
	} \; ,
\end{align}
with flavour symmetry and Higgs branch dimension
\begin{align}
 G_F = \surm(4)\times \uo^{n+1}
 \qquad
 \text{and}
 \qquad
 \dim\,\Higgs^{6d} = n+10
 \,.
\end{align}
Passing to the phase of the brane system for the magnetic quiver yields
\begin{align}
   \raisebox{-.5\height}{
  %
	} 
\end{align}
and symmetry and Coulomb branch dimension are as follows
\begin{align}
 G_J = \surm(4)\times \uo^{n+1}
 \qquad
 \text{and}
 \qquad
 \dim\,\Coulomb^{3d} = n+10
 \,.
\end{align}
%
%
%
\subsubsection{Symmetry \texorpdfstring{$\surm(4)\times (\uo \times
\sorm(14))$}{SU(4)xSO(14)} 
--- 
case \texorpdfstring{$m_1=2, m'_2=1$}{m1=2,mp2=1}}
\label{sec:k=4_m1=2_mp2=1}
Here, the linking numbers are $l= (2,0^7)$ such that the brane system looks like
\begin{subequations}
\begin{align}
 \raisebox{-.5\height}{
 %
 
 }
\end{align}
\end{subequations}
with corresponding electric quiver
\begin{align}
 \raisebox{-.5\height}{
 	%
	} \; ,
\end{align}
which has flavour symmetry and Higgs branch dimension
\begin{align}
 G_F= \surm(4)\times \uo^{n+1} \times \sorm(14)
 \qquad
 \text{and}
 \qquad
 \dim\, \Higgs^{6d}=n+23
 \,.
\end{align}
Note that the 6d theory is anomaly free as $\usprm(N_c)$ is equipped with $N_f 
=N_c +8$ flavours. 
Next, the phase of the brane system that allows to deduce the magnetic quiver 
reads
\begin{align}
   \raisebox{-.5\height}{
  %
	} 
\end{align}
and Coulomb branch dimension and symmetry are 
\begin{align}
 G_J= \surm(4)\times \uo^{n+1} \times \sorm(14)
 \qquad
 \text{and}
 \qquad
 \dim\, \Coulomb^{3d}=n+23
 \,.
\end{align}
%
%
%
%
\subsubsection{Symmetry \texorpdfstring{$\surm(4)\times(\uo\times 
\surm(2)\times \sorm(12))$}{SU(4)xU(1)xSU(2)xSO(12)} 
--- 
case \texorpdfstring{$m_2=m'_2=1$}{m2=mp2=1}}
\label{sec:k=4_m2=mp2=1}
One computes to linking numbers to be $l= (1^2,0^6)$ such that the brane 
picture becomes
\begin{subequations}
\begin{align}
 \raisebox{-.5\height}{
 %
 
 }
\end{align}
\end{subequations}
such that the electric quiver reads
\begin{align}
 \raisebox{-.5\height}{
 	%
	} \; ,
\end{align}
with flavour symmetry and Higgs branch dimension
\begin{align}
 G_F = \surm(4)\times \uo^{n+1} \times \surm(2) \times \sorm(12) 
\qquad
\text{and}
\qquad
\dim\,\Higgs^{6d} =n+24
\;.
\end{align}
Note that the 6d theory is anomaly free as $\usprm(N_c)$ is equipped with $N_f 
=N_c +8$ flavours.
Passing to the brane system for the magnetic quiver yields
\begin{align}
   \raisebox{-.5\height}{
  %
	} 
\end{align}
where the Coulomb branch dimension and symmetry are 
\begin{align}
 G_J = \surm(4)\times \uo^{n+1} \times \surm(2) \times \sorm(12) 
\qquad
\text{and}
\qquad
\dim\,\Coulomb^{3d} =n+24
\;.
\end{align}
%
%
%
\subsubsection{Symmetry \texorpdfstring{$\surm(4)\times 
(\surm(2)\times E_7)$}{SU(4)xSU(2)xE7} 
--- 
case \texorpdfstring{$m_2=2$}{m2=2}}
\label{sec:k=4_m2=2}
The linking numbers $l= (2^2,0^6)$ imply the following brane system:
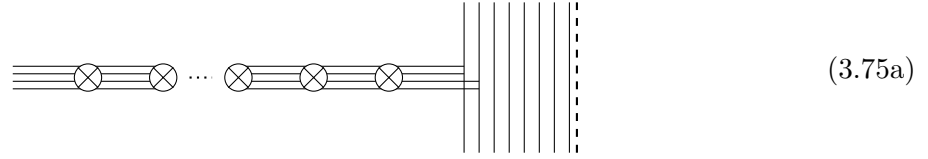
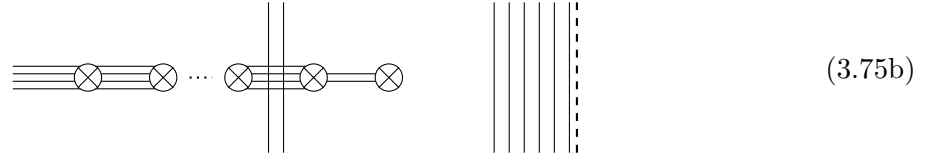
\begin{subequations}
\begin{align}
 \raisebox{-.5\height}{
 %
 }
\end{align}
\end{subequations}
such that the electric quiver reads
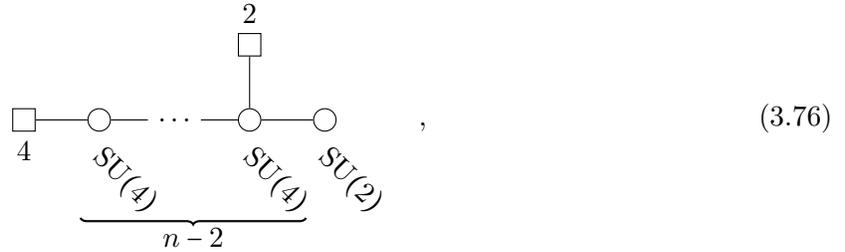
\begin{align}
 \raisebox{-.5\height}{
 	%
	} \; ,
\end{align}
which has flavour symmetry and Higgs branch dimension
\begin{align}
 G_F = \surm(4) \times \uo^n \times \surm(2)  
\qquad 
\text{and}
\qquad
\dim\, \Higgs^{6d} = n+11
\;.
\end{align}
Passing to the phase of the brane system for the magnetic quiver results in
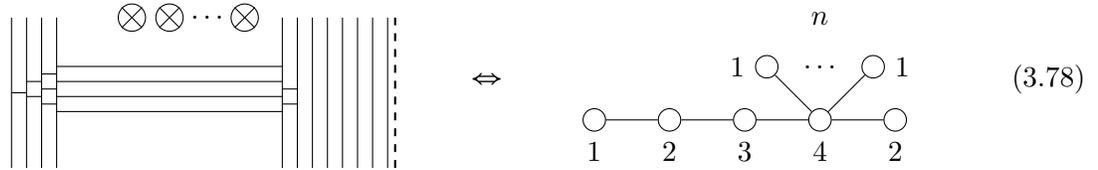
\begin{align}
   \raisebox{-.5\height}{
  %
	} 
\end{align}
where the Coulomb branch dimension and symmetry are 
\begin{align}
 G_J = \surm(4) \times \uo^n \times \surm(2)  
\qquad 
\text{and}
\qquad
\dim\, \Coulomb^{3d} = n+11
\;.
\end{align}
%
%
%
\subsubsection{Symmetry \texorpdfstring{$\surm(4)\times 
\sorm(16)$}{SU(4)xSO(16)} 
--- 
case \texorpdfstring{$m'_2=2$}{mp2=2}}
\label{sec:k=4_mp2=2}
For linking numbers $l= (0^8)$  the corresponding brane system becomes
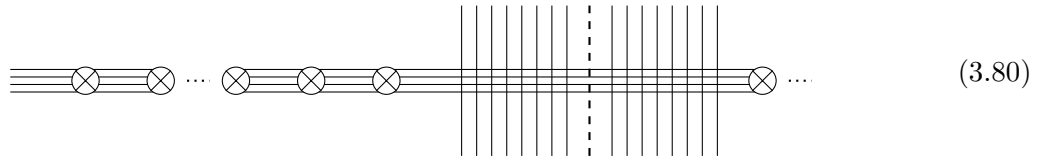
\begin{align}
 \raisebox{-.5\height}{
 %
 
 }
\end{align}
and the electric quiver reads
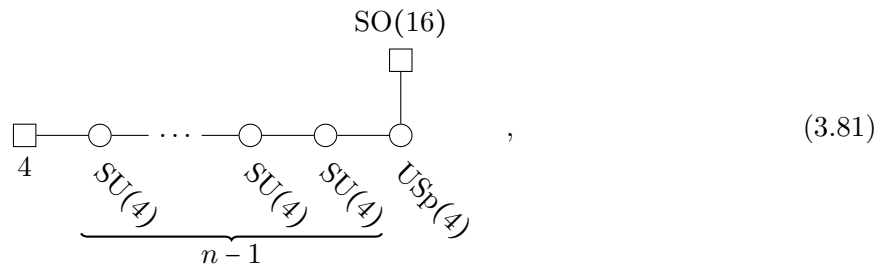
\begin{align}
 \raisebox{-.5\height}{
 	%
	} \; ,
\end{align}
with flavour symmetry and Higgs branch dimension
\begin{align}
 G_F =   \surm(4)\times \uo^{n} \times \sorm(16) 
 \qquad
 \text{and}
\qquad
\dim\,\Higgs^{6d} = n+37
 \; .
\end{align}
Note that the 6d theory is anomaly free as $\usprm(N_c)$ is equipped with $N_f 
=N_c +8$ flavours.
Next, changing to the brane system for the magnetic quiver results in
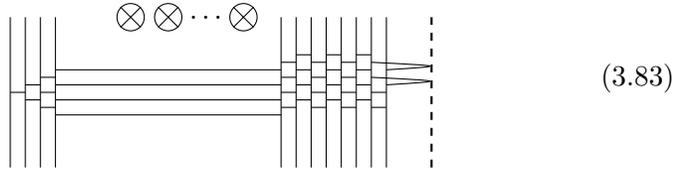
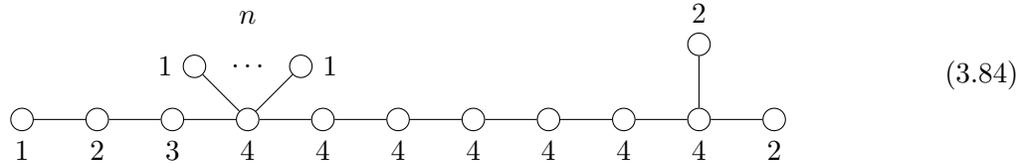
\begin{align}
   \raisebox{-.5\height}{
  %
	} 
\end{align}
where the Coulomb branch dimension and symmetry are 
\begin{align}
 G_J =   \surm(4)\times \uo^{n} \times \sorm(16) 
 \qquad
 \text{and}
\qquad
\dim\,\Coulomb^{3d} = n+37
 \; .
\end{align}
%
%
%
\subsubsection{Symmetry \texorpdfstring{$\surm(4)\times 
(\uo\times \surm(2)\times E_6)$}{SU(4)xU(1)xSU(2)xE6} 
--- 
case \texorpdfstring{$m_1=m_3=1$}{m1=m3=1}}
\label{sec:k=4_m1=m3=1}
Here, the linking numbers read $l= (2,1^2,0^5)$ and imply the following system
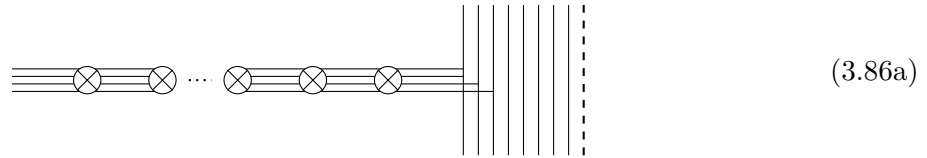
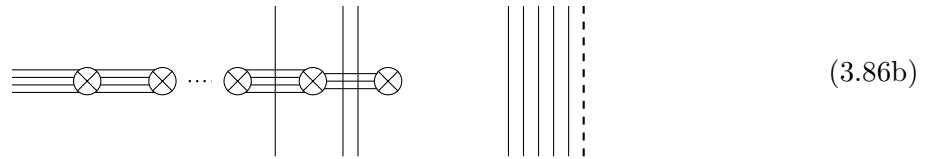
\begin{subequations}
\begin{align}
 \raisebox{-.5\height}{
 %
 }
\end{align}
\end{subequations}
with associated electric quiver
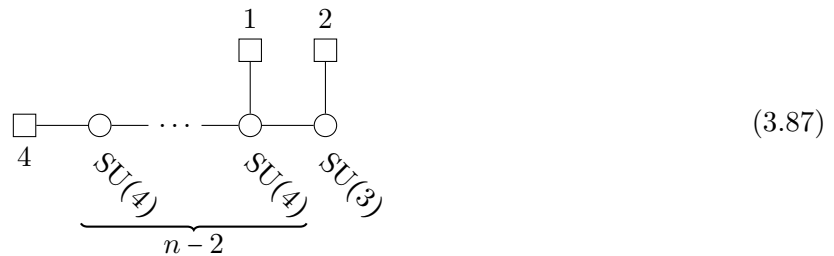
\begin{align}
 \raisebox{-.5\height}{
 	%
	} 
\end{align}
which has flavour symmetry and Higgs branch dimension
\begin{align}
 G_F = \surm(4)\times \uo^{n+1} \times \surm(2)  
\qquad
\text{and}
\qquad
\dim\,\Higgs^{6d}=n+12
\;.
\end{align}
Changing the phase of the brane system to be able to read off the magnetic 
quiver results in
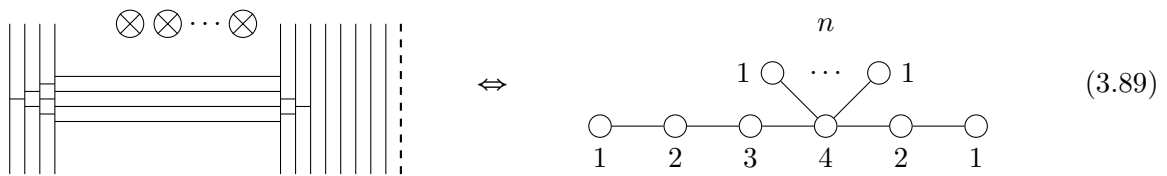
\begin{align}
   \raisebox{-.5\height}{
  %
	} 
\end{align}
where Coulomb branch dimension and symmetry are 
\begin{align}
 G_J = \surm(4)\times \uo^{n+1} \times \surm(2)  
\qquad
\text{and}
\qquad
\dim\,\Coulomb^{3d}=n+12
\;.
\end{align}
%
%
%
\subsubsection{Symmetry \texorpdfstring{$\surm(4)\times 
(\uo \times \surm(8))$}{SU(4)xU(1)xSU(8)} 
--- 
case \texorpdfstring{$m_1=m'_3=1$}{m1=mp3=1}}
\label{sec:k=4_m1=mp3=1}
In this case, the linking numbers are $l= (\frac{3}{2},\frac{1}{2}^7)$  and the 
Type IIA brane realisation becomes
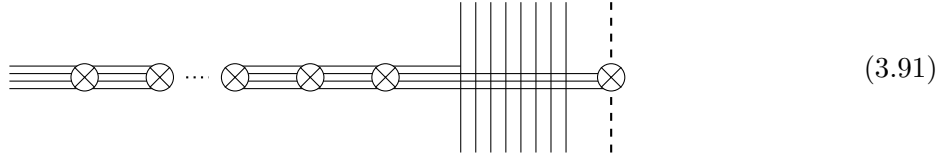
\begin{align}
 \raisebox{-.5\height}{
 %
 
 } 
\end{align}
As before, the linking number of $\frac{1}{2}$ has been realised by a stuck 
half \NS\ on the orientifold.
The brane system allows to derive the electric quiver
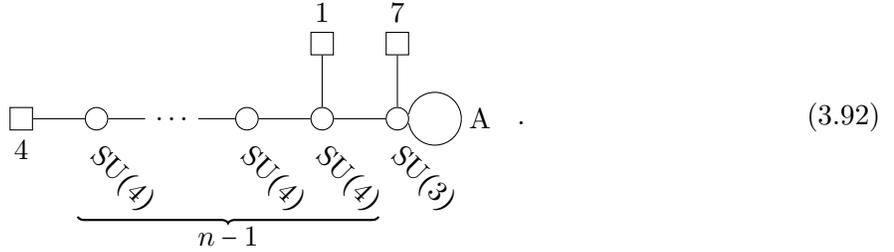
\begin{align}
 \raisebox{-.5\height}{
 	%
	} \; .
\end{align}
Here, the anti-symmetric loop at the last gauge node is necessary for anomaly 
cancellation in $6$ dimensions, but clearly follows from the brane 
construction, cf.\ 
\cite[Sec.\ 2.1]{Hanany:1997gh}. Nonetheless, for 
$\surm(3)$ one has $\Lambda^2[1,0]=[0,1]\cong 
[1,0]$. Consequently, the electric quiver is equivalent to
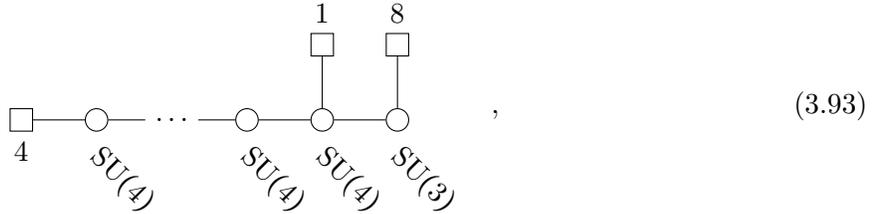
\begin{align}
 \raisebox{-.5\height}{
 	%
	} \; ,
\end{align}
with flavour symmetry and Higgs branch dimension
\begin{align}
 G_F = \surm(4)\times \uo^{n+2} \times \surm(8) 
\qquad
\text{and}
\qquad
\dim\,\Higgs^{6d} =n+31
\; .
\end{align}
Passing to the brane system for the magnetic quiver proceeds as
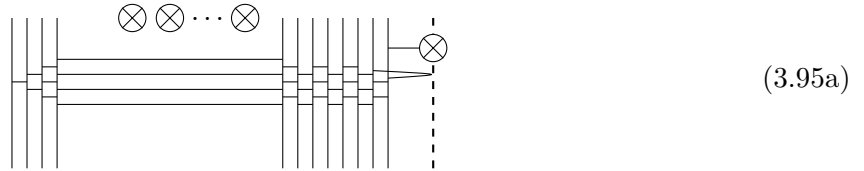
\begin{subequations}
\begin{align}
    \raisebox{-.5\height}{
  %
 } 
 \end{align}
 such that the magnetic quiver is read off to be
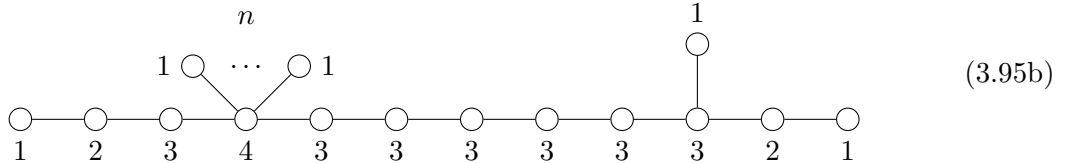
\begin{align}
  \raisebox{-.5\height}{
 	%
	} 
\end{align}
\end{subequations}
and the Coulomb branch dimension and symmetry are \begin{align}
 G_J = \surm(4)\times \uo^{n+2} \times \surm(8) 
\qquad
\text{and}
\qquad
\dim\,\Coulomb^{3d} =n+31
\; .
\end{align}
%
%
%
\subsubsection{Symmetry \texorpdfstring{$\surm(4)\times 
(\surm(4)\times \sorm(10))$}{SU(4)xSU(4)xSO(10)} 
--- 
case \texorpdfstring{$m_4=1$}{m4=1}}
\label{sec:k=4_m4=1}
One computes the linking numbers as $l= (1^4,0^4)$ such that the brane 
configuration becomes
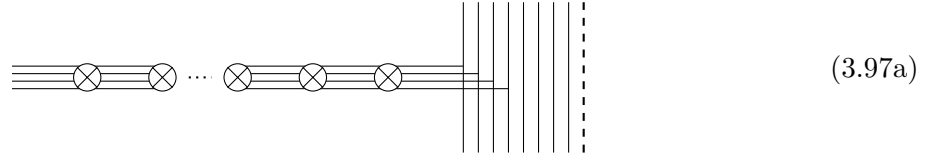
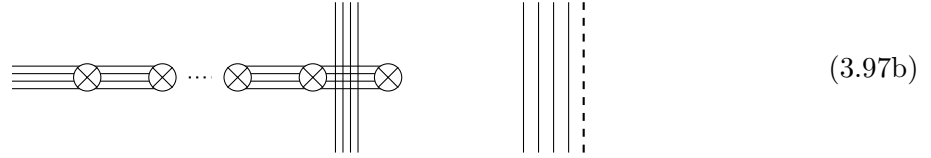
\begin{subequations}
\begin{align}
 \raisebox{-.5\height}{
 %
 }
\end{align}
\end{subequations}
and the electric quiver reads
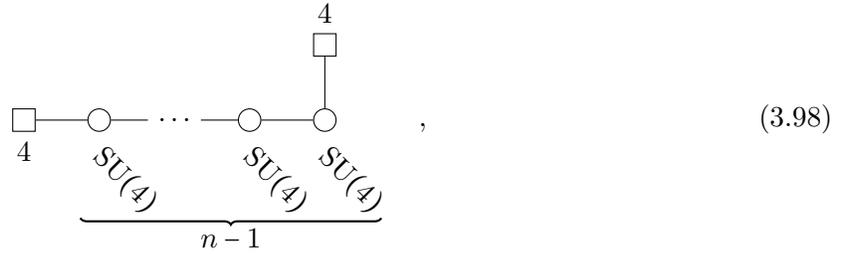
\begin{align}
 \raisebox{-.5\height}{
 	%
	} \; ,
\end{align}
with  flavour symmetry and Higgs branch dimension
\begin{align}
 G_F = \surm(4)^2\times \uo^{n}
 \qquad
 \text{and}
 \qquad
\dim\,\Higgs^{6d} =n+15
 \; .
\end{align}
Proceeding to the brane system for magnetic quiver leads to
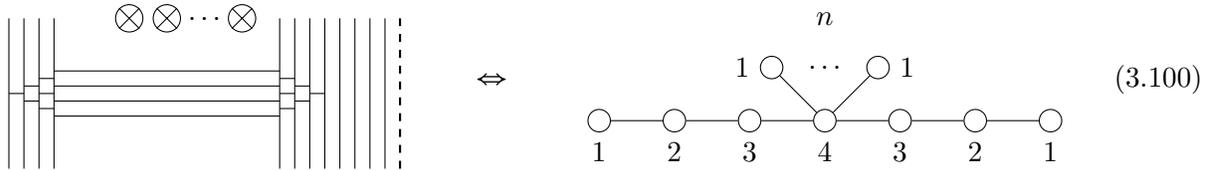
\begin{align}
   \raisebox{-.5\height}{
  %
	} 
\end{align}
and Coulomb branch dimension and symmetry are 
\begin{align}
 G_J = \surm(4)^2\times \uo^{n}
 \qquad
 \text{and}
 \qquad
\dim\,\Coulomb^{3d} =n+15
 \; .
\end{align}
%
%
%
\subsubsection{Symmetry \texorpdfstring{$\surm(4)\times 
(\surm(8)\times \surm(2))$}{SU(4)xSU(8)xSU(2)} 
--- 
case \texorpdfstring{$m'_4=1$}{mp4=1}}
\label{sec:k=4_mp4=1}
The linking numbers  read $l= (\frac{1}{2}^7,-\frac{1}{2})$, and we observe 
that $\lambda_8$ is negative, which can be realised in a Type IIA brane 
configuration as follows:
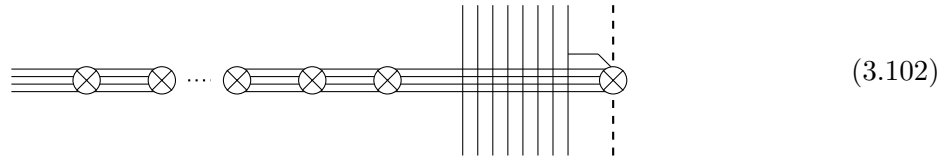
\begin{align}
 \raisebox{-.5\height}{
 %
 
 } 
\end{align}
Here, the linking number of $\frac{1}{2}$ has been realised by a stuck half 
\NS\ on the orientifold.
However, one may perform a transition with brane creation / annihilation by 
pushing the 8th D8 through the 
orientifold and the stuck half \NS . Then the mirror of the original \De\ 
appears, but has no \Ds\ ending on it, i.e.\ the brane configuration becomes
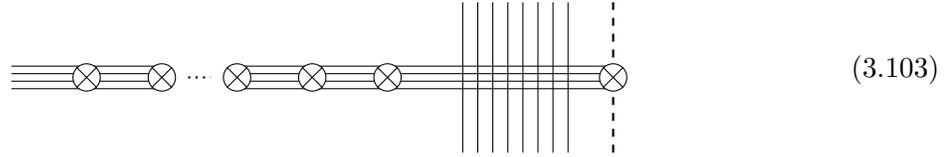
\begin{align}
 \raisebox{-.5\height}{
 %
 
 } 
\end{align}
The linking numbers of this configuration are $l'=(\frac{1}{2}^8)$, but the 
system is equivalent to the original. The corresponding electric quiver reads
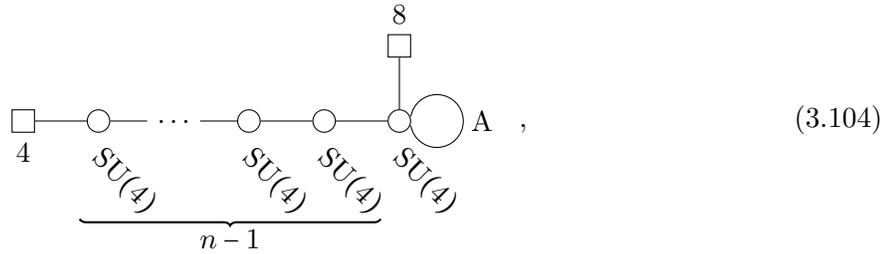
\begin{align}
 \raisebox{-.5\height}{
 	%
	} \; ,
\end{align}
with flavour symmetry and Higgs branch dimension
\begin{align}
G_F = \surm(4)\times \uo^{n+1} \times \surm(8)\times \surm(2) 
\qquad
\text{and}
\qquad
\dim\,\Higgs^{6d} =n+38
\;.
\end{align}
Here, the anti-symmetric loop at the last gauge node is necessary for anomaly 
cancellation, but is clearly derived from the brane system, cf.\ 
\cite[Sec.\ 2.1]{Hanany:1997gh}. The corresponding hypermultiplet in the second 
anti-symmetric representation of $\surm(4)$, which is a real representation, 
contributes a $\usprm(2)\cong \surm(2)$ flavour 
symmetry.
Passing to the brane system for the magnetic quiver yields
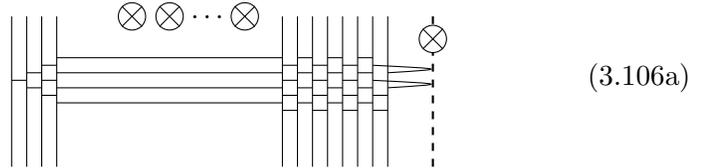
\begin{subequations}
\begin{align}
    \raisebox{-.5\height}{
  \begin{tikzpicture}
      \DeightMany{4}{-0.6}{2}
      \DsixFree{-0.6}{0.15}{-0.4}{0.15}
      \DsixSomeMore{2}{-0.4}{0.2}{0}
      \DsixSomeMore{3}{-0.2}{0.2}{-0.05}
      \DsixSomeMore{4}{0}{3}{-0.1}
      \ns{1,1}
      \ns{1.5,1}
      \draw (2,1) node {$\ldots$};
      \ns{2.5,1}
      \DeightMany{8}{3}{2}
      \Oeight{5}
      \DsixSomeMore{4}{3}{0.2}{-0.2}
      \DsixSomeMore{4}{3.2}{0.2}{-0.1}
      \DsixSomeMore{4}{3.4}{0.2}{-0.2}
      \DsixSomeMore{4}{3.6}{0.2}{-0.1}
      \DsixSomeMore{4}{3.8}{0.2}{-0.2}
      \DsixSomeMore{4}{4.0}{0.2}{-0.1}     
      \draw (4.2,0.35) .. controls (5.2,0.3) .. (4.4,0.25);
      \draw (4.2,0.15) .. controls (5.2,0.1) .. (4.4,0.05);
      \DsixSomeMore{2}{4.2}{0.2}{-0.3}
       \ns{5,0.7}
 \end{tikzpicture}
 } 
 \\
  \raisebox{-.5\height}
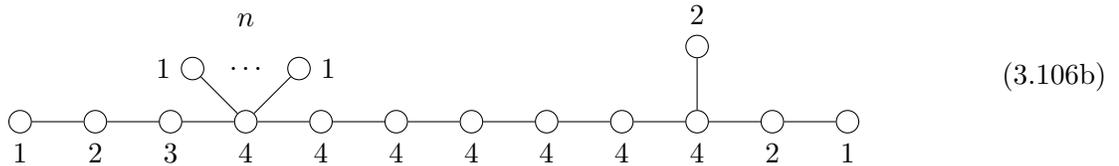
{
 	\begin{tikzpicture}
	\tikzstyle{gauge} = [circle, draw,inner sep=3pt];
	\tikzstyle{flavour} = [regular polygon,regular polygon sides=4,inner 
sep=3pt, draw];
	\node (g1) [gauge,label=below:{$1$}] {};
	\node (g2) [gauge,right of=g1,label=below:{$2$}] {};
	\node (g3) [gauge,right of=g2,label=below:{$3$}] {};
	\node (g4) [gauge,right of=g3,label=below:{$4$}] {};
	\node (g5) [gauge,right of=g4,label=below:{$4$}] {};
	\node (g6) [gauge,right of=g5,label=below:{$4$}] {};
	\node (g7) [gauge,right of=g6,label=below:{$4$}] {};
	\node (g8) [gauge,right of=g7,label=below:{$4$}] {};
	\node (g9) [gauge,right of=g8,label=below:{$4$}] {};
	\node (g10) [gauge,right of=g9,label=below:{$4$}] {};
	\node (g11) [gauge,right of=g10,label=below:{$2$}] {};
	\node (g12) [gauge,right of=g11,label=below:{$1$}] {};
	\node (g13) [gauge,above of=g10,label=above:{$2$}] {};
	\node (b1) [gauge,above left of=g4,label=left:{$1$}] {};
	\node (b0) [above of=g4,label=below:{$\ldots$},label=above:{$n$}] {} ;
	\node (b2) [gauge,above right of=g4,label=right:{$1$}] {};
	\draw (g1)--(g2) (g2)--(g3) (g3)--(g4) (g4)--(g5) (g5)--(g6) 
(g6)--(g7) (g7)--(g8) (g8)--(g9) (g9)--(g10) (g10)--(g11) (g11)--(g12) 
(g10)--(g13) (g4)--(b1) (g4)--(b2);
	\end{tikzpicture}
	} 
\end{align}
\end{subequations}
and the Coulomb branch dimension and symmetry are 
\begin{align}
G_J = \surm(4)\times \uo^{n+1} \times \surm(8)\times \surm(2) 
\qquad
\text{and}
\qquad
\dim\,\Coulomb^{3d} =n+38
\;.
\end{align}
As a remark, pulling up the stuck \NS\ on the \Oe\ orientifold and reconnecting 
the \Ds\ branes with their mirrors reduces the system the $m'_2=2$ 
configuration. On the field theory side, one Higgses the anti-symmetric 
hypermultipet of the $\surm(4)$ such that the gauge group is broken to 
$\usprm(4)$.
%
%
\subsection{General case}
\label{sec:general_case}
Having discussed multiple examples, one can approach the general construction. 
From the examples considered, a case study of the linking numbers seems the best 
way to 
proceed.
To begin with, inverting the relations \eqref{eq:fluxes_via_lambda} yields
\begin{equation}
\begin{aligned}
 \lambda_i &= m_i + \lambda_{i+1} = \sum_{j=i}^{6} m_j + \lambda_7 \;, \qquad 
i=1,\ldots,6 \\
\lambda_7&= \frac{1}{2}(m'_3+m'_4) \; , \qquad 
\lambda_8= \frac{1}{2}(m'_3-m'_4) \; , \qquad 
\lambda_9= \frac{1}{2}(m'_3-m'_4-2m'_2) \; , \qquad 
\end{aligned}
\label{eq:lambda_via_fluxes}
\end{equation}
from which one observes 
\begin{compactenum}[(i)]
 \item $\lambda_1 \geq \lambda_2 \geq \ldots \geq \lambda_6 \geq \lambda_7 
\geq0$,
 \item $\lambda_7 \geq \lambda_8 \geq \lambda_9$, but $\lambda_8$ and / or 
$\lambda_9$ may become negative,
\item either $\lambda_i \in \Z$ for all $i=1,\ldots,9$ or $\lambda_i \in 
\Z+\frac{1}{2}$ for all $i=1,\ldots,9$.
\end{compactenum}
For reasons that become clear later, define the following quantities
\begin{align}
 p \coloneqq \min \left\{ \lfloor \frac{m'_3+m'_4}{2} \rfloor , \lfloor 
\frac{m'_2 +m'_3 +2m'_4}{3} \rfloor \right\}  
=\min \left\{ \lfloor \lambda_7 \rfloor, \lfloor 
\lambda_7-\frac{1}{3}\lambda_9 \rfloor \right\} \; , \qquad  
r \coloneqq \lambda_7 -p  \; .
\end{align}
Therefore, as long as $\lambda_9\leq0$ it follows that $p=\lfloor \lambda_7 
\rfloor$ and $r$ is either zero or a half.
Consequently, the discussion is split in several cases, cf.\ 
\cite{Mekareeya:2017jgc},
\begin{compactenum}[(1)]
 \item For $m'_4\geq m'_3, m'_4\pm m'_3=\text{even}$ it follows 
 \begin{align}
  \lambda_1 \geq \ldots \geq \lambda_7 \geq 0 \geq \lambda_8 \geq \lambda_9 \in 
\Z 
\; , \qquad
p= \lambda_7, \; r=0 \; .
\label{eq:linking_no_1}
 \end{align}
 \item For $m'_4\geq m'_3, m'_4\pm m'_3=\text{odd}$ one finds
 \begin{align}
  \lambda_1 \geq \ldots \geq \lambda_7 \geq 0 \geq \lambda_8\geq \lambda_9 \in 
\Z+\frac{1}{2} 
\; , \qquad 
p=\lfloor \lambda_7 \rfloor \; , r=\frac{1}{2} \;.
\label{eq:linking_no_2}
 \end{align}
  \item For $m'_3\geq m'_4,m'_3-m'_4\leq 2 m'_2 , m'_3\pm m'_4=\text{even}$ one 
has 
\begin{align}
  \lambda_1 \geq \ldots \geq \lambda_7 \geq  \lambda_8 \geq 0 \geq \lambda_9 
\in \Z
\; , \qquad
p= \lambda_7 \; , r=0\; .
\label{eq:linking_no_3}
 \end{align}
\item For $m'_3\geq m'_4,m'_3-m'_4\leq 2 m'_2 , m'_3\pm m'_4=\text{odd}$ one 
obtains 
\begin{align}
  \lambda_1 \geq \ldots \geq \lambda_7 \geq  \lambda_8 \geq 0 \geq \lambda_9 
\in \Z +\frac{1}{2}
\; , \qquad 
p=\lfloor \lambda_7 \rfloor , \; r=\frac{1}{2} \;.
\label{eq:linking_no_4}
 \end{align}
\item For $m'_3\geq m'_4,m'_3-m'_4\geq 2 m'_2 $ it follows 
\begin{align}
  \lambda_1 \geq \ldots \geq \lambda_7 \geq  \lambda_8  \geq \lambda_9 \geq 0 
\; , \qquad
  p=\lfloor \lambda_7 -\frac{1}{3} \lambda_9 \rfloor \,.
  \label{eq:linking_no_5}
 \end{align}
\end{compactenum}
\paragraph{Strategy.}
For cases $(1)$--$(4)$, one may employ the known Type IIA constructions 
\cite{Hanany:1997gh,Brunner:1997gk,Hanany:1997sa} with non-vanishing 
cosmological constant that yield $6$-dimensional $\Ncal=(1,0)$ theories.

The numbers $2p$ and $2r$ are interpreted as total number of stuck half \NS\ 
branes on the \Oe. The difference is that the $2p$ \NS\ can leave the \Oe\ in 
pairs, while the $2r$ \NS\ cannot. 
Note that $2r$ can be larger than one due to the cosmological constant outside 
the orientifold plane.
These numbers are determined from the brane 
picture by charge conservation, in the sense that the RR-charge and 
cosmological constant determine how many \Ds\ branes are in each interval. The 
number of stuck \NS\ branes is then determined by the linking numbers.

The remaining case $(5)$ can be treated by Type I$^\prime$ constructions 
\cite{Morrison:1996xf,Douglas:1996xp,Gorbatov:2001pw} which includes an \Ost\ 
instead of an \Oe\ orientifold, see also \cite{Ohmori:2015tka}. The arguments 
imply that there are two ways to split a single \De\ from out of a system of 
coincident \Oe\ and \De. Firstly, the stack of \Oe\ and \De\ can turn into a 
separate \Oe\ and \De. Secondly, a stack of \Oe\ and \De\ can emanate a 
additional \De\ while tuning into a stack of coincident \Ost\ and \De. The 
latter can then be separated as usual such that there are a single \Ost\ and 
two separate \De s.

In all cases, once the brane configuration is known for the electric theory, 
one can straightforwardly apply the rules of Conjecture 
\ref{conj:magnetic_quiver} to derive the associated magnetic quiver.
%
%
\subsubsection{\texorpdfstring{$m'_4\geq m'_3, m'_4\pm 
m'_3=\text{even}$}{mp4>m3p,mp4-mp3=even}}
\label{sec:case_1}
\paragraph{Electric quiver.}
Construct a Type IIA brane realisation for the linking number 
\eqref{eq:linking_no_1} and interpret $2p$ as the number half \NS\ branes that 
are stuck on the \Oe\ orientifold.
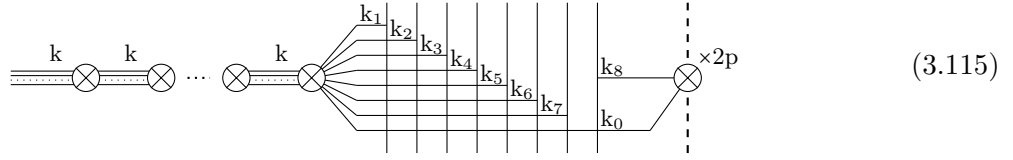
\begin{align}
 \raisebox{-.5\height}{
 \begin{tikzpicture}
      \DsixMany{k}{0}
      \DsixMany{k}{1}
      \ns{1,0}
      \DsixEmpty{2}
      \ns{2,0}
      \DsixMany{k}{3}
      \ns{3,0}
      \DeightMany{8}{5}{4}
      \Oeight{9}
      \DsixFree{4.6}{0.7}{5}{0.7}
      \DsixFree{4.6}{0.5}{5.4}{0.5}
      \DsixFree{4.6}{0.3}{5.8}{0.3}
      \DsixFree{4.6}{0.1}{6.2}{0.1}
      \DsixFree{4.6}{-0.1}{6.6}{-0.1}
      \DsixFree{4.6}{-0.3}{7}{-0.3}
      \DsixFree{4.6}{-0.5}{7.4}{-0.5}
      \DsixFree{4.6}{0.7}{4}{0}
      \DsixFree{4.6}{0.5}{4}{0}
      \DsixFree{4.6}{0.3}{4}{0}
      \DsixFree{4.6}{0.1}{4}{0}
      \DsixFree{4.6}{-0.1}{4}{0}
      \DsixFree{4.6}{-0.3}{4}{0}
      \DsixFree{4.6}{-0.5}{4}{0}
      \DsixFree{4.6}{-0.7}{4}{0}
      \ns{4,0}
      \draw (4.8,0.85) node {$\text{\footnotesize{k$_1$}}$};
      \draw (5.2,0.65) node {$\text{\footnotesize{k$_2$}}$};
      \draw (5.6,0.45) node {$\text{\footnotesize{k$_3$}}$};
      \draw (6.0,0.25) node {$\text{\footnotesize{k$_4$}}$};
      \draw (6.4,0.05) node {$\text{\footnotesize{k$_5$}}$};
      \draw (6.8,-0.15) node {$\text{\footnotesize{k$_6$}}$};
      \draw (7.2,-0.35) node {$\text{\footnotesize{k$_7$}}$};
      \DsixFree{7.8}{0}{9}{0}
      \draw (8,0.15) node {$\text{\footnotesize{k$_8$}}$};
      \DsixFree{4.6}{-0.7}{8.5}{-0.7}
      \DsixFree{8.5}{-0.7}{9}{0}
      \draw (8,-0.55) node {$\text{\footnotesize{k$_0$}}$};
      \ns{9,0}
      \draw (9.35,0.25) node {$\text{\footnotesize{ $\times$2p}}$};
 \end{tikzpicture} 
 } 
 \label{eq:branes_case_1a}
\end{align}
From the linking numbers \eqref{eq:linking_no_1} one readily computes 
\begin{align}
 k_j &= \sum_{j=i}^6 m_i \;, \qquad j=1,\ldots,6 \; ,\qquad
 k_7 =0 \; , \qquad k_8=m'_4 \; , \qquad 
 k_0 =2m'_2 +3m'_3 +4m'_4 \;,
 \label{eq:linking_aux_1}\\
p&= \frac{1}{2}(m'_3 +m'_4) \;.
\notag
\end{align}
However, one may perform a brane transition of the last \De\ through the \Oe\ 
to obtain a brane configuration with non-negative linking numbers only. In 
other words, pushing the \De\ with linking number $\lambda_8 \leq0$ through the 
\Oe\, the mirror \De\ reappears with linking number 
$|\lambda_8|$. The effects of brane creation and annihilation modify 
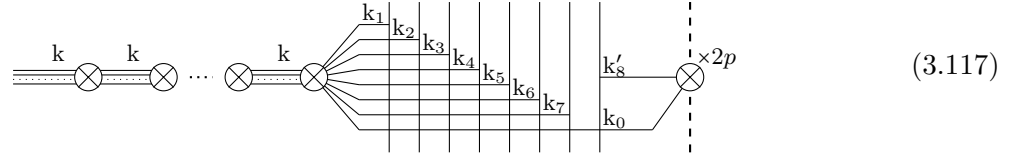
\eqref{eq:branes_case_1a} to
\begin{align}
 \raisebox{-.5\height}{
 \begin{tikzpicture}
      \DsixMany{k}{0}
      \DsixMany{k}{1}
      \ns{1,0}
      \DsixEmpty{2}
      \ns{2,0}
      \DsixMany{k}{3}
      \ns{3,0}
      \DeightMany{8}{5}{4}
      \Oeight{9}
      \DsixFree{4.6}{0.7}{5}{0.7}
      \DsixFree{4.6}{0.5}{5.4}{0.5}
      \DsixFree{4.6}{0.3}{5.8}{0.3}
      \DsixFree{4.6}{0.1}{6.2}{0.1}
      \DsixFree{4.6}{-0.1}{6.6}{-0.1}
      \DsixFree{4.6}{-0.3}{7}{-0.3}
      \DsixFree{4.6}{-0.5}{7.4}{-0.5}
      \DsixFree{4.6}{0.7}{4}{0}
      \DsixFree{4.6}{0.5}{4}{0}
      \DsixFree{4.6}{0.3}{4}{0}
      \DsixFree{4.6}{0.1}{4}{0}
      \DsixFree{4.6}{-0.1}{4}{0}
      \DsixFree{4.6}{-0.3}{4}{0}
      \DsixFree{4.6}{-0.5}{4}{0}
      \DsixFree{4.6}{-0.7}{4}{0}
      \ns{4,0}
      \draw (4.8,0.85) node {$\text{\footnotesize{k$_1$}}$};
      \draw (5.2,0.65) node {$\text{\footnotesize{k$_2$}}$};
      \draw (5.6,0.45) node {$\text{\footnotesize{k$_3$}}$};
      \draw (6,0.25) node {$\text{\footnotesize{k$_4$}}$};
      \draw (6.4,0.05) node {$\text{\footnotesize{k$_5$}}$};
      \draw (6.8,-0.15) node {$\text{\footnotesize{k$_6$}}$};
      \draw (7.2,-0.35) node {$\text{\footnotesize{k$_7$}}$};
%
      \draw (7.8,0)--(9,0);
      \draw (8,0.15) node {$\text{\footnotesize{k$'_8$}}$};
      \DsixFree{4.6}{-0.7}{8.5}{-0.7}
      \DsixFree{8.5}{-0.7}{9}{0}
      \draw (8,-0.55) node {$\text{\footnotesize{k$_0$}}$};
      \ns{9,0}
      \draw (9.35,0.25) node {$\text{\footnotesize{$\times 2p$}}$};
 \end{tikzpicture} 
 }
  \label{eq:branes_case_1b}
\end{align}
Such that the linking numbers of the \De\ become 
$(\lambda_1,\ldots,\lambda_7,|\lambda_8|)$, which are ordered and non-negative 
integer numbers. A computation shows:
\begin{align}
 k'_8 = 2p - k_8 = m'_3 +m'_4 -m'_4 = m'_3 \; .
\end{align}
Next, one may remove the $2p$ stuck half \NS\ branes from the \Oe\ pairwise, 
i.e.\ there will be $p$ pairs. This leads to
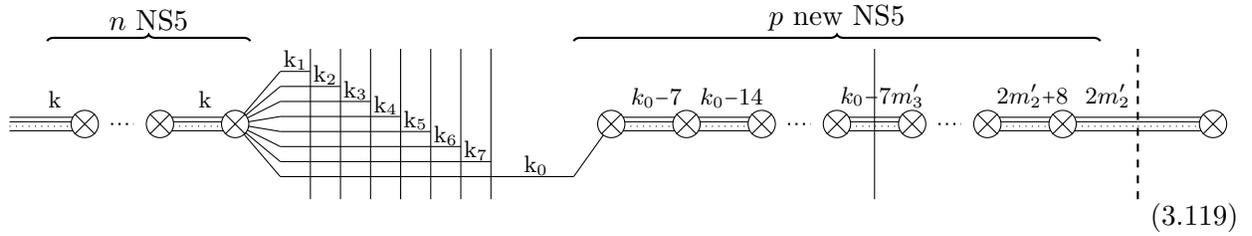
\begin{align}
 \raisebox{-.5\height}{
 \begin{tikzpicture}
      \DsixMany{k}{1}
      \DsixEmpty{2}
      \ns{2,0}
      \DsixMany{k}{3}
      \ns{3,0}
\draw[decoration={brace,mirror,raise=20pt},decorate,thick]
  (4.2,0.4) -- node[above=20pt] {$n$ \NS\ } (1.5,0.4);
      \DeightMany{7}{5}{4}
      \Oeight{16}
      \DsixFree{4.6}{0.7}{5}{0.7}
      \DsixFree{4.6}{0.5}{5.4}{0.5}
      \DsixFree{4.6}{0.3}{5.8}{0.3}
      \DsixFree{4.6}{0.1}{6.2}{0.1}
      \DsixFree{4.6}{-0.1}{6.6}{-0.1}
      \DsixFree{4.6}{-0.3}{7}{-0.3}
      \DsixFree{4.6}{-0.5}{7.4}{-0.5}
      \DsixFree{4.6}{0.7}{4}{0}
      \DsixFree{4.6}{0.5}{4}{0}
      \DsixFree{4.6}{0.3}{4}{0}
      \DsixFree{4.6}{0.1}{4}{0}
      \DsixFree{4.6}{-0.1}{4}{0}
      \DsixFree{4.6}{-0.3}{4}{0}
      \DsixFree{4.6}{-0.5}{4}{0}
      \DsixFree{4.6}{-0.7}{4}{0}
      \ns{4,0}
      \draw (4.8,0.85) node {$\text{\footnotesize{k$_1$}}$};
      \draw (5.2,0.65) node {$\text{\footnotesize{k$_2$}}$};
      \draw (5.6,0.45) node {$\text{\footnotesize{k$_3$}}$};
      \draw (6,0.25) node {$\text{\footnotesize{k$_4$}}$};
      \draw (6.4,0.05) node {$\text{\footnotesize{k$_5$}}$};
      \draw (6.8,-0.15) node {$\text{\footnotesize{k$_6$}}$};
      \draw (7.2,-0.35) node {$\text{\footnotesize{k$_7$}}$};
      \DsixFree{4.6}{-0.7}{8.5}{-0.7}
      \DsixFree{8.5}{-0.7}{9}{0}
      \draw (8,-0.55) node {$\text{\footnotesize{k$_0$}}$};
      \DsixMany{$k_0{-}7$}{9}
      \ns{9,0}
      \DsixMany{$k_0{-}14$}{10}
      \ns{10,0}
      \DsixEmpty{11}
      \ns{11,0}
      \DsixMany{$k_0{-}7m'_3$}{12}
      \Deight{12.5}
      \ns{12,0}
      \DsixEmpty{13}
      \ns{13,0}
      \DsixMany{$2m'_2{+}8$}{14}
      \ns{14,0}
      \DsixMany{$2m'_2$}{15}
      \ns{15,0}
      \DsixMany{}{16}
      \ns{17,0}
\draw[decoration={brace,mirror,raise=20pt},decorate,thick]
  (15.5,0.4) -- node[above=20pt] {$p$ new \NS\ } (8.5,0.4);
 \end{tikzpicture} 
 }
 \label{eq:branes_case_1b_aux}
\end{align}
The tail of the resulting electric quiver looks like
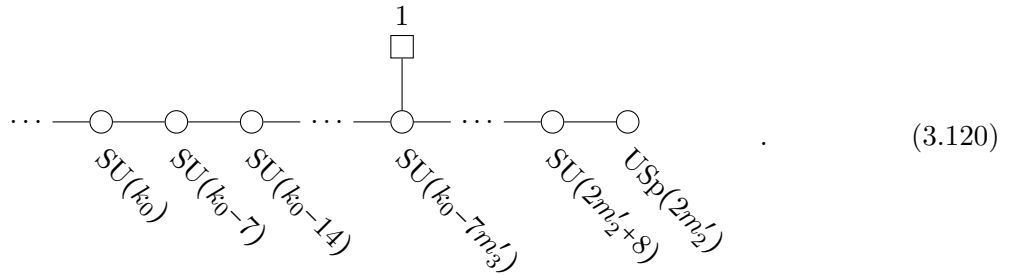
\begin{align}
   \raisebox{-.5\height}{
 	\begin{tikzpicture}
	\tikzstyle{gauge} = [circle, draw,inner sep=3pt];
	\tikzstyle{flavour} = [regular polygon,regular polygon sides=4,inner 
sep=3pt, draw];
\node (g0) [] {$\ldots$};
\node (g1) [gauge,right of=g0,label={[rotate=-45]below right:{$\surm(k_0)$}}] 
{};
\node (g2) [gauge,right of=g1,label={[rotate=-45]below 
right:{$\surm(k_0{-}7)$}}] {};
\node (g3) [gauge,right of=g2,label={[rotate=-45]below 
right:{$\surm(k_0{-}14)$}}] {};
\node (g4) [right of=g3] {$\ldots$};
\node (g5) [gauge,right of=g4,label={[rotate=-45]below 
right:{$\surm(k_0{-}7m'_3)$}}] 
{};
\node (g6) [right of=g5] {$\ldots$};
\node (g7) [gauge,right of=g6,label={[rotate=-45]below 
right:{$\surm(2m'_2{+}8)$}}] {};
\node (g8) [gauge,right of=g7,label={[rotate=-45]below 
right:{$\usprm(2m'_2)$}}] {};
	\node (f5) [flavour,above of=g5,label=above:{$1$}] {};
	\draw (g0)--(g1) (g1)--(g2) (g2)--(g3) (g3)--(g4) (g4)--(g5) (g5)--(g6) 
(g6)--(g7) (g7)--(g8)  (g5)--(f5);
	\end{tikzpicture}
	}  
	\;.
	\label{eq:6d_quiver_case_1}
\end{align}
\paragraph{Magnetic quiver.}
Moreover, the brane picture \eqref{eq:branes_case_1b} allows to change to the 
phase where all \Ds s are suspended between \De s in order to read off the 
magnetic quiver. The brane picture in this phase becomes
\begin{align}
 \raisebox{-.5\height}{
 \begin{tikzpicture}
 \Deight{0}
 \Deight{0.4}
 \draw (0.8,0) node {$\cdots$};
 \Deight{1.2}
 \Deight{1.6}
 \ns{2.2,0.75}
 \draw (2.6,0.75) node {$\cdots$};
 \ns{3.0,0.75}
 \DsixFree{0}{0}{0.4}{0} 
 \DsixFree{1.2}{0.4}{1.6}{0.4}
 \DsixFree{1.2}{0.2}{1.6}{0.2}
 \DsixFree{1.2}{-0.4}{1.6}{-0.4}
 \draw (1.4,-0.1) node {$\vdots$};
\DsixMany{k}{1.6}
\DsixMany{}{2.6}
      \DeightMany{8}{3.6}{4}
  \DsixFree{3.6}{-0.1}{4.0}{-0.1}
  \DsixFree{4.0}{0.1}{4.4}{0.1}
  \DsixFree{4.4}{-0.1}{4.8}{-0.1}
  \DsixFree{4.8}{0.1}{5.2}{0.1}
  \DsixFree{5.2}{-0.1}{5.6}{-0.1}
  \DsixFree{5.6}{0.1}{6.0}{0.1}
      \draw (3.8,0.05) node {$\text{\footnotesize{$d_1$}}$};
      \draw (4.2,0.25) node {$\text{\footnotesize{$d_2$}}$};
      \draw (4.6,0.05) node {$\text{\footnotesize{$d_3$}}$};
      \draw (5.0,0.25) node {$\text{\footnotesize{$d_4$}}$};
      \draw (5.4,0.05) node {$\text{\footnotesize{$d_5$}}$};
      \draw (5.8,0.25) node {$\text{\footnotesize{$d_6$}}$};
    \DsixFree{6.0}{0.3}{6.4}{0.3}
     \draw (6.2,0.45) node {$\text{\footnotesize{$a$}}$};
\draw (6.0,-0.3) .. controls (8.5,-0.4) .. (6.4,-0.5);
     \draw (6.2,-0.15) node {$\text{\footnotesize{$b$}}$};
      \Oeight{8}
      \ns{8,0}
      \draw (8.35,0.25) node {$\text{\footnotesize{$\times 2p$}}$};
 \end{tikzpicture} 
 }
 \;.
  \label{eq:branes_case_1c}
\end{align}
From the linking numbers \eqref{eq:linking_no_1} or 
\eqref{eq:linking_aux_1} one computes the number of \Ds\ branes to be
\begin{subequations}
\label{eq:numbers_3d_brane}
\begin{align}
d_j &= \sum_{i=1}^{6-j} i\, m_{i+j} +k_0 \,, \qquad \text{for} \quad 
j=1,\ldots,6  \,,\\
 a&=m'_2 +m'_3 +2m'_4 \,, \qquad
 b=m'_2 + 2m'_3 +2m'_4 \,.
\end{align}
\end{subequations}
In particular, note that $k_0 =a+b$ and $a+m'_3=b$, which allows to rearrange 
the \Ds\ branes in the last two segments compared to \eqref{eq:branes_case_1b}. 
Hence, the magnetic quiver becomes
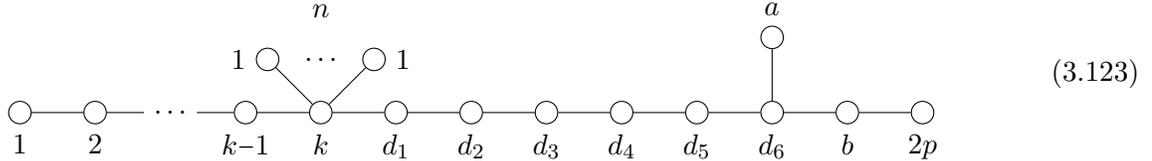
\begin{align}
   \raisebox{-.5\height}{
 	\begin{tikzpicture}
	\tikzstyle{gauge} = [circle, draw,inner sep=3pt];
	\tikzstyle{flavour} = [regular polygon,regular polygon sides=4,inner 
sep=3pt, draw];
	\node (g0) [gauge,label=below:{$1$}] {};
	\node (g1) [gauge,right of=g0,label=below:{$2$}] {};
	\node (g2) [right of=g1] {$\ldots$};
	\node (g3) [gauge,right of=g2,label=below:{$k{-}1$}] {};
	\node (g4) [gauge,right of=g3,label=below:{$k$}] {};
	\node (g5) [gauge,right of=g4,label=below:{$d_1$}] {};
	\node (g6) [gauge,right of=g5,label=below:{$d_2$}] {};
	\node (g7) [gauge,right of=g6,label=below:{$d_3$}] {};
	\node (g8) [gauge,right of=g7,label=below:{$d_4$}] {};
	\node (g9) [gauge,right of=g8,label=below:{$d_5$}] {};
	\node (g10) [gauge,right of=g9,label=below:{$d_6$}] {};
	\node (g11) [gauge,right of=g10,label=below:{$b$}] {};
	\node (g12) [gauge,right of=g11,label=below:{$2p$}] {};
	\node (g13) [gauge,above of=g10,label=above:{$a$}] {};
	\node (b1) [gauge,above left of=g4,label=left:{$1$}] {};
	\node (b0) [above of=g4,label=below:{$\ldots$},label=above:{$n$}] {} ;
	\node (b2) [gauge,above right of=g4,label=right:{$1$}] {};
	\draw (g0)--(g1) (g1)--(g2) (g2)--(g3) (g3)--(g4) (g4)--(g5) (g5)--(g6) 
(g6)--(g7) (g7)--(g8) (g8)--(g9) (g9)--(g10) (g10)--(g11) (g11)--(g12) 
(g10)--(g13) (g4)--(b1) (g4)--(b2);
	\end{tikzpicture}
	}
	\label{eq:general_3d_quiver}
\end{align}
and it is apparent that one can perform $p$ additional small $E_8$ instanton 
transitions.

This additional $E_8$ transitions should not come as a surprise, because the brane phase \eqref{eq:branes_case_1c} corresponds to the original brane configuration \eqref{eq:branes_case_1b}. The electric quiver, however, is associated to the brane configuration \eqref{eq:branes_case_1b_aux} in which the additional $p$ \NS\ branes have been pulled off the orientifold. Inspecting the linking numbers in \eqref{eq:branes_case_1b_aux} shows that the $p$ new half \NS\ branes need to be moved to the left of all eight \De\ of the \Mn\ system such that the \Ds\ branes can be suspended between \De\ branes only. Fortunately, this is nothing else than the brane transition associated to the $E_8$ transition displayed in \eqref{eq:1M5_in_M9} and \eqref{eq:1M5_in_M9_alternative}. Hence, one arrives at the following magnetic quiver
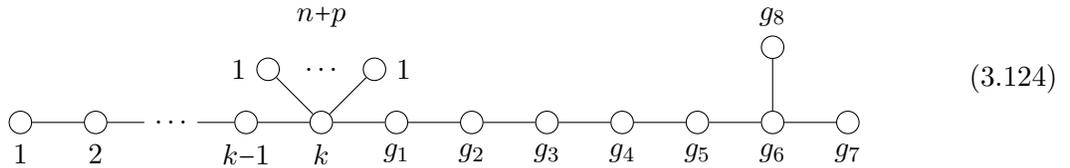
\begin{align}
   \raisebox{-.5\height}{
 	\begin{tikzpicture}
	\tikzstyle{gauge} = [circle, draw,inner sep=3pt];
	\tikzstyle{flavour} = [regular polygon,regular polygon sides=4,inner 
sep=3pt, draw];
	\node (g0) [gauge,label=below:{$1$}] {};
	\node (g1) [gauge,right of=g0,label=below:{$2$}] {};
	\node (g2) [right of=g1] {$\ldots$};
	\node (g3) [gauge,right of=g2,label=below:{$k{-}1$}] {};
	\node (g4) [gauge,right of=g3,label=below:{$k$}] {};
	\node (g5) [gauge,right of=g4,label=below:{$g_1$}] {};
	\node (g6) [gauge,right of=g5,label=below:{$g_2$}] {};
	\node (g7) [gauge,right of=g6,label=below:{$g_3$}] {};
	\node (g8) [gauge,right of=g7,label=below:{$g_4$}] {};
	\node (g9) [gauge,right of=g8,label=below:{$g_5$}] {};
	\node (g10) [gauge,right of=g9,label=below:{$g_6$}] {};
	\node (g11) [gauge,right of=g10,label=below:{$g_7$}] {};
	\node (g13) [gauge,above of=g10,label=above:{$g_8$}] {};
	\node (b1) [gauge,above left of=g4,label=left:{$1$}] {};
	\node (b0) [above of=g4,label=below:{$\ldots$},label=above:{$n{+}p$}] {} ;
	\node (b2) [gauge,above right of=g4,label=right:{$1$}] {};
	\draw (g0)--(g1) (g1)--(g2) (g2)--(g3) (g3)--(g4) (g4)--(g5) (g5)--(g6) 
(g6)--(g7) (g7)--(g8) (g8)--(g9) (g9)--(g10) (g10)--(g11) 
(g10)--(g13) (g4)--(b1) (g4)--(b2);
	\end{tikzpicture}
	}
	\label{eq:general_3d_quiver_mod1}
\end{align}
with unitary nodes of rank
\begin{subequations}
\label{eq:ranks_case1}
\begin{align}
g_j&= \sum_{i=1}^{6-j} i\ m_{i+j} +2m'_2 +m'_4 +\frac{6-j}{2}(m'_3+m'_4) \,, \; \text{for } j=1,\ldots,6 \\
  g_7 &=  m'_2\,, \qquad  g_8 = m'_2 + \frac{1}{2} (m'_4-m'_3)  \,.
\end{align}
\end{subequations}
Note that all ranks are non-negative integers by definition of the considered 
case.
As a special case, consider $m_i=0$ for all $i=1,\ldots,6$ as well as $m'_3=m'_4=0$ then one recognises the $SO(16)$ Dynkin diagram from the balanced nodes of the quiver.  Finally, the important result is
\begin{align}
 \Higgs^{6d}\left( \substack{\text{electric} \\ \text{quiver}} \eqref{eq:6d_quiver_case_1} \right)
 =
  \Coulomb^{3d}\left( \substack{\text{magnetic} \\ \text{quiver}} \eqref{eq:general_3d_quiver_mod1} \right) \,.
\end{align}
%
%
\subsubsection{\texorpdfstring{$m'_4\geq m'_3, m'_4\pm 
m'_3=\text{odd}$}{mp4>m3p,mp4-mp3=odd}}
\label{sec:case_2}
\paragraph{Electric quiver.}
Construct a Type IIA brane realisation for the linking number 
\eqref{eq:linking_no_2} and interpret $2p+1$ as the number half \NS\ branes 
that are stuck on the \Oe\ orientifold. The half-integer character of the 
linking numbers is a consequence of the odd number of half \NS\ branes on the 
orientifold.
Then much of the analysis from 
\eqref{eq:branes_case_1a} and \eqref{eq:branes_case_1b} carries over, with the 
suitable replacement of $2p$ to $2p+1$. (Note that 
$p=\frac{1}{2}(m'_3+m'_4-1)$.) Hence, the negative linking number of the 8th 
\De\ can be traded for its positive version via pushing the 8th \De\ (and its 
mirror) through the orientifold and the stuck \NS. 
As a consequence, one can again remove $2p$ of the stuck half \NS\ branes from 
the \Oe, but one half \NS\ inevitably remains on the orientifold. This leads to 
the brane picture
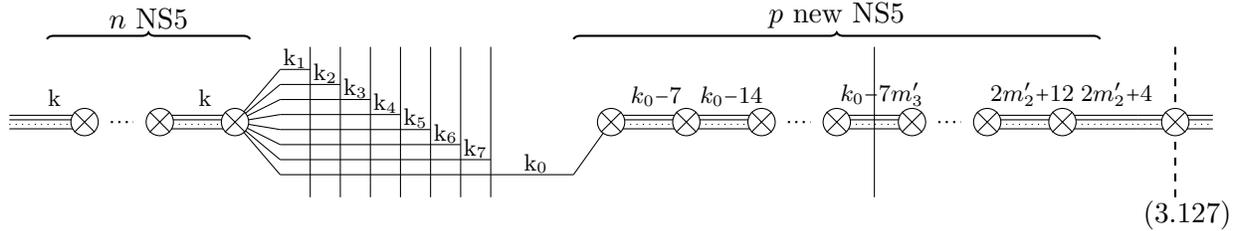
\begin{align}
 \raisebox{-.5\height}{
 \begin{tikzpicture}
      \DsixMany{k}{1}
      \DsixEmpty{2}
      \ns{2,0}
      \DsixMany{k}{3}
      \ns{3,0}
\draw[decoration={brace,mirror,raise=20pt},decorate,thick]
  (4.2,0.4) -- node[above=20pt] {$n$ \NS\ } (1.5,0.4);
      \DeightMany{7}{5}{4}
      \Oeight{16.5}
      \DsixFree{4.6}{0.7}{5}{0.7}
      \DsixFree{4.6}{0.5}{5.4}{0.5}
      \DsixFree{4.6}{0.3}{5.8}{0.3}
      \DsixFree{4.6}{0.1}{6.2}{0.1}
      \DsixFree{4.6}{-0.1}{6.6}{-0.1}
      \DsixFree{4.6}{-0.3}{7}{-0.3}
      \DsixFree{4.6}{-0.5}{7.4}{-0.5}
      \DsixFree{4.6}{0.7}{4}{0}
      \DsixFree{4.6}{0.5}{4}{0}
      \DsixFree{4.6}{0.3}{4}{0}
      \DsixFree{4.6}{0.1}{4}{0}
      \DsixFree{4.6}{-0.1}{4}{0}
      \DsixFree{4.6}{-0.3}{4}{0}
      \DsixFree{4.6}{-0.5}{4}{0}
      \DsixFree{4.6}{-0.7}{4}{0}
      \ns{4,0}
      \draw (4.8,0.85) node {$\text{\footnotesize{k$_1$}}$};
      \draw (5.2,0.65) node {$\text{\footnotesize{k$_2$}}$};
      \draw (5.6,0.45) node {$\text{\footnotesize{k$_3$}}$};
      \draw (6,0.25) node {$\text{\footnotesize{k$_4$}}$};
      \draw (6.4,0.05) node {$\text{\footnotesize{k$_5$}}$};
      \draw (6.8,-0.15) node {$\text{\footnotesize{k$_6$}}$};
      \draw (7.2,-0.35) node {$\text{\footnotesize{k$_7$}}$};
      \DsixFree{4.6}{-0.7}{8.5}{-0.7}
      \DsixFree{8.5}{-0.7}{9}{0}
      \draw (8,-0.55) node {$\text{\footnotesize{k$_0$}}$};
      \DsixMany{$k_0{-}7$}{9}
      \ns{9,0}
      \DsixMany{$k_0{-}14$}{10}
      \ns{10,0}
      \DsixEmpty{11}
      \ns{11,0}
      \DsixMany{$k_0{-}7m'_3$}{12}
      \Deight{12.5}
      \ns{12,0}
      \DsixEmpty{13}
      \ns{13,0}
      \DsixMany{$2m'_2{+}12$}{14}
      \ns{14,0}
      \DsixMany{$\; \; \; 2m'_2{+}4$}{15}
      \ns{15,0}
      \DsixMany{}{16}
      \ns{16.5,0}
\draw[decoration={brace,mirror,raise=20pt},decorate,thick]
  (15.5,0.4) -- node[above=20pt] {$p$ new \NS\ } (8.5,0.4);
 \end{tikzpicture} 
 }
 \label{eq:branes_case_2a}
\end{align}
The tail of the resulting electric quiver looks like
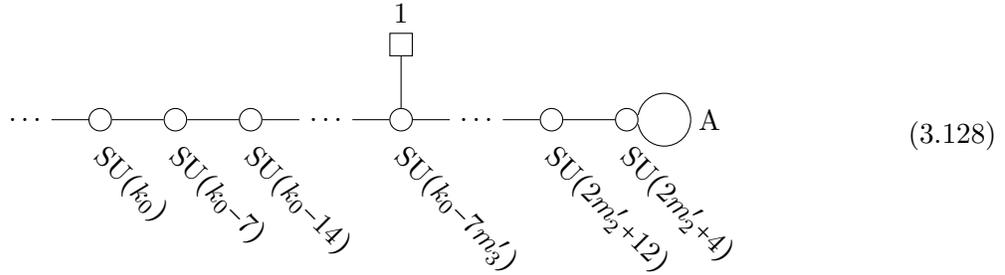
\begin{align}
   \raisebox{-.5\height}{
 	\begin{tikzpicture}
	\tikzstyle{gauge} = [circle, draw,inner sep=3pt];
	\tikzstyle{flavour} = [regular polygon,regular polygon sides=4,inner 
sep=3pt, draw];
\node (g0) [] {$\ldots$};
\node (g1) [gauge,right of=g0,label={[rotate=-45]below right:{$\surm(k_0)$}}] 
{};
\node (g2) [gauge,right of=g1,label={[rotate=-45]below 
right:{$\surm(k_0{-}7)$}}] {};
\node (g3) [gauge,right of=g2,label={[rotate=-45]below 
right:{$\surm(k_0{-}14)$}}] {};
\node (g4) [right of=g3] {$\ldots$};
\node (g5) [gauge,right of=g4,label={[rotate=-45]below 
right:{$\surm(k_0{-}7m'_3)$}}] 
{};
\node (g6) [right of=g5] {$\ldots$};
\node (g7) [gauge,right of=g6,label={[rotate=-45]below 
right:{$\surm(2m'_2{+}12)$}}] 
{};
\node (g8) [gauge,right of=g7,label={[rotate=-45]below 
right:{$\surm(2m'_2{+}4)$}}] {};
	\node (f5) [flavour,above of=g5,label=above:{$1$}] {};
	\draw [-] (8.85,0) arc (0:170:10pt);
	\draw [-] (8.85,-0) arc (360:190:10pt);
	\draw (9.1,0) node {$\mathrm{A}$};
	\draw (g0)--(g1) (g1)--(g2) (g2)--(g3) (g3)--(g4) (g4)--(g5) (g5)--(g6) 
(g6)--(g7) (g7)--(g8)  (g5)--(f5);
	\end{tikzpicture}
	} 
	\label{eq:6d_quiver_case_2}
\end{align}
\paragraph{Magnetic quiver.}
Analogously to the previous case, above brane picture \eqref{eq:branes_case_2a} 
allows to change to the phase where all \Ds\ are suspended between \De\ in order 
to read off the magnetic quiver. The brane picture in this phase becomes a 
small adaptation of \eqref{eq:branes_case_1c}, i.e.\
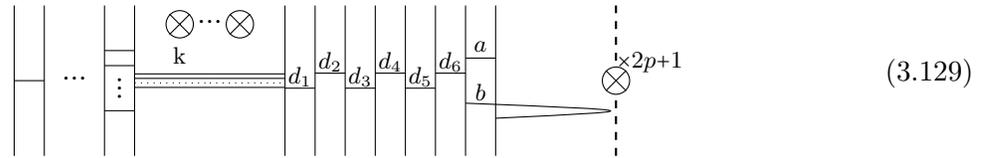
\begin{align}
 \raisebox{-.5\height}{
 \begin{tikzpicture}
 \Deight{0}
 \Deight{0.4}
 \draw (0.8,0) node {$\cdots$};
 \Deight{1.2}
 \Deight{1.6}
 \ns{2.2,0.75}
 \draw (2.6,0.75) node {$\cdots$};
 \ns{3.0,0.75}
 \DsixFree{0}{0}{0.4}{0} 
 \DsixFree{1.2}{0.4}{1.6}{0.4}
 \DsixFree{1.2}{0.2}{1.6}{0.2}
 \DsixFree{1.2}{-0.4}{1.6}{-0.4}
 \draw (1.4,-0.1) node {$\vdots$};
\DsixMany{k}{1.6}
\DsixMany{}{2.6}
      \DeightMany{8}{3.6}{4}
  \DsixFree{3.6}{-0.1}{4.0}{-0.1}
  \DsixFree{4.0}{0.1}{4.4}{0.1}
  \DsixFree{4.4}{-0.1}{4.8}{-0.1}
  \DsixFree{4.8}{0.1}{5.2}{0.1}
  \DsixFree{5.2}{-0.1}{5.6}{-0.1}
  \DsixFree{5.6}{0.1}{6.0}{0.1}
      \draw (3.8,0.05) node {$\text{\footnotesize{$d_1$}}$};
      \draw (4.2,0.25) node {$\text{\footnotesize{$d_2$}}$};
      \draw (4.6,0.05) node {$\text{\footnotesize{$d_3$}}$};
      \draw (5.0,0.25) node {$\text{\footnotesize{$d_4$}}$};
      \draw (5.4,0.05) node {$\text{\footnotesize{$d_5$}}$};
      \draw (5.8,0.25) node {$\text{\footnotesize{$d_6$}}$};
    \DsixFree{6.0}{0.3}{6.4}{0.3}
     \draw (6.2,0.45) node {$\text{\footnotesize{$a$}}$};
\draw (6.0,-0.3) .. controls (8.5,-0.4) .. (6.4,-0.5);
     \draw (6.2,-0.15) node {$\text{\footnotesize{$b$}}$};
      \Oeight{8}
      \ns{8,0}
      \draw (8.45,0.25) node {$\text{\footnotesize{$\times 2p{+}1$}}$};
 \end{tikzpicture} 
 }
  \label{eq:branes_case_2c}
\end{align}
From the linking numbers \eqref{eq:linking_no_2} one computes the number of 
\Ds s 
to be as in \eqref{eq:numbers_3d_brane}.
As before, relations $k_0 =a+b$ and $a+m'_3=b$ allow to rearrange the \Ds\ 
branes in the last two segments compared to \eqref{eq:branes_case_2a}. 
Hence, the magnetic quiver becomes
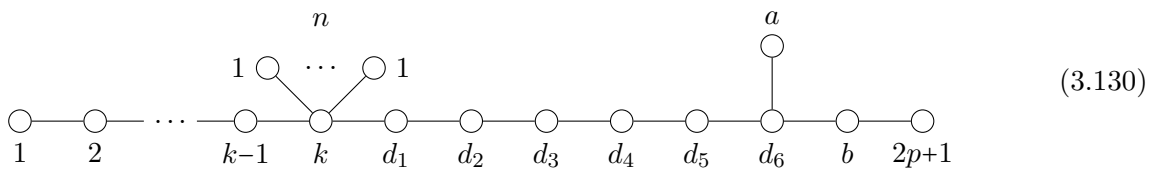
\begin{align}
   \raisebox{-.5\height}{
 	\begin{tikzpicture}
	\tikzstyle{gauge} = [circle, draw,inner sep=3pt];
	\tikzstyle{flavour} = [regular polygon,regular polygon sides=4,inner 
sep=3pt, draw];
	\node (g0) [gauge,label=below:{$1$}] {};
	\node (g1) [gauge,right of=g0,label=below:{$2$}] {};
	\node (g2) [right of=g1] {$\ldots$};
	\node (g3) [gauge,right of=g2,label=below:{$k{-}1$}] {};
	\node (g4) [gauge,right of=g3,label=below:{$k$}] {};
	\node (g5) [gauge,right of=g4,label=below:{$d_1$}] {};
	\node (g6) [gauge,right of=g5,label=below:{$d_2$}] {};
	\node (g7) [gauge,right of=g6,label=below:{$d_3$}] {};
	\node (g8) [gauge,right of=g7,label=below:{$d_4$}] {};
	\node (g9) [gauge,right of=g8,label=below:{$d_5$}] {};
	\node (g10) [gauge,right of=g9,label=below:{$d_6$}] {};
	\node (g11) [gauge,right of=g10,label=below:{$b$}] {};
	\node (g12) [gauge,right of=g11,label=below:{$2p{+}1$}] {};
	\node (g13) [gauge,above of=g10,label=above:{$a$}] {};
	\node (b1) [gauge,above left of=g4,label=left:{$1$}] {};
	\node (b0) [above of=g4,label=below:{$\ldots$},label=above:{$n$}] {} ;
	\node (b2) [gauge,above right of=g4,label=right:{$1$}] {};
	\draw (g0)--(g1) (g1)--(g2) (g2)--(g3) (g3)--(g4) (g4)--(g5) (g5)--(g6) 
(g6)--(g7) (g7)--(g8) (g8)--(g9) (g9)--(g10) (g10)--(g11) (g11)--(g12) 
(g10)--(g13) (g4)--(b1) (g4)--(b2);
	\end{tikzpicture}
	}
\end{align}
and it is apparent that one can perform $p$ additional small $E_8$ instanton transitions. Again, this quiver results from the magnetic brane configuration given by \eqref{eq:branes_case_2c}. In order to obtain the magnetic quiver for the electric quiver \eqref{eq:6d_quiver_case_2} with brane configuration \eqref{eq:branes_case_2a}, one needs to reverse the $p$ additional $E_8$ instanton transitions. The resulting magnetic quiver looks as follows:
\begin{align}
   \raisebox{-.5\height}{
 	\begin{tikzpicture}
	\tikzstyle{gauge} = [circle, draw,inner sep=3pt];
	\tikzstyle{flavour} = [regular polygon,regular polygon sides=4,inner 
sep=3pt, draw];
	\node (g0) [gauge,label=below:{$1$}] {};
	\node (g1) [gauge,right of=g0,label=below:{$2$}] {};
	\node (g2) [right of=g1] {$\ldots$};
	\node (g3) [gauge,right of=g2,label=below:{$k{-}1$}] {};
	\node (g4) [gauge,right of=g3,label=below:{$k$}] {};
	\node (g5) [gauge,right of=g4,label=below:{$g_1$}] {};
	\node (g6) [gauge,right of=g5,label=below:{$g_2$}] {};
	\node (g7) [gauge,right of=g6,label=below:{$g_3$}] {};
	\node (g8) [gauge,right of=g7,label=below:{$g_4$}] {};
	\node (g9) [gauge,right of=g8,label=below:{$g_5$}] {};
	\node (g10) [gauge,right of=g9,label=below:{$g_6$}] {};
	\node (g11) [gauge,right of=g10,label=below:{$g_7$}] {};
	\node (g12) [gauge,right of=g11,label=below:{$1$}] {};
	\node (g13) [gauge,above of=g10,label=above:{$g_8$}] {};
	\node (b1) [gauge,above left of=g4,label=left:{$1$}] {};
	\node (b0) [above of=g4,label=below:{$\ldots$},label=above:{$n{+}p$}] {} ;
	\node (b2) [gauge,above right of=g4,label=right:{$1$}] {};
	\draw (g0)--(g1) (g1)--(g2) (g2)--(g3) (g3)--(g4) (g4)--(g5) (g5)--(g6) 
(g6)--(g7) (g7)--(g8) (g8)--(g9) (g9)--(g10) (g10)--(g11) (g11)--(g12) 
(g10)--(g13) (g4)--(b1) (g4)--(b2);
	\end{tikzpicture}
	}
	\label{eq:general_3d_quiver_mod2}
\end{align}
with unitary nodes of rank
\begin{subequations}
\label{eq:ranks_case2}
\begin{align}
g_j&= \sum_{i=1}^{6-j} i\ m_{i+j} +2m'_2 +m'_4 +\frac{6-j}{2}(m'_3+m'_4-1) +3 
\,, \; \text{for } j=1,\ldots,6 \,,\\
  g_7 &=  m'_2+2\,, \qquad  g_8= m'_2 +1+ \frac{1}{2} (m'_4-m'_3+1)  \,.
\end{align}
\end{subequations}
All ranks are non-negative integers by definition of the considered case. Finally, the important result is
\begin{align}
 \Higgs^{6d}\left( \substack{\text{electric} \\ \text{quiver}} \eqref{eq:6d_quiver_case_2} \right)
 =
  \Coulomb^{3d}\left( \substack{\text{magnetic} \\ \text{quiver}} \eqref{eq:general_3d_quiver_mod2} \right) \,.
\end{align}
%
%
\subsubsection{\texorpdfstring{$m'_3\geq m'_4,m'_3-m'_4\leq 2 m'_2 , m'_3\pm 
m'_4=\text{even}$}{mp3>m4p,mp3-mp4<2mp2,mp3-mp4=even}}
\label{sec:case_3}
\paragraph{Electric quiver.}
Construct a Type IIA brane realisation for the linking number 
\eqref{eq:linking_no_3} and interpret $2p$ as the number half \NS\ branes that 
are stuck on the \Oe\ orientifold.
\begin{align}
 \raisebox{-.5\height}{
 \begin{tikzpicture}
      \DsixMany{k}{0}
      \DsixMany{k}{1}
      \ns{1,0}
      \DsixEmpty{2}
      \ns{2,0}
      \DsixMany{k}{3}
      \ns{3,0}
      \DeightMany{8}{5}{4}
      \Oeight{9}
      \DsixFree{4.6}{0.7}{5}{0.7}
      \DsixFree{4.6}{0.5}{5.4}{0.5}
      \DsixFree{4.6}{0.3}{5.8}{0.3}
      \DsixFree{4.6}{0.1}{6.2}{0.1}
      \DsixFree{4.6}{-0.1}{6.6}{-0.1}
      \DsixFree{4.6}{-0.3}{7}{-0.3}
      \DsixFree{4.6}{-0.5}{7.4}{-0.5}
      \DsixFree{4.6}{0.7}{4}{0}
      \DsixFree{4.6}{0.5}{4}{0}
      \DsixFree{4.6}{0.3}{4}{0}
      \DsixFree{4.6}{0.1}{4}{0}
      \DsixFree{4.6}{-0.1}{4}{0}
      \DsixFree{4.6}{-0.3}{4}{0}
      \DsixFree{4.6}{-0.5}{4}{0}
      \DsixFree{4.6}{-0.7}{4}{0}
      \ns{4,0}
      \draw (4.8,0.85) node {$\text{\footnotesize{k$_1$}}$};
      \draw (5.2,0.65) node {$\text{\footnotesize{k$_2$}}$};
      \draw (5.6,0.45) node {$\text{\footnotesize{k$_3$}}$};
      \draw (6.0,0.25) node {$\text{\footnotesize{k$_4$}}$};
      \draw (6.4,0.05) node {$\text{\footnotesize{k$_5$}}$};
      \draw (6.8,-0.15) node {$\text{\footnotesize{k$_6$}}$};
      \draw (7.2,-0.35) node {$\text{\footnotesize{k$_7$}}$};
      \DsixFree{7.8}{0}{9}{0}
      \draw (8,0.15) node {$\text{\footnotesize{k$_8$}}$};
      \DsixFree{4.6}{-0.7}{8.5}{-0.7}
      \DsixFree{8.5}{-0.7}{9}{0}
      \draw (8,-0.55) node {$\text{\footnotesize{k$_0$}}$};
      \ns{9,0}
      \draw (9.35,0.25) node {$\text{\footnotesize{ $\times$2p}}$};
 \end{tikzpicture} 
 } 
 \label{eq:branes_case_3a}
\end{align}
From the linking numbers \eqref{eq:linking_no_3} one readily computes the 
number of \Ds s to be as in \eqref{eq:linking_aux_1}. 
Next, one may remove the $2p$ stuck half \NS\ branes from the \Oe\ pairwise, 
i.e.\ there will be $p$ pairs. This leads to
\begin{align}
 \raisebox{-.5\height}{
 \begin{tikzpicture}
      \DsixMany{k}{1}
      \DsixEmpty{2}
      \ns{2,0}
      \DsixMany{k}{3}
      \ns{3,0}
\draw[decoration={brace,mirror,raise=20pt},decorate,thick]
  (4.2,0.4) -- node[above=20pt] {$n$ \NS\ } (1.5,0.4);
      \DeightMany{7}{5}{4}
      \Oeight{16}
      \DsixFree{4.6}{0.7}{5}{0.7}
      \DsixFree{4.6}{0.5}{5.4}{0.5}
      \DsixFree{4.6}{0.3}{5.8}{0.3}
      \DsixFree{4.6}{0.1}{6.2}{0.1}
      \DsixFree{4.6}{-0.1}{6.6}{-0.1}
      \DsixFree{4.6}{-0.3}{7}{-0.3}
      \DsixFree{4.6}{-0.5}{7.4}{-0.5}
      \DsixFree{4.6}{0.7}{4}{0}
      \DsixFree{4.6}{0.5}{4}{0}
      \DsixFree{4.6}{0.3}{4}{0}
      \DsixFree{4.6}{0.1}{4}{0}
      \DsixFree{4.6}{-0.1}{4}{0}
      \DsixFree{4.6}{-0.3}{4}{0}
      \DsixFree{4.6}{-0.5}{4}{0}
      \DsixFree{4.6}{-0.7}{4}{0}
      \ns{4,0}
      \draw (4.8,0.85) node {$\text{\footnotesize{k$_1$}}$};
      \draw (5.2,0.65) node {$\text{\footnotesize{k$_2$}}$};
      \draw (5.6,0.45) node {$\text{\footnotesize{k$_3$}}$};
      \draw (6,0.25) node {$\text{\footnotesize{k$_4$}}$};
      \draw (6.4,0.05) node {$\text{\footnotesize{k$_5$}}$};
      \draw (6.8,-0.15) node {$\text{\footnotesize{k$_6$}}$};
      \draw (7.2,-0.35) node {$\text{\footnotesize{k$_7$}}$};
%
%
      \DsixFree{4.6}{-0.7}{8.5}{-0.7}
      \DsixFree{8.5}{-0.7}{9}{0}
      \draw (8,-0.55) node {$\text{\footnotesize{k$_0$}}$};
      \DsixMany{$k_0{-}7$}{9}
      \ns{9,0}
      \DsixMany{$k_0{-}14$}{10}
      \ns{10,0}
      \DsixEmpty{11}
      \ns{11,0}
      \DsixMany{$k_0{-}7m'_4$}{12}
      \Deight{12.5}
      \ns{12,0}
      \DsixEmpty{13}
      \ns{13,0}
      \DsixMany{$a_0{+}8$}{14}
      \ns{14,0}
      \DsixMany{$a_0$}{15}
      \ns{15,0}
      \DsixMany{}{16}
      \ns{17,0}
\draw[decoration={brace,mirror,raise=20pt},decorate,thick]
  (15.5,0.4) -- node[above=20pt] {$p$ new \NS\ } (8.5,0.4);
%
 \end{tikzpicture} 
 }
\end{align}
with $a_0=2m'_2-(m'_3-m'_4)$. 
The tail of the resulting electric quiver becomes
\begin{align}
   \raisebox{-.5\height}{
 	\begin{tikzpicture}
	\tikzstyle{gauge} = [circle, draw,inner sep=3pt];
	\tikzstyle{flavour} = [regular polygon,regular polygon sides=4,inner 
sep=3pt, draw];
\node (g0) [] {$\ldots$};
\node (g1) [gauge,right of=g0,label={[rotate=-45]below right:{$\surm(k_0)$}}] 
{};
\node (g2) [gauge,right of=g1,label={[rotate=-45]below 
right:{$\surm(k_0{-}7)$}}] {};
\node (g3) [gauge,right of=g2,label={[rotate=-45]below 
right:{$\surm(k_0{-}14)$}}] {};
\node (g4) [right of=g3] {$\ldots$};
\node (g5) [gauge,right of=g4,label={[rotate=-45]below 
right:{$\surm(k_0{-}7m'_4)$}}] 
{};
\node (g6) [right of=g5] {$\ldots$};
\node (g7) [gauge,right of=g6,label={[rotate=-45]below 
right:{$\surm(a_0{+}8)$}}] {};
\node (g8) [gauge,right of=g7,label={[rotate=-45]below 
right:{$\usprm(a_0)$}}] {};
	\node (f5) [flavour,above of=g5,label=above:{$1$}] {};
	\draw (g0)--(g1) (g1)--(g2) (g2)--(g3) (g3)--(g4) (g4)--(g5) (g5)--(g6) 
(g6)--(g7) (g7)--(g8)  (g5)--(f5);
	\end{tikzpicture}
	} 
	\label{eq:6d_quiver_case_3}
\end{align}
\paragraph{Magnetic quiver.}
Next, the brane configuration \eqref{eq:branes_case_3a} 
allows to change to the phase where all \Ds\ are suspended between \De\ in 
order to derive the magnetic quiver. The brane configuration in this phase 
becomes 
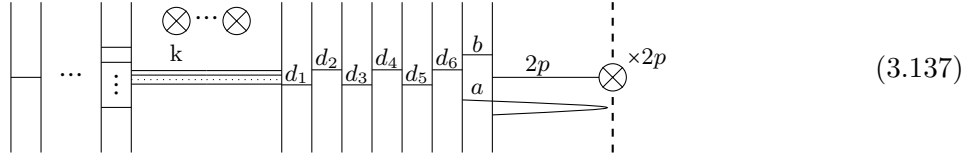
\begin{align}
 \raisebox{-.5\height}{
 \begin{tikzpicture}
 \Deight{0}
 \Deight{0.4}
 \draw (0.8,0) node {$\cdots$};
 \Deight{1.2}
 \Deight{1.6}
 \ns{2.2,0.75}
 \draw (2.6,0.75) node {$\cdots$};
 \ns{3.0,0.75}
 \DsixFree{0}{0}{0.4}{0} 
 \DsixFree{1.2}{0.4}{1.6}{0.4}
 \DsixFree{1.2}{0.2}{1.6}{0.2}
 \DsixFree{1.2}{-0.4}{1.6}{-0.4}
 \draw (1.4,-0.1) node {$\vdots$};
\DsixMany{k}{1.6}
\DsixMany{}{2.6}
      \DeightMany{8}{3.6}{4}
  \DsixFree{3.6}{-0.1}{4.0}{-0.1}
  \DsixFree{4.0}{0.1}{4.4}{0.1}
  \DsixFree{4.4}{-0.1}{4.8}{-0.1}
  \DsixFree{4.8}{0.1}{5.2}{0.1}
  \DsixFree{5.2}{-0.1}{5.6}{-0.1}
  \DsixFree{5.6}{0.1}{6.0}{0.1}
      \draw (3.8,0.05) node {$\text{\footnotesize{$d_1$}}$};
      \draw (4.2,0.25) node {$\text{\footnotesize{$d_2$}}$};
      \draw (4.6,0.05) node {$\text{\footnotesize{$d_3$}}$};
      \draw (5.0,0.25) node {$\text{\footnotesize{$d_4$}}$};
      \draw (5.4,0.05) node {$\text{\footnotesize{$d_5$}}$};
      \draw (5.8,0.25) node {$\text{\footnotesize{$d_6$}}$};
    \DsixFree{6.0}{0.3}{6.4}{0.3}
     \draw (6.2,0.45) node {$\text{\footnotesize{$b$}}$};
\draw (6.0,-0.3) .. controls (8.5,-0.4) .. (6.4,-0.5);
     \draw (6.2,-0.15) node {$\text{\footnotesize{$a$}}$};
      \Oeight{8}
\DsixFree{6.4}{0}{8}{0}
      \draw (7.0,0.15) node {$\text{\footnotesize{$2p$}}$};
      \ns{8,0}
      \draw (8.45,0.25) node {$\text{\footnotesize{$\times 2p$}}$};
 \end{tikzpicture} 
 }
  \label{eq:branes_case_3b}
\end{align}
From the linking numbers \eqref{eq:linking_no_3} one computes the number of 
\Ds\ to be as in \eqref{eq:numbers_3d_brane}.
Here, the relations $k_0 =2a+m'_3$ and $2p=m'_3+m'_4$ allow to rearrange the 
\Ds\ branes in the last two segments compared to \eqref{eq:branes_case_3a}. 
Thus, the magnetic quiver becomes
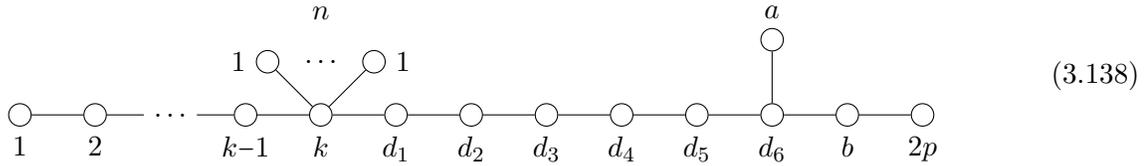
\begin{align}
   \raisebox{-.5\height}{
 	\begin{tikzpicture}
	\tikzstyle{gauge} = [circle, draw,inner sep=3pt];
	\tikzstyle{flavour} = [regular polygon,regular polygon sides=4,inner 
sep=3pt, draw];
	\node (g0) [gauge,label=below:{$1$}] {};
	\node (g1) [gauge,right of=g0,label=below:{$2$}] {};
	\node (g2) [right of=g1] {$\ldots$};
	\node (g3) [gauge,right of=g2,label=below:{$k{-}1$}] {};
	\node (g4) [gauge,right of=g3,label=below:{$k$}] {};
	\node (g5) [gauge,right of=g4,label=below:{$d_1$}] {};
	\node (g6) [gauge,right of=g5,label=below:{$d_2$}] {};
	\node (g7) [gauge,right of=g6,label=below:{$d_3$}] {};
	\node (g8) [gauge,right of=g7,label=below:{$d_4$}] {};
	\node (g9) [gauge,right of=g8,label=below:{$d_5$}] {};
	\node (g10) [gauge,right of=g9,label=below:{$d_6$}] {};
	\node (g11) [gauge,right of=g10,label=below:{$b$}] {};
	\node (g12) [gauge,right of=g11,label=below:{$2p$}] {};
	\node (g13) [gauge,above of=g10,label=above:{$a$}] {};
	\node (b1) [gauge,above left of=g4,label=left:{$1$}] {};
	\node (b0) [above of=g4,label=below:{$\ldots$},label=above:{$n$}] {} ;
	\node (b2) [gauge,above right of=g4,label=right:{$1$}] {};
	\draw (g0)--(g1) (g1)--(g2) (g2)--(g3) (g3)--(g4) (g4)--(g5) (g5)--(g6) 
(g6)--(g7) (g7)--(g8) (g8)--(g9) (g9)--(g10) (g10)--(g11) (g11)--(g12) 
(g10)--(g13) (g4)--(b1) (g4)--(b2);
	\end{tikzpicture}
	}
\end{align}
and it is apparent that one can perform $p$ additional small $E_8$ instanton 
transitions. 
The possibility of $p$ additional small instanton transition is the unsurprising indication that \eqref{eq:branes_case_3b} is not the magnetic phase of \eqref{eq:branes_case_3a}. As before, to obtain the magnetic quiver associated to the electric quiver \eqref{eq:6d_quiver_case_3} one needs to reverse the $p$ $E_8$ transitions. The result is readily obtained as 
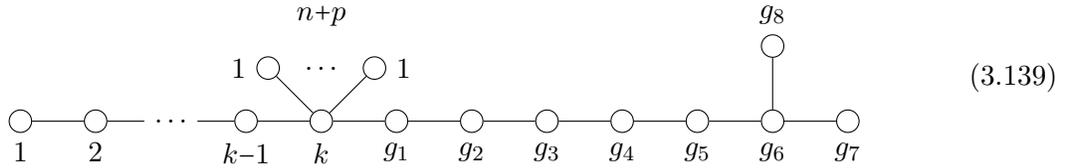
\begin{align}
   \raisebox{-.5\height}{
 	\begin{tikzpicture}
	\tikzstyle{gauge} = [circle, draw,inner sep=3pt];
	\tikzstyle{flavour} = [regular polygon,regular polygon sides=4,inner 
sep=3pt, draw];
	\node (g0) [gauge,label=below:{$1$}] {};
	\node (g1) [gauge,right of=g0,label=below:{$2$}] {};
	\node (g2) [right of=g1] {$\ldots$};
	\node (g3) [gauge,right of=g2,label=below:{$k{-}1$}] {};
	\node (g4) [gauge,right of=g3,label=below:{$k$}] {};
	\node (g5) [gauge,right of=g4,label=below:{$g_1$}] {};
	\node (g6) [gauge,right of=g5,label=below:{$g_2$}] {};
	\node (g7) [gauge,right of=g6,label=below:{$g_3$}] {};
	\node (g8) [gauge,right of=g7,label=below:{$g_4$}] {};
	\node (g9) [gauge,right of=g8,label=below:{$g_5$}] {};
	\node (g10) [gauge,right of=g9,label=below:{$g_6$}] {};
	\node (g11) [gauge,right of=g10,label=below:{$g_7$}] {};
	\node (g13) [gauge,above of=g10,label=above:{$g_8$}] {};
	\node (b1) [gauge,above left of=g4,label=left:{$1$}] {};
	\node (b0) [above of=g4,label=below:{$\ldots$},label=above:{$n{+}p$}] {} ;
	\node (b2) [gauge,above right of=g4,label=right:{$1$}] {};
	\draw (g0)--(g1) (g1)--(g2) (g2)--(g3) (g3)--(g4) (g4)--(g5) (g5)--(g6) 
(g6)--(g7) (g7)--(g8) (g8)--(g9) (g9)--(g10) (g10)--(g11) 
(g10)--(g13) (g4)--(b1) (g4)--(b2);
	\end{tikzpicture}
	}
	\label{eq:general_3d_quiver_mod3}
\end{align}
where the ranks of the unitary nodes are idential to \eqref{eq:ranks_case1}.
Note that all ranks are non-negative integer by definition of the considered case. Finally, the important result is
\begin{align}
 \Higgs^{6d}\left( \substack{\text{electric} \\ \text{quiver}} \eqref{eq:6d_quiver_case_3} \right)
 =
  \Coulomb^{3d}\left( \substack{\text{magnetic} \\ \text{quiver}} \eqref{eq:general_3d_quiver_mod3} \right) \,.
\end{align}
%
%
\subsubsection{\texorpdfstring{$m'_3\geq m'_4,m'_3-m'_4\leq 2 m'_2 , m'_3\pm 
m'_4=\text{odd}$}{mp3>m4p,mp3-mp4<2mp2,mp3-mp4=odd}}
\label{sec:case_4}
\paragraph{Electric quiver.}
Construct a Type IIA brane realisation for the linking number 
\eqref{eq:linking_no_4} and interpret $2p+1$ as the number half \NS\ branes 
that are stuck on the \Oe\ orientifold. The half-integer character of the 
linking numbers is a consequence of the odd number of half \NS\ branes on the 
orientifold.
Then much of the analysis from 
\eqref{eq:branes_case_3a} carries over, with the 
suitable replacement of $2p$ to $2p+1$. (Note that 
$p=\frac{1}{2}(m'_3+m'_4-1)$.)  
As a consequence, one can again remove $2p$ of the stuck half \NS\ branes from 
the \Oe, but one half \NS\ inevitably remains on the orientifold. This leads to 
the brane picture
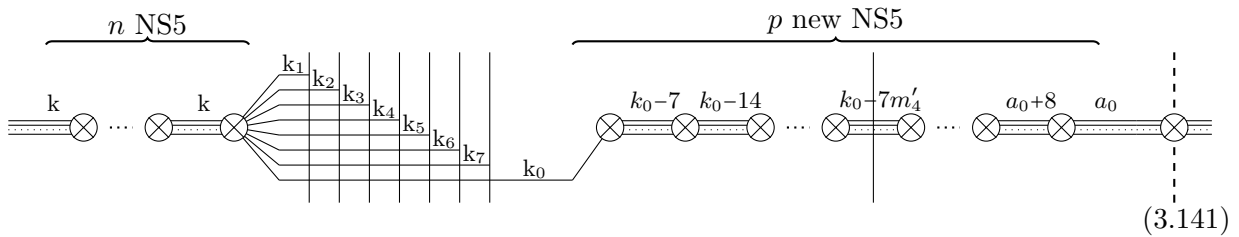
\begin{align}
 \raisebox{-.5\height}{
 \begin{tikzpicture}
      \DsixMany{k}{1}
      \DsixEmpty{2}
      \ns{2,0}
      \DsixMany{k}{3}
      \ns{3,0}
\draw[decoration={brace,mirror,raise=20pt},decorate,thick]
  (4.2,0.4) -- node[above=20pt] {$n$ \NS\ } (1.5,0.4);
      \DeightMany{7}{5}{4}
      \Oeight{16.5}
      \DsixFree{4.6}{0.7}{5}{0.7}
      \DsixFree{4.6}{0.5}{5.4}{0.5}
      \DsixFree{4.6}{0.3}{5.8}{0.3}
      \DsixFree{4.6}{0.1}{6.2}{0.1}
      \DsixFree{4.6}{-0.1}{6.6}{-0.1}
      \DsixFree{4.6}{-0.3}{7}{-0.3}
      \DsixFree{4.6}{-0.5}{7.4}{-0.5}
      \DsixFree{4.6}{0.7}{4}{0}
      \DsixFree{4.6}{0.5}{4}{0}
      \DsixFree{4.6}{0.3}{4}{0}
      \DsixFree{4.6}{0.1}{4}{0}
      \DsixFree{4.6}{-0.1}{4}{0}
      \DsixFree{4.6}{-0.3}{4}{0}
      \DsixFree{4.6}{-0.5}{4}{0}
      \DsixFree{4.6}{-0.7}{4}{0}
      \ns{4,0}
      \draw (4.8,0.85) node {$\text{\footnotesize{k$_1$}}$};
      \draw (5.2,0.65) node {$\text{\footnotesize{k$_2$}}$};
      \draw (5.6,0.45) node {$\text{\footnotesize{k$_3$}}$};
      \draw (6,0.25) node {$\text{\footnotesize{k$_4$}}$};
      \draw (6.4,0.05) node {$\text{\footnotesize{k$_5$}}$};
      \draw (6.8,-0.15) node {$\text{\footnotesize{k$_6$}}$};
      \draw (7.2,-0.35) node {$\text{\footnotesize{k$_7$}}$};
      \DsixFree{4.6}{-0.7}{8.5}{-0.7}
      \DsixFree{8.5}{-0.7}{9}{0}
      \draw (8,-0.55) node {$\text{\footnotesize{k$_0$}}$};
      \DsixMany{$k_0{-}7$}{9}
      \ns{9,0}
      \DsixMany{$k_0{-}14$}{10}
      \ns{10,0}
      \DsixEmpty{11}
      \ns{11,0}
      \DsixMany{$k_0{-}7m'_4$}{12}
      \Deight{12.5}
      \ns{12,0}
      \DsixEmpty{13}
      \ns{13,0}
      \DsixMany{$a_0{+}8$}{14}
      \ns{14,0}
      \DsixMany{$a_0$}{15}
      \ns{15,0}
      \DsixMany{}{16}
      \ns{16.5,0}
\draw[decoration={brace,mirror,raise=20pt},decorate,thick]
  (15.5,0.4) -- node[above=20pt] {$p$ new \NS\ } (8.5,0.4);
 \end{tikzpicture} 
 }
 \label{eq:branes_case_4a} 
\end{align}
with $a_0=2m'_2 -(m'_3-m'_4)+4$.
The tail of the resulting electric quiver looks as follows:
\begin{align}
   \raisebox{-.5\height}{
 	\begin{tikzpicture}
	\tikzstyle{gauge} = [circle, draw,inner sep=3pt];
	\tikzstyle{flavour} = [regular polygon,regular polygon sides=4,inner 
sep=3pt, draw];
\node (g0) [] {$\ldots$};
\node (g1) [gauge,right of=g0,label={[rotate=-45]below right:{$\surm(k_0)$}}] 
{};
\node (g2) [gauge,right of=g1,label={[rotate=-45]below 
right:{$\surm(k_0{-}7)$}}] {};
\node (g3) [gauge,right of=g2,label={[rotate=-45]below 
right:{$\surm(k_0{-}14)$}}] {};
\node (g4) [right of=g3] {$\ldots$};
\node (g5) [gauge,right of=g4,label={[rotate=-45]below 
right:{$\surm(k_0{-}7m'_3)$}}] 
{};
\node (g6) [right of=g5] {$\ldots$};
\node (g7) [gauge,right of=g6,label={[rotate=-45]below right:{$\surm(a_0+8)$}}] 
{};
\node (g8) [gauge,right of=g7,label={[rotate=-45]below 
right:{$\surm(a_0)$}}] {};
	\node (f5) [flavour,above of=g5,label=above:{$1$}] {};
	\draw [-] (8.85,0) arc (0:170:10pt);
	\draw [-] (8.85,-0) arc (360:190:10pt);
	\draw (9.1,0) node {$\mathrm{A}$};
	\draw (g0)--(g1) (g1)--(g2) (g2)--(g3) (g3)--(g4) (g4)--(g5) (g5)--(g6) 
(g6)--(g7) (g7)--(g8)  (g5)--(f5);
	\end{tikzpicture}
	} 
	\label{eq:6d_quiver_case_4}
\end{align}
\paragraph{Magnetic quiver.}
Next, the brane picture \eqref{eq:branes_case_4a} 
allows to change to the phase where all \Ds\ are suspended between \De\ in 
order to read off the magnetic quiver. The brane configuration in this phase 
becomes an adaptation of \eqref{eq:branes_case_3b}
\begin{align}
 \raisebox{-.5\height}{
 \begin{tikzpicture}
 \Deight{0}
 \Deight{0.4}
 \draw (0.8,0) node {$\cdots$};
 \Deight{1.2}
 \Deight{1.6}
 \ns{2.2,0.75}
 \draw (2.6,0.75) node {$\cdots$};
 \ns{3.0,0.75}
 \DsixFree{0}{0}{0.4}{0} 
 \DsixFree{1.2}{0.4}{1.6}{0.4}
 \DsixFree{1.2}{0.2}{1.6}{0.2}
 \DsixFree{1.2}{-0.4}{1.6}{-0.4}
 \draw (1.4,-0.1) node {$\vdots$};
\DsixMany{k}{1.6}
\DsixMany{}{2.6}
      \DeightMany{8}{3.6}{4}
  \DsixFree{3.6}{-0.1}{4.0}{-0.1}
  \DsixFree{4.0}{0.1}{4.4}{0.1}
  \DsixFree{4.4}{-0.1}{4.8}{-0.1}
  \DsixFree{4.8}{0.1}{5.2}{0.1}
  \DsixFree{5.2}{-0.1}{5.6}{-0.1}
  \DsixFree{5.6}{0.1}{6.0}{0.1}
      \draw (3.8,0.05) node {$\text{\footnotesize{$d_1$}}$};
      \draw (4.2,0.25) node {$\text{\footnotesize{$d_2$}}$};
      \draw (4.6,0.05) node {$\text{\footnotesize{$d_3$}}$};
      \draw (5.0,0.25) node {$\text{\footnotesize{$d_4$}}$};
      \draw (5.4,0.05) node {$\text{\footnotesize{$d_5$}}$};
      \draw (5.8,0.25) node {$\text{\footnotesize{$d_6$}}$};
    \DsixFree{6.0}{0.3}{6.4}{0.3}
     \draw (6.2,0.45) node {$\text{\footnotesize{$b$}}$};
\draw (6.0,-0.3) .. controls (8.5,-0.4) .. (6.4,-0.5);
     \draw (6.2,-0.15) node {$\text{\footnotesize{$a$}}$};
      \Oeight{8}
\DsixFree{6.4}{0}{8}{0}
      \draw (7.0,0.15) node {$\text{\footnotesize{$2p{+}1$}}$};
      \ns{8,0}
      \draw (8.45,0.25) node {$\text{\footnotesize{$\times 2p{+}1$}}$};
 \end{tikzpicture} 
 }
  \label{eq:branes_case_4b}
\end{align}
From the linking nunbers \eqref{eq:linking_no_4} one computes the number of 
\Ds\ to be as in \eqref{eq:numbers_3d_brane}.
Here, the relations $k_0 =2a+m'_3$ and $2p=m'_3+m'_4$ allow to rearrange the 
\Ds\ 
branes in the last two segments compared to \eqref{eq:branes_case_4a}. 
Hence, the magnetic quiver becomes
\begin{align}
   \raisebox{-.5\height}{
 	\begin{tikzpicture}
	\tikzstyle{gauge} = [circle, draw,inner sep=3pt];
	\tikzstyle{flavour} = [regular polygon,regular polygon sides=4,inner 
sep=3pt, draw];
	\node (g0) [gauge,label=below:{$1$}] {};
	\node (g1) [gauge,right of=g0,label=below:{$2$}] {};
	\node (g2) [right of=g1] {$\ldots$};
	\node (g3) [gauge,right of=g2,label=below:{$k{-}1$}] {};
	\node (g4) [gauge,right of=g3,label=below:{$k$}] {};
	\node (g5) [gauge,right of=g4,label=below:{$d_1$}] {};
	\node (g6) [gauge,right of=g5,label=below:{$d_2$}] {};
	\node (g7) [gauge,right of=g6,label=below:{$d_3$}] {};
	\node (g8) [gauge,right of=g7,label=below:{$d_4$}] {};
	\node (g9) [gauge,right of=g8,label=below:{$d_5$}] {};
	\node (g10) [gauge,right of=g9,label=below:{$d_6$}] {};
	\node (g11) [gauge,right of=g10,label=below:{$b$}] {};
	\node (g12) [gauge,right of=g11,label=below:{$2p{+}1$}] {};
	\node (g13) [gauge,above of=g10,label=above:{$a$}] {};
	\node (b1) [gauge,above left of=g4,label=left:{$1$}] {};
	\node (b0) [above of=g4,label=below:{$\ldots$},label=above:{$n$}] {} ;
	\node (b2) [gauge,above right of=g4,label=right:{$1$}] {};
	\draw (g0)--(g1) (g1)--(g2) (g2)--(g3) (g3)--(g4) (g4)--(g5) (g5)--(g6) 
(g6)--(g7) (g7)--(g8) (g8)--(g9) (g9)--(g10) (g10)--(g11) (g11)--(g12) 
(g10)--(g13) (g4)--(b1) (g4)--(b2);
	\end{tikzpicture}
	}
\end{align}
and it is apparent that one can perform $p$ additional small $E_8$ instanton 
transitions.
Reversing the $p$ small instanton transition reveals the magnetic quiver associated to the electric quiver \eqref{eq:6d_quiver_case_4}. A straightforward computations yields
\begin{align}
   \raisebox{-.5\height}{
 	\begin{tikzpicture}
	\tikzstyle{gauge} = [circle, draw,inner sep=3pt];
	\tikzstyle{flavour} = [regular polygon,regular polygon sides=4,inner 
sep=3pt, draw];
	\node (g0) [gauge,label=below:{$1$}] {};
	\node (g1) [gauge,right of=g0,label=below:{$2$}] {};
	\node (g2) [right of=g1] {$\ldots$};
	\node (g3) [gauge,right of=g2,label=below:{$k{-}1$}] {};
	\node (g4) [gauge,right of=g3,label=below:{$k$}] {};
	\node (g5) [gauge,right of=g4,label=below:{$g_1$}] {};
	\node (g6) [gauge,right of=g5,label=below:{$g_2$}] {};
	\node (g7) [gauge,right of=g6,label=below:{$g_3$}] {};
	\node (g8) [gauge,right of=g7,label=below:{$g_4$}] {};
	\node (g9) [gauge,right of=g8,label=below:{$g_5$}] {};
	\node (g10) [gauge,right of=g9,label=below:{$g_6$}] {};
	\node (g11) [gauge,right of=g10,label=below:{$g_7$}] {};
	\node (g12) [gauge,right of=g11,label=below:{$1$}] {};
	\node (g13) [gauge,above of=g10,label=above:{$g_8$}] {};
	\node (b1) [gauge,above left of=g4,label=left:{$1$}] {};
	\node (b0) [above of=g4,label=below:{$\ldots$},label=above:{$n{+}p$}] {} ;
	\node (b2) [gauge,above right of=g4,label=right:{$1$}] {};
	\draw (g0)--(g1) (g1)--(g2) (g2)--(g3) (g3)--(g4) (g4)--(g5) (g5)--(g6) 
(g6)--(g7) (g7)--(g8) (g8)--(g9) (g9)--(g10) (g10)--(g11) (g11)--(g12) 
(g10)--(g13) (g4)--(b1) (g4)--(b2);
	\end{tikzpicture}
	}
	\label{eq:general_3d_quiver_mod4}
\end{align}
and the ranks of the unitary nodes are given by \eqref{eq:ranks_case2}. All ranks are non-negative integers by definition of the considered case. Finally, the important result is
\begin{align}
 \Higgs^{6d}\left( \substack{\text{electric} \\ \text{quiver}} \eqref{eq:6d_quiver_case_4} \right)
 =
  \Coulomb^{3d}\left( \substack{\text{magnetic} \\ \text{quiver}} \eqref{eq:general_3d_quiver_mod4} \right) \,.
\end{align}
%
%
\subsubsection{\texorpdfstring{$m'_3\geq m'_4,m'_3-m'_4\geq 2 m'_2 
$}{mp3>m4p,mp3-mp4>2mp2}}
\label{sec:case_5}
\paragraph{Electric quiver.}
For convenience of computing the quiver, one works with $9$ \De\ branes and one 
\Ost\ plane.
Moreover, choose the parametrisation
\begin{align}
 m'_3 -m'_4 -2m'_2 \equiv 3x +l \; , \qquad \text{for } x\in \NN_{>0} \,, \; 
l\in\{0,1,2\} \; ,
\end{align}
such that 
\begin{align}
 p=m'_2 + m'_4 +x \,, \qquad r=\frac{1}{2}(x+l) \,.
\end{align}
Then the Type I$^\prime$ brane set-up becomes
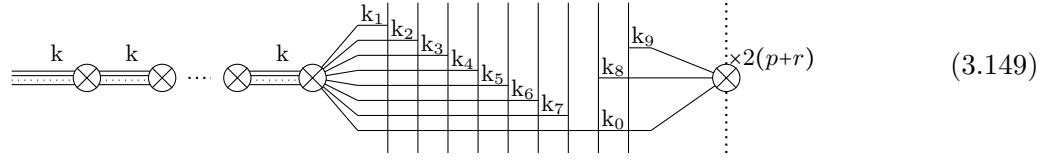
\begin{align}
 \raisebox{-.5\height}{
 \begin{tikzpicture}
      \DsixMany{k}{0}
      \DsixMany{k}{1}
      \ns{1,0}
      \DsixEmpty{2}
      \ns{2,0}
      \DsixMany{k}{3}
      \ns{3,0}
      \DeightMany{9}{5}{4}
      \Ostar{9.5}
      \DsixFree{4.6}{0.7}{5}{0.7}
      \DsixFree{4.6}{0.5}{5.4}{0.5}
      \DsixFree{4.6}{0.3}{5.8}{0.3}
      \DsixFree{4.6}{0.1}{6.2}{0.1}
      \DsixFree{4.6}{-0.1}{6.6}{-0.1}
      \DsixFree{4.6}{-0.3}{7}{-0.3}
      \DsixFree{4.6}{-0.5}{7.4}{-0.5}
      \DsixFree{4.6}{0.7}{4}{0}
      \DsixFree{4.6}{0.5}{4}{0}
      \DsixFree{4.6}{0.3}{4}{0}
      \DsixFree{4.6}{0.1}{4}{0}
      \DsixFree{4.6}{-0.1}{4}{0}
      \DsixFree{4.6}{-0.3}{4}{0}
      \DsixFree{4.6}{-0.5}{4}{0}
      \DsixFree{4.6}{-0.7}{4}{0}
      \ns{4,0}
      \draw (4.8,0.85) node {$\text{\footnotesize{k$_1$}}$};
      \draw (5.2,0.65) node {$\text{\footnotesize{k$_2$}}$};
      \draw (5.6,0.45) node {$\text{\footnotesize{k$_3$}}$};
      \draw (6.0,0.25) node {$\text{\footnotesize{k$_4$}}$};
      \draw (6.4,0.05) node {$\text{\footnotesize{k$_5$}}$};
      \draw (6.8,-0.15) node {$\text{\footnotesize{k$_6$}}$};
      \draw (7.2,-0.35) node {$\text{\footnotesize{k$_7$}}$};
      \DsixFree{7.8}{0}{9.5}{0}
      \draw (8,0.15) node {$\text{\footnotesize{k$_8$}}$};
      \DsixFree{8.2}{0.4}{8.5}{0.4}
      \DsixFree{8.5}{0.4}{9.5}{0.0}
      \draw (8.4,0.55) node {$\text{\footnotesize{k$_9$}}$};
      \DsixFree{4.6}{-0.7}{8.5}{-0.7}
      \DsixFree{8.5}{-0.7}{9.5}{0}
      \draw (8,-0.55) node {$\text{\footnotesize{k$_0$}}$};
      \ns{9.5,0}
      \draw (10.05,0.25) node {$\text{\footnotesize{ $\times2(p{+}r)$}}$};
 \end{tikzpicture} 
 } 
 \label{eq:branes_case_5a}
\end{align}
where the dotted vertical line denotes the \Ost\ plane.
From the linking numbers \eqref{eq:linking_no_5} one computes
\begin{subequations}
\begin{align}
 k_9 &= \lambda_7-\lambda_9 = m'_2 +m'_4 
 \; ,\qquad
 k_8 = \lambda_7 -\lambda_8 = m'_4
 \; ,\qquad
 k_7= \lambda_7-\lambda_7=0
 \; ,\\
 k_i &= \lambda_i -\lambda_7 = \sum_{j=i}^6 m_i \; , \quad i=1,\ldots,6 
 \; ,\qquad
 k_0 = \sum_{i=2,3,4}a'_i m'_i \;.
\end{align}
\end{subequations}
As in the cases above, one can remove $p$ pairs of half \NS\ from the \Ost\ and 
obtains
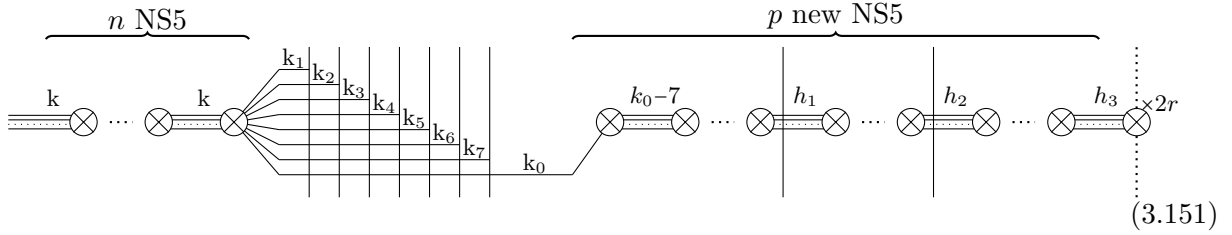
\begin{align}
 \raisebox{-.5\height}{
 \begin{tikzpicture}
      \DsixMany{k}{1}
      \DsixEmpty{2}
      \ns{2,0}
      \DsixMany{k}{3}
      \ns{3,0}
\draw[decoration={brace,mirror,raise=20pt},decorate,thick]
  (4.2,0.4) -- node[above=20pt] {$n$ \NS\ } (1.5,0.4);
      \DeightMany{7}{5}{4}
      \Ostar{16}
      \DsixFree{4.6}{0.7}{5}{0.7}
      \DsixFree{4.6}{0.5}{5.4}{0.5}
      \DsixFree{4.6}{0.3}{5.8}{0.3}
      \DsixFree{4.6}{0.1}{6.2}{0.1}
      \DsixFree{4.6}{-0.1}{6.6}{-0.1}
      \DsixFree{4.6}{-0.3}{7}{-0.3}
      \DsixFree{4.6}{-0.5}{7.4}{-0.5}
      \DsixFree{4.6}{0.7}{4}{0}
      \DsixFree{4.6}{0.5}{4}{0}
      \DsixFree{4.6}{0.3}{4}{0}
      \DsixFree{4.6}{0.1}{4}{0}
      \DsixFree{4.6}{-0.1}{4}{0}
      \DsixFree{4.6}{-0.3}{4}{0}
      \DsixFree{4.6}{-0.5}{4}{0}
      \DsixFree{4.6}{-0.7}{4}{0}
      \ns{4,0}
      \draw (4.8,0.85) node {$\text{\footnotesize{k$_1$}}$};
      \draw (5.2,0.65) node {$\text{\footnotesize{k$_2$}}$};
      \draw (5.6,0.45) node {$\text{\footnotesize{k$_3$}}$};
      \draw (6,0.25) node {$\text{\footnotesize{k$_4$}}$};
      \draw (6.4,0.05) node {$\text{\footnotesize{k$_5$}}$};
      \draw (6.8,-0.15) node {$\text{\footnotesize{k$_6$}}$};
      \draw (7.2,-0.35) node {$\text{\footnotesize{k$_7$}}$};
      \DsixFree{4.6}{-0.7}{8.5}{-0.7}
      \DsixFree{8.5}{-0.7}{9}{0}
      \draw (8,-0.55) node {$\text{\footnotesize{k$_0$}}$};
      \DsixMany{$k_0{-}7$}{9}
      \ns{9,0}
      \DsixEmpty{10}
      \ns{10,0}
      \DsixMany{$h_1$}{11}
      \ns{11,0}
      \Deight{11.3}
      \DsixEmpty{12}
      \ns{12,0}
      \DsixMany{$h_2$}{13}
      \ns{13,0}
      \Deight{13.3}
      \DsixEmpty{14}
      \ns{14,0}
      \DsixMany{$h_3$}{15}
      \ns{15,0}
      \ns{16,0}
     \draw (16.25,0.25) node {$\text{\footnotesize{ $\times 2 r$}}$};
\draw[decoration={brace,mirror,raise=20pt},decorate,thick]
  (15.5,0.4) -- node[above=20pt] {$p$ new \NS\ } (8.5,0.4);
 \end{tikzpicture} 
 }
 \label{eq:branes_case_5b} 
\end{align}
The numbers $h_1$, $h_2$, $h_3$ of \Ds\ spanned between neighbouring \NS\ 
branes can be computed to
\begin{subequations}
\begin{align}
 h_1 &= k_0 - 7 k_8 = 2m'_2 +3(m'_3 -m'_4) \geq0 \; ,\\
 h_2 &= h_1 -8 (k_9-k_8) = 3 (m'_3 -m'_4 -2m'_2)  = 9x+3l\geq 0 \; ,\\
 h_3 &= 3l    \; .
\end{align}
\end{subequations}
Therefore, the tail of the electric quiver looks like 
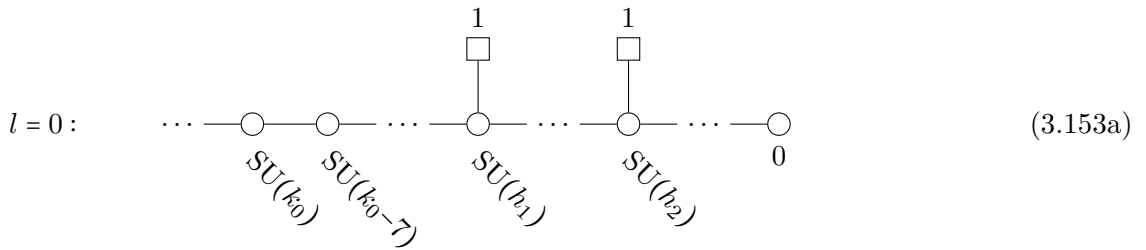
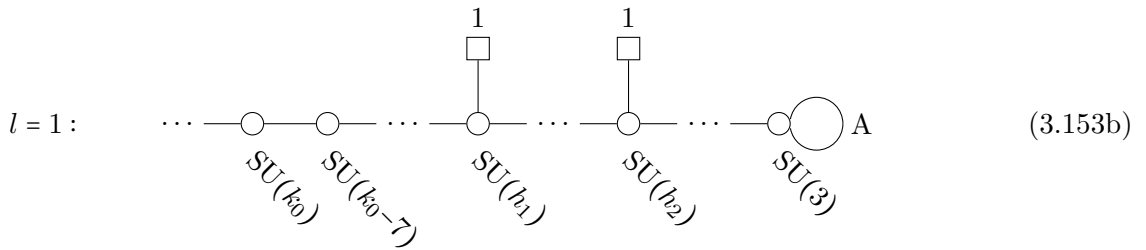
\cite{Zafrir:2015rga,Hayashi:2015zka}
\begin{subequations}
\label{eq:6d_quiver_case_5}
\begin{align}
l&=0: \qquad
   \raisebox{-.5\height}{
 	\begin{tikzpicture}
	\tikzstyle{gauge} = [circle, draw,inner sep=3pt];
	\tikzstyle{flavour} = [regular polygon,regular polygon sides=4,inner 
sep=3pt, draw];
\node (g0) [] {$\ldots$};
\node (g1) [gauge,right of=g0,label={[rotate=-45]below right:{$\surm(k_0)$}}] 
{};
\node (g2) [gauge,right of=g1,label={[rotate=-45]below 
right:{$\surm(k_0{-7})$}}] {};
\node (g3) [right of=g2] {$\ldots$};
\node (g4) [gauge,right of=g3,label={[rotate=-45]below right:{$\surm(h_1)$}}] 
{};
\node (g5) [right of=g4] {$\ldots$};
\node (g6) [gauge,right of=g5,label={[rotate=-45]below right:{$\surm(h_2)$}}] 
{};
\node (g7) [right of=g6] {$\ldots$};
\node (g8) [gauge,right of=g7,label=below:{$0$}] {};
	\node (f4) [flavour,above of=g4,label=above:{$1$}] {};
	\node (f6) [flavour,above of=g6,label=above:{$1$}] {};
%
%
%
	\draw (g0)--(g1) (g1)--(g2) (g2)--(g3) (g3)--(g4) (g4)--(g5) (g5)--(g6) 
(g6)--(g7) (g7)--(g8)  (g4)--(f4) (g6)--(f6);
	\end{tikzpicture}
	} \\
l&=1: \qquad
   \raisebox{-.5\height}{
 	\begin{tikzpicture}
	\tikzstyle{gauge} = [circle, draw,inner sep=3pt];
	\tikzstyle{flavour} = [regular polygon,regular polygon sides=4,inner 
sep=3pt, draw];
\node (g0) [] {$\ldots$};
\node (g1) [gauge,right of=g0,label={[rotate=-45]below right:{$\surm(k_0)$}}] 
{};
\node (g2) [gauge,right of=g1,label={[rotate=-45]below 
right:{$\surm(k_0{-}7)$}}] {};
\node (g3) [right of=g2] {$\ldots$};
\node (g4) [gauge,right of=g3,label={[rotate=-45]below right:{$\surm(h_1)$}}] 
{};
\node (g5) [right of=g4] {$\ldots$};
\node (g6) [gauge,right of=g5,label={[rotate=-45]below right:{$\surm(h_2)$}}] 
{};
\node (g7) [right of=g6] {$\ldots$};
\node (g8) [gauge,right of=g7,label={[rotate=-45]below right:{$\surm(3)$}}] 
{};
	\node (f4) [flavour,above of=g4,label=above:{$1$}] {};
	\node (f6) [flavour,above of=g6,label=above:{$1$}] {};
	\draw [-] (8.85,0) arc (0:170:10pt);
	\draw [-] (8.85,-0) arc (360:190:10pt);
	\draw (9.1,0) node {$\mathrm{A}$};
	\draw (g0)--(g1) (g1)--(g2) (g2)--(g3) (g3)--(g4) (g4)--(g5) (g5)--(g6) 
(g6)--(g7) (g7)--(g8)  (g4)--(f4) (g6)--(f6);
	\end{tikzpicture}
	} \\
l&=2: \qquad
   \raisebox{-.5\height}{
 	\begin{tikzpicture}
	\tikzstyle{gauge} = [circle, draw,inner sep=3pt];
	\tikzstyle{flavour} = [regular polygon,regular polygon sides=4,inner 
sep=3pt, draw];
\node (g0) [] {$\ldots$};
\node (g1) [gauge,right of=g0,label={[rotate=-45]below right:{$\surm(k_0)$}}] 
{};
\node (g2) [gauge,right of=g1,label={[rotate=-45]below 
right:{$\surm(k_0{-}7)$}}] {};
\node (g3) [right of=g2] {$\ldots$};
\node (g4) [gauge,right of=g3,label={[rotate=-45]below right:{$\surm(h_1)$}}] 
{};
\node (g5) [right of=g4] {$\ldots$};
\node (g6) [gauge,right of=g5,label={[rotate=-45]below right:{$\surm(h_2)$}}] 
{};
\node (g7) [right of=g6] {$\ldots$};
\node (g8) [gauge,right of=g7,label={[rotate=-45]below right:{$\surm(6)$}}] 
{};
	\node (f4) [flavour,above of=g4,label=above:{$1$}] {};
	\node (f6) [flavour,above of=g6,label=above:{$1$}] {};
	\draw [-] (8.85,0) arc (0:170:10pt);
	\draw [-] (8.85,-0) arc (360:190:10pt);
	\draw (9.2,0) node {$\frac{1}{2}\Lambda^3$};
	\draw (g0)--(g1) (g1)--(g2) (g2)--(g3) (g3)--(g4) (g4)--(g5) (g5)--(g6) 
(g6)--(g7) (g7)--(g8)  (g4)--(f4) (g6)--(f6);
	\end{tikzpicture}
	} 
\end{align}
\end{subequations}
Note that the extra matter content is both required by anomaly cancellation and it is 
consistent with the brane configuration, see \cite{Gorbatov:2001pw}.
\paragraph{Magnetic quiver.}
As in the cases of Sections \ref{sec:case_1}--\ref{sec:case_4}, the strategy is to first derive the magnetic quiver for the phase where the $2p$ \NS\ branes are stuck at the orientifold and then to compute the quiver corresponding to the electric quiver by reversing $p$ additional small $E_8$ instanton transitions. 
To begin with, one may try to deduce the magnetic quiver from a phase of the 
brane system 
in which all \Ds\ are suspended between \De\ branes. 
\begin{align}
 \raisebox{-.5\height}{
 \begin{tikzpicture}
 \Deight{0}
 \Deight{0.4}
 \draw (0.8,0) node {$\cdots$};
 \Deight{1.2}
 \Deight{1.6}
 \ns{2.2,0.75}
 \draw (2.6,0.75) node {$\cdots$};
 \ns{3.0,0.75}
 \DsixFree{0}{0}{0.4}{0} 
 \DsixFree{1.2}{0.4}{1.6}{0.4}
 \DsixFree{1.2}{0.2}{1.6}{0.2}
 \DsixFree{1.2}{-0.4}{1.6}{-0.4}
 \draw (1.4,-0.1) node {$\vdots$};
\DsixMany{k}{1.6}
\DsixMany{}{2.6}
      \DeightMany{9}{3.6}{4}
  \DsixFree{3.6}{-0.1}{4.0}{-0.1}
  \DsixFree{4.0}{0.1}{4.4}{0.1}
  \DsixFree{4.4}{-0.1}{4.8}{-0.1}
  \DsixFree{4.8}{0.1}{5.2}{0.1}
  \DsixFree{5.2}{-0.1}{5.6}{-0.1}
  \DsixFree{5.6}{0.1}{6.0}{0.1}
      \draw (3.8,0.05) node {$\text{\footnotesize{$d_1$}}$};
      \draw (4.2,0.25) node {$\text{\footnotesize{$d_2$}}$};
      \draw (4.6,0.05) node {$\text{\footnotesize{$d_3$}}$};
      \draw (5.0,0.25) node {$\text{\footnotesize{$d_4$}}$};
      \draw (5.4,0.05) node {$\text{\footnotesize{$d_5$}}$};
      \draw (5.8,0.25) node {$\text{\footnotesize{$d_6$}}$};
    \DsixFree{6.0}{0.3}{6.4}{0.3}
     \draw (6.2,0.45) node {$\text{\footnotesize{$b$}}$};
\draw (6.0,-0.3) .. controls (9.25,-0.4) .. (6.4,-0.5);
     \draw (6.2,-0.15) node {$\text{\footnotesize{$a$}}$};
      \Ostar{8.5}
\DsixFree{6.4}{0}{8.5}{0}
      \draw (7.4,0.15) node {$\text{\footnotesize{$2(p{+}r)$}}$};
      \ns{8.5,0}
      \draw (9.25,0.25) node {$\text{\footnotesize{$\times 2(p{+}r)$}}$};
\DsixFree{6.8}{0.5}{8.5}{0.5}
\draw (7.6,0.65) node {$\text{\footnotesize{$m'_2{+}m'_4$}}$};
 \end{tikzpicture} 
 }
  \label{eq:branes_case_5c}
\end{align}
where the $d_i$ are as in \eqref{eq:numbers_3d_brane}. Next, one can 
change to a different parameter region\cite{Morrison:1996xf} in the 
brane configuration 
\eqref{eq:branes_case_5c} in which the 9th \De\ merges again with the \Ost\ such 
that the magnetic quiver can be read off from the following Type IIA 
configuration:
\begin{align}
 \raisebox{-.5\height}{
 \begin{tikzpicture}
 \Deight{0}
 \Deight{0.4}
 \draw (0.8,0) node {$\cdots$};
 \Deight{1.2}
 \Deight{1.6}
 \ns{2.2,0.75}
 \draw (2.6,0.75) node {$\cdots$};
 \ns{3.0,0.75}
 \DsixFree{0}{0}{0.4}{0} 
 \DsixFree{1.2}{0.4}{1.6}{0.4}
 \DsixFree{1.2}{0.2}{1.6}{0.2}
 \DsixFree{1.2}{-0.4}{1.6}{-0.4}
 \draw (1.4,-0.1) node {$\vdots$};
\DsixMany{k}{1.6}
\DsixMany{}{2.6}
      \DeightMany{8}{3.6}{4}
  \DsixFree{3.6}{-0.1}{4.0}{-0.1}
  \DsixFree{4.0}{0.1}{4.4}{0.1}
  \DsixFree{4.4}{-0.1}{4.8}{-0.1}
  \DsixFree{4.8}{0.1}{5.2}{0.1}
  \DsixFree{5.2}{-0.1}{5.6}{-0.1}
  \DsixFree{5.6}{0.1}{6.0}{0.1}
      \draw (3.8,0.05) node {$\text{\footnotesize{$d_1$}}$};
      \draw (4.2,0.25) node {$\text{\footnotesize{$d_2$}}$};
      \draw (4.6,0.05) node {$\text{\footnotesize{$d_3$}}$};
      \draw (5.0,0.25) node {$\text{\footnotesize{$d_4$}}$};
      \draw (5.4,0.05) node {$\text{\footnotesize{$d_5$}}$};
      \draw (5.8,0.25) node {$\text{\footnotesize{$d_6$}}$};
    \DsixFree{6.0}{0.3}{6.4}{0.3}
     \draw (6.2,0.45) node {$\text{\footnotesize{$b$}}$};
\draw (6.0,-0.3) .. controls (9.25,-0.4) .. (6.4,-0.5);
     \draw (6.2,-0.15) node {$\text{\footnotesize{$a$}}$};
      \Oeight{8.5}
\DsixFree{6.4}{0}{8.5}{0}
      \draw (7.4,0.15) node {$\text{\footnotesize{$2(p{+}r)$}}$};
      \ns{8.5,0}
      \draw (9.25,0.25) node {$\text{\footnotesize{$\times 2(p{+}r)$}}$};
%
 \end{tikzpicture} 
 }
  \label{eq:branes_case_5d}
\end{align}
such that the magnetic quiver becomes
\begin{align}
   \raisebox{-.5\height}{
 	\begin{tikzpicture}
	\tikzstyle{gauge} = [circle, draw,inner sep=3pt];
	\tikzstyle{flavour} = [regular polygon,regular polygon sides=4,inner 
sep=3pt, draw];
	\node (g0) [gauge,label=below:{$1$}] {};
	\node (g1) [gauge,right of=g0,label=below:{$2$}] {};
	\node (g2) [right of=g1] {$\ldots$};
	\node (g3) [gauge,right of=g2,label=below:{$k{-}1$}] {};
	\node (g4) [gauge,right of=g3,label=below:{$k$}] {};
	\node (g5) [gauge,right of=g4,label=below:{$d_1$}] {};
	\node (g6) [gauge,right of=g5,label=below:{$d_2$}] {};
	\node (g7) [gauge,right of=g6,label=below:{$d_3$}] {};
	\node (g8) [gauge,right of=g7,label=below:{$d_4$}] {};
	\node (g9) [gauge,right of=g8,label=below:{$d_5$}] {};
	\node (g10) [gauge,right of=g9,label=below:{$d_6$}] {};
	\node (g11) [gauge,right of=g10,label=below:{$b$}] {};
	\node (g12) [gauge,right of=g11,label=below:{$2(p{+}r)$}] {};
	\node (g13) [gauge,above of=g10,label=above:{$a$}] {};
	\node (b1) [gauge,above left of=g4,label=left:{$1$}] {};
	\node (b0) [above of=g4,label=below:{$\ldots$},label=above:{$n$}] {} ;
	\node (b2) [gauge,above right of=g4,label=right:{$1$}] {};
	\draw (g0)--(g1) (g1)--(g2) (g2)--(g3) (g3)--(g4) (g4)--(g5) (g5)--(g6) 
(g6)--(g7) (g7)--(g8) (g8)--(g9) (g9)--(g10) (g10)--(g11) (g11)--(g12) 
(g10)--(g13) (g4)--(b1) (g4)--(b2);
	\end{tikzpicture}
	}
\end{align}
and it is apparent that one can perform $p$ additional small $E_8$ instanton 
transitions. As before, to derive the magnetic quiver associated to the electric quiver \eqref{eq:6d_quiver_case_5} one needs to reverse $p$ $E_8$ transitions, which results in
\begin{align}
   \raisebox{-.5\height}{
 	\begin{tikzpicture}
	\tikzstyle{gauge} = [circle, draw,inner sep=3pt];
	\tikzstyle{flavour} = [regular polygon,regular polygon sides=4,inner 
sep=3pt, draw];
	\node (g0) [gauge,label=below:{$1$}] {};
	\node (g1) [gauge,right of=g0,label=below:{$2$}] {};
	\node (g2) [right of=g1] {$\ldots$};
	\node (g3) [gauge,right of=g2,label=below:{$k{-}1$}] {};
	\node (g4) [gauge,right of=g3,label=below:{$k$}] {};
	\node (g5) [gauge,right of=g4,label=below:{$g_1$}] {};
	\node (g6) [gauge,right of=g5,label=below:{$g_2$}] {};
	\node (g7) [gauge,right of=g6,label=below:{$g_3$}] {};
	\node (g8) [gauge,right of=g7,label=below:{$g_4$}] {};
	\node (g9) [gauge,right of=g8,label=below:{$g_5$}] {};
	\node (g10) [gauge,right of=g9,label=below:{$g_6$}] {};
	\node (g11) [gauge,right of=g10,label=below:{$g_7$}] {};
	\node (g12) [gauge,right of=g11,label=below:{$2r$}] {};
	\node (g13) [gauge,above of=g10,label=above:{$g_8$}] {};
	\node (b1) [gauge,above left of=g4,label=left:{$1$}] {};
	\node (b0) [above of=g4,label=below:{$\ldots$},label=above:{$n{+}p$}] {} ;
	\node (b2) [gauge,above right of=g4,label=right:{$1$}] {};
	\draw (g0)--(g1) (g1)--(g2) (g2)--(g3) (g3)--(g4) (g4)--(g5) (g5)--(g6) 
(g6)--(g7) (g7)--(g8) (g8)--(g9) (g9)--(g10) (g10)--(g11) (g11)--(g12) 
(g10)--(g13) (g4)--(b1) (g4)--(b2);
	\end{tikzpicture}
	}
	\label{eq:general_3d_quiver_mod5}
\end{align}
with unitary gauge node ranks given by
\begin{subequations}
\label{eq:ranks_case_4}
 \begin{align}
  g_j&= \sum_{i=1}^{6-j} i \ m_{i+j} +(9-j)x +(8-j)m'_2 +(7-j)m'_4 +3l \,, \; \text{for } j=1,\ldots,6 \,,   \\
g_7&= m'_2 +2l +2x \, , \qquad 
g_8 =l \, , \qquad 
2r= x+l
\,.
 \end{align}
\end{subequations}
Then the statement becomes
\begin{align}
 \Higgs^{6d}\left( \substack{\text{electric} \\ \text{quiver}} \eqref{eq:6d_quiver_case_5} \right)
 =
  \Coulomb^{3d}\left( \substack{\text{magnetic} \\ \text{quiver}} \eqref{eq:general_3d_quiver_mod5} \right) \,.
\end{align}
%
%
\subsection{Observations}
\label{sec:observation}
After establishing the general case in Section \ref{sec:general_case}, there 
are some observations to be addressed. To begin with, consider the 
similarities between the $\lambda_i \in \Z$ cases:
\begin{compactenum}[(i)]
\item The electric quiver for $\lambda'_1 \geq \ldots 
\geq \lambda'_7 \geq 0 \geq  \lambda'_8 \geq \lambda'_9$ in 
\eqref{eq:6d_quiver_case_1}
 \item The electric quiver for $\lambda_1 \geq \ldots 
\geq \lambda_7 \geq \lambda_8 \geq 0 \geq \lambda_9$ in 
\eqref{eq:6d_quiver_case_3}
\end{compactenum}
Note in particular that the final gauge node is symplectic for both 
configurations.
Similarly, one may inspect the $\lambda_i \in \Z +\tfrac{1}{2}$ cases:
\begin{compactenum}[(i)]
\item The 6d quiver for $\lambda'_1 \geq \ldots 
\geq \lambda'_7 \geq 0 \geq  \lambda'_8 \geq \lambda'_9$ in 
\eqref{eq:6d_quiver_case_2}
 \item The 6d quiver for $\lambda_1 \geq \ldots 
\geq \lambda_7 \geq \lambda_8 \geq 0 \geq \lambda_9$ in 
\eqref{eq:6d_quiver_case_4}
\end{compactenum}
Here, the common feature is the special unitary gauge node with the 
antisymmetric hyper.
The immediate question is whether these electric quivers can coincide, and if 
so, what does this imply for the magnetic quiver.
\subsubsection{6d Theta angle}
The objective is to analyse flux configurations which yield identical electric 
quivers. For this consider two families of fluxes
\begin{compactenum}[(A)]
  \item $(\{m_i\}_{i=1}^6,m'_2,m'_3,m'_4)$ such that $\lambda_1 \geq \ldots 
\geq \lambda_7 \geq \lambda_8 \geq 0 \geq \lambda_9 \in \Z$, cf. Section 
\ref{sec:case_3}
  \item $(\{M_i\}_{i=1}^6,M'_2,M'_3,M'_4)$ such that $\lambda'_1 \geq \ldots 
\geq \lambda'_7 \geq 0 \geq  \lambda'_8 \geq \lambda'_9 \in \Z$, cf. Section 
\ref{sec:case_1}
\end{compactenum}
such that
\begin{align}
 \lambda_i = \lambda'_i \; , \quad \forall i=1,\ldots,7 
 \; ,\qquad
 \lambda_8 = - \lambda'_8  
 \; , \qquad  
k= \sum_{i=1}^6 a_i m_i + \sum_{i=2,3,4} a'_i m'_i
= \sum_{i=1}^6 a_i M_i + \sum_{i=2,3,4} a'_i M'_i \,,
\label{eq:condition_theta}
\end{align}
which, by the consideration of Section \ref{sec:general_case}, implies that 
the electric quiver theories are identical. 
One straightforwardly solves \eqref{eq:condition_theta} and obtains
\begin{align}
 M_i = m_i \; , \quad \forall i=1,\ldots,6
 \; , \quad
 M'_3 = m'_4 \; , \quad M'_4 = m'_3 
 \; , \quad 
 M'_2=-\frac{1}{2}(m'_3 -m'_4 -2m'_2) = - \lambda_9 \geq0\,.
 \label{eq:map_case_1_case_3}
\end{align}
Since $\lambda_i\in\Z$, $M'_2 \in \NN$ is well-defined.
In particular, this map $(m_i,m'_j)\mapsto 
(M_i,M'_j)$ provides an identification between the electric quivers of 
\eqref{eq:6d_quiver_case_3} 
(with $(m_i,m'_j)$) and \eqref{eq:6d_quiver_case_1} (with $(M_i,M'_j)$).
Moreover, this map yields two different magnetic quivers for each phase of the 
corresponding 6d system. Here, the magnetic quivers for two phases are 
illustrated.
To begin with, consider the magnetic quivers obtained from 
\eqref{eq:general_3d_quiver} by 
specifying the fluxes as in \eqref{eq:numbers_3d_brane}. Recall, this represents 
to phase in which the $p$ additional \NS\ branes are still within the 
orientifold.
One obtains
\begin{align}
   \raisebox{-.5\height}{
 	\begin{tikzpicture}
	\tikzstyle{gauge} = [circle, draw,inner sep=3pt];
	\tikzstyle{flavour} = [regular polygon,regular polygon sides=4,inner 
sep=3pt, draw];
	\node (g0) [gauge,label=below:{$1$}] {};
	\node (g1) [gauge,right of=g0,label=below:{$2$}] {};
	\node (g2) [right of=g1] {$\ldots$};
	\node (g3) [gauge,right of=g2,label=below:{$k{-}1$}] {};
	\node (g4) [gauge,right of=g3,label=below:{$k$}] {};
	\node (g5) [gauge,right of=g4,label=below:{$d_1^{I}$}] {};
	\node (g6) [gauge,right of=g5,label=below:{$d_2^{I}$}] {};
	\node (g7) [gauge,right of=g6,label=below:{$d_3^{I}$}] {};
	\node (g8) [gauge,right of=g7,label=below:{$d_4^{I}$}] {};
	\node (g9) [gauge,right of=g8,label=below:{$d_5^{I}$}] {};
	\node (g10) [gauge,right of=g9,label=below:{$d_6^{I}$}] {};
	\node (g11) [gauge,right of=g10,label=below:{$b^{I}$}] {};
	\node (g12) [gauge,right of=g11,label=below:{$2p^{I}$}] {};
	\node (g13) [gauge,above of=g10,label=above:{$a^{I}$}] {};
	\node (b1) [gauge,above left of=g4,label=left:{$1$}] {};
	\node (b0) [above of=g4,label=above:{$n$}] {$\ldots$};
	\node (b2) [gauge,above right of=g4,label=right:{$1$}] {};
	\draw (g0)--(g1) (g1)--(g2) (g2)--(g3) (g3)--(g4) (g4)--(g5) (g5)--(g6) 
(g6)--(g7) (g7)--(g8) (g8)--(g9) (g9)--(g10) (g10)--(g11) (g11)--(g12) 
(g10)--(g13) (g4)--(b1) (g4)--(b2);
	\end{tikzpicture}
	}
\end{align}
for $I\in\{A,B\}$ and with
\begin{subequations}
\begin{alignat}{2}
 d_i^{A}&= d_i^{B} \; , \quad \forall i=1,\ldots 6
 \;, \qquad  &
 2p^{A} &= m'_3 +m'_4 = 2p^{B} \;,\\
a^A &= m'_2 +m'_3 +2m'_4 \;, \qquad & b^A &= m'_2 +2m'_3 +2m'_4  \;, \\
a^B &= m'_2 +\frac{3}{2}\left( m'_3 +m'_4 \right) \;, \qquad & b^B &= m'_2 
+\frac{3}{2}m'_3 +\frac{5}{2}m'_4  \;.
\end{alignat}
\end{subequations}
Hence, $a^B -a^A = \tfrac{1}{2} \left(m'_3 -m'_4 \right) = b^A -b^B \geq 0$, and 
 really only $a$ and $b$ do change.
Similarly, the magnetic quivers \eqref{eq:general_3d_quiver_mod1} or 
\eqref{eq:general_3d_quiver_mod3} associated to the electric quiver read as 
follows:
\begin{align}
   \raisebox{-.5\height}{
 	\begin{tikzpicture}
	\tikzstyle{gauge} = [circle, draw,inner sep=3pt];
	\tikzstyle{flavour} = [regular polygon,regular polygon sides=4,inner 
sep=3pt, draw];
	\node (g0) [gauge,label=below:{$1$}] {};
	\node (g1) [gauge,right of=g0,label=below:{$2$}] {};
	\node (g2) [right of=g1] {$\ldots$};
	\node (g3) [gauge,right of=g2,label=below:{$k{-}1$}] {};
	\node (g4) [gauge,right of=g3,label=below:{$k$}] {};
	\node (g5) [gauge,right of=g4,label=below:{$g_1^I$}] {};
	\node (g6) [gauge,right of=g5,label=below:{$g_2^I$}] {};
	\node (g7) [gauge,right of=g6,label=below:{$g_3^I$}] {};
	\node (g8) [gauge,right of=g7,label=below:{$g_4^I$}] {};
	\node (g9) [gauge,right of=g8,label=below:{$g_5^I$}] {};
	\node (g10) [gauge,right of=g9,label=below:{$g_6^I$}] {};
	\node (g11) [gauge,right of=g10,label=below:{$g_7^I$}] {};
	\node (g13) [gauge,above of=g10,label=above:{$g_8^I$}] {};
	\node (b1) [gauge,above left of=g4,label=left:{$1$}] {};
	\node (b0) [above of=g4,label=below:{$\ldots$},label=above:{$n{+}p^I$}] {} ;
	\node (b2) [gauge,above right of=g4,label=right:{$1$}] {};
	\draw (g0)--(g1) (g1)--(g2) (g2)--(g3) (g3)--(g4) (g4)--(g5) (g5)--(g6) 
(g6)--(g7) (g7)--(g8) (g8)--(g9) (g9)--(g10) (g10)--(g11) 
(g10)--(g13) (g4)--(b1) (g4)--(b2);
	\end{tikzpicture}
	}
\end{align}
for $I\in \{A,B\}$ and with 
\begin{subequations}
\begin{alignat}{2}
 g_i^{A}&= g_i^{B} \; , \quad \forall i=1,\ldots 6
 \;, \qquad  &
 2p^{A} &= m'_3 +m'_4 = 2p^{B} \;,\\
g_7^A &= m'_2  \;, \qquad & g_8^A &= m'_2 +\frac{1}{2}(m'_4-m'_3)  \;, \\
g_7^B &= m'_2 +\frac{1}{2}\left( m'_4 -m'_3 \right) \;, \qquad & g_8^B &= m'_2 
\;. 
\end{alignat}
\end{subequations}
In this phase, the difference in the magnetic quiver is particularly visible as 
only the node $g_7$ and $g_8$ are interchanged.

As a remark, since the derivation has employed linear algebra, the 
solution found is unique. Therefore, there is a one-to-one correspondences 
between the theories of Section \ref{sec:case_1} and Section \ref{sec:case_3}. 
Hence, a (physical) explanation is desirable.
In \cite[Sec.\ 5.4]{Mekareeya:2017jgc} two 1-parameter families of fluxes 
have been considered that correspond to the same electric quiver, but different 
magnetic quivers. The discussed cases are a subset of the general solution 
\eqref{eq:map_case_1_case_3}. According to \cite[Sec.\ 3.3]{Mekareeya:2017jgc}, 
the different resulting magnetic quivers are due to different embeddings of 
$\surm(2N+8)$ into $\usprm(2N)$. 
In detail, the tail of the electric quivers \eqref{eq:6d_quiver_case_1}, 
\eqref{eq:6d_quiver_case_3} is a $\usprm(2N)$ gauge node connected to a 
$\surm(2N+8)$ node. A $\usprm(2N)$ gauge group with $2n$ half-hypers in the 
fundamental representation has a classically enhanced flavour symmetry of 
$\orm(2n)$, which is reduced to $\sorm(2n)$ by the action of the parity inside 
$\orm(2n)$. Consequently, the link (in the quiver diagram) between the 
$\surm(2N+8)$ node and the $\usprm(2N)$ requires a choice of embedding 
$\surm(2N+8)\hookrightarrow \sorm(4N+16)$. In other words, the choice which 
$\sormL(4N+16)$ spinor node contributes to $\surmL(2N+8)$. It follows that the 
two choices are related by the parity in $\orm(4N+16)$.
The different embeddings are a manifestation of the non-trivial discrete 
6d theta angle for $\usprm(2N)$ due to $\pi_5(\usprm(2N)) = \Z_2$, see 
\cite{Bott:1959}.
%
%
\subsubsection{Comments}
Analogously, one could consider the two families of fluxes
\begin{compactenum}[(A)]
  \item $(\{m_i\}_{i=1}^6,m'_2,m'_3,m'_4)$ such that $\lambda_1 \geq \ldots 
\geq \lambda_7 \geq \lambda_8 > 0 > \lambda_9 \in \Z+\tfrac{1}{2}$, cf.\ 
Section \ref{sec:case_2}
  \item $(\{M_i\}_{i=1}^6,M'_2,M'_3,M'_4)$ such that $\lambda'_1 \geq \ldots 
\geq \lambda'_7 > 0 >  \lambda'_8 \geq \lambda'_9 \in \Z+\tfrac{1}{2}$, 
cf.\ Section \ref{sec:case_4}
\end{compactenum}
such that \eqref{eq:condition_theta} holds again. However, the solution to 
these equations
\begin{align}
{m}_i^{B}= {m}_i^{A} \; \forall i
\; , \quad 
 {m'}_3^{B}= {m'}_3^{A}
 \; , \quad 
 {m'}_4^{B}= {m'}_4^{A} +\frac{1}{4}({m'}_3^{A}-{m'}_4^{A}) 
 \; , \quad
 {m'}_2^{B}= {m'}_2^{A} -\frac{1}{2}({m'}_3^{A}-{m'}_4^{A})
\end{align}
never yields integer fluxes, as ${m'}_3^{A}\pm {m'}_4^{A} =$ odd by 
construction. Therefore, it is not possible to obtain the same electric quivers 
from both scenarios.
%
%
\subsection{From finite coupling to infinite coupling}
\label{sec:finite_to_infinite}
For $n$ \Mf\ branes with a chosen embedding $\Z_k \hookrightarrow E_8$, the 
magnetic quivers for the finite coupling phase have been derived in Section 
\ref{sec:general_case}. The Higgs branches for \emph{any} singular loci on the 
tensor branch can be computed straightforwardly by the techniques presented in 
Section \ref{sec:Preliminaries}.  Generically, the number of Higgs branch phases 
is rather large, but one can restrict to contrasting the two extreme phases: 
finite coupling, i.e.\ generic point of the tensor branch, versus infinite 
coupling, i.e.\ origin of the tensor branch.

\begin{table}[!ht]
 \centering
 \begin{tabular}{p{1.5cm}|c|p{2cm}}
 \toprule
 \multicolumn{3}{c}{Finite coupling: generic point on tensor branch}
 \\ \midrule
electric phase 
& 
 \raisebox{-.5\height}{
 \begin{tikzpicture}
      \DsixMany{k}{0}
      \DsixMany{k}{1}
      \ns{1,0}
      \DsixEmpty{2}
      \ns{2,0}
      \DsixMany{k}{3}
      \ns{3,0}
      \DeightMany{8}{4.5}{1}
      \Oeight{9}
      \DsixFree{4}{0.03}{9}{0.03}
      \DsixFree{4}{0.09}{9}{0.09}
      \draw[dotted] (4,-0.03)--(9,-0.03);
      \DsixFree{4}{-0.09}{9}{-0.09}
        \ns{4,0}
        \ns{6,0}
        \ns{7,0}
        \ns{8,0}
      \ns{9,0}
      \draw (9.35,0.25) node {$\text{\footnotesize{ $\times$2r}}$};
      \draw[decoration={brace,mirror,raise=20pt},decorate,thick]
  (4.2,0.4) -- node[above=20pt] {$n$ \NS\ } (0.5,0.4);
  \draw[decoration={brace,mirror,raise=20pt},decorate,thick]
  (8.5,0.4) -- node[above=20pt] {$p$ \NS\ } (5.5,0.4);
 \end{tikzpicture} 
 } 
& 
\multirow{2}{*}{
$\substack{ \Higgs^{6d}_{\mathrm{finite}} \\ \boldsymbol{=} 
\\ \text{classical} \\ \text{hK-quotient} } $ 
} 
\\
& $\rightarrow$ 6d $\Ncal=(1,0)$ quiver gauge theory determined by 
fluxes $(m_i,m'_j)$ & \\ \midrule
magnetic phase & 
  \raisebox{-.5\height}{
 \begin{tikzpicture}
 \Deight{0}
 \Deight{0.4}
 \draw (0.8,0) node {$\cdots$};
 \Deight{1.2}
 \Deight{1.6}
 \ns{2.2,0.75}
 \draw (2.6,0.75) node {$\cdots$};
 \ns{3.0,0.75}
 \DsixFree{0}{0}{0.4}{0} 
 \DsixFree{1.2}{0.4}{1.6}{0.4}
 \DsixFree{1.2}{0.2}{1.6}{0.2}
 \DsixFree{1.2}{-0.4}{1.6}{-0.4}
 \draw (1.4,-0.1) node {$\vdots$};
\DsixMany{k}{1.6}
\DsixMany{}{2.6}
      \DeightMany{8}{3.6}{4}
  \DsixFree{3.6}{-0.1}{4.0}{-0.1}
  \DsixFree{4.0}{0.1}{4.4}{0.1}
  \DsixFree{4.4}{-0.1}{4.8}{-0.1}
  \DsixFree{4.8}{0.1}{5.2}{0.1}
  \DsixFree{5.2}{-0.1}{5.6}{-0.1}
  \DsixFree{5.6}{0.1}{6.0}{0.1}
    \DsixFree{6.0}{0.3}{6.4}{0.3}
\draw (6.0,-0.3) .. controls (9.25,-0.4) .. (6.4,-0.5);
      \Oeight{8.5}
%
      \ns{8.5,0}
      \draw (9.05,0.25) node {$\text{\footnotesize{$\times 2r$}}$};
\draw[decoration={brace,mirror,raise=20pt},decorate,thick]
  (3.4,0.4) -- node[above=20pt] {$n{+}p$ \NS\ } (1.8,0.4);
 \end{tikzpicture} 
 } 
 & 
 $\substack{ \Higgs^{6d}_{\mathrm{finite}} \\ \boldsymbol{=} \\ \text{3d 
$\Ncal=4$ } \\ \text{Coulomb branch}} $ 
  \\
&
   \raisebox{-.5\height}{
 	\begin{tikzpicture}
 	\tikzset{node distance = 0.75cm}
	\tikzstyle{gauge} = [circle, draw,inner sep=3pt];
	\tikzstyle{flavour} = [regular polygon,regular polygon sides=4,inner 
sep=3pt, draw];
	\node (g0) [gauge,label=below:{$1$}] {};
	\node (g1) [gauge,right of=g0,label=below:{$2$}] {};
	\node (g2) [right of=g1] {$\ldots$};
	\node (g3) [gauge,right of=g2,label=below:{$k{-}1$}] {};
	\node (g4) [gauge,right of=g3,label=below:{$k$}] {};
	\node (g5) [gauge,right of=g4,label=below:{$g_1$}] {};
	\node (g6) [gauge,right of=g5,label=below:{$g_2$}] {};
	\node (g7) [gauge,right of=g6,label=below:{$g_3$}] {};
	\node (g8) [gauge,right of=g7,label=below:{$g_4$}] {};
	\node (g9) [gauge,right of=g8,label=below:{$g_5$}] {};
	\node (g10) [gauge,right of=g9,label=below:{$g_6$}] {};
	\node (g11) [gauge,right of=g10,label=below:{$g_7$}] {};
	\node (g12) [gauge,right of=g11,label=below:{$2r$}] {};
	\node (g13) [gauge,above of=g10,label=above:{$g_8$}] {};
	\node (b1) [gauge,above left of=g4,label=left:{$1$}] {};
	\node (b0) [above of=g4,label=below:{$\ldots$},label=above:{$n{+}p$}] {} ;
	\node (b2) [gauge,above right of=g4,label=right:{$1$}] {};
	\draw (g0)--(g1) (g1)--(g2) (g2)--(g3) (g3)--(g4) (g4)--(g5) (g5)--(g6) 
(g6)--(g7) (g7)--(g8) (g8)--(g9) (g9)--(g10) (g10)--(g11) (g11)--(g12) 
(g10)--(g13) (g4)--(b1) (g4)--(b2);
	\end{tikzpicture}
	}
& 
$\substack{\{g_1, \ldots, g_8, r,p\} \\\text{determined by} \\ \text{ fluxes 
$(m_i,m'_j)$}} $
\\ \midrule
%
\multicolumn{3}{c}{Infinite coupling: origin of tensor branch}\\
 \midrule
electric phase &
\raisebox{-.5\height}{
 \begin{tikzpicture}
      \DsixMany{k}{3}
      \DeightMany{8}{4.5}{1}
      \Oeight{5.4}
      \DsixFree{4}{0.03}{5.4}{0.03}
      \DsixFree{4}{0.09}{5.4}{0.09}
      \draw[dotted] (4,-0.03)--(5.4,-0.03);
      \DsixFree{4}{-0.09}{5.4}{-0.09}
      \ns{5.4,0}
      \draw (6.25,0.25) node {$\text{\footnotesize{ $\times$2($r{+}p{+}n$)}}$};
%
 \end{tikzpicture} 
 } 
 & 
 \multirow{2}{*}{$ \Higgs^{6d}_{\infty}=\text{???} $}
 \\
 & $\rightarrow$ 6d $\Ncal=(1,0)$ SCFT determined by fluxes $(m_i,m'_j)$ &  \\ 
\midrule
%
%
magnetic phase & 
  \raisebox{-.5\height}{
 \begin{tikzpicture}
 \Deight{0}
 \Deight{0.4}
 \draw (0.8,0) node {$\cdots$};
 \Deight{1.2}
 \Deight{1.6}
 \DsixFree{0}{0}{0.4}{0} 
 \DsixFree{1.2}{0.4}{1.6}{0.4}
 \DsixFree{1.2}{0.2}{1.6}{0.2}
 \DsixFree{1.2}{-0.4}{1.6}{-0.4}
 \draw (1.4,-0.1) node {$\vdots$};
\DsixMany{ }{1.6}
\DsixMany{ }{2.6}
      \DeightMany{8}{3.6}{4}
  \DsixFree{3.6}{-0.1}{4.0}{-0.1}
  \DsixFree{4.0}{0.1}{4.4}{0.1}
  \DsixFree{4.4}{-0.1}{4.8}{-0.1}
  \DsixFree{4.8}{0.1}{5.2}{0.1}
  \DsixFree{5.2}{-0.1}{5.6}{-0.1}
  \DsixFree{5.6}{0.1}{6.0}{0.1}
    \DsixFree{6.0}{0.3}{6.4}{0.3}
\draw (6.0,-0.3) .. controls (9.25,-0.4) .. (6.4,-0.5);
      \Oeight{8.5}
      \ns{8.5,0}
      \draw (9.25,0.25) node {$\text{\footnotesize{$\times 2(r{+}p{+}n)$}}$};
 \end{tikzpicture} 
 } 
 &
 $\substack{\Higgs^{6d}_{\infty} \\ \boldsymbol{=} \\ \text{3d $\Ncal=4$ } 
\\ \text{Coulomb branch}} $ 
  \\
&
   \raisebox{-.5\height}{
 	\begin{tikzpicture}
 	\tikzset{node distance = 0.75cm}
	\tikzstyle{gauge} = [circle, draw,inner sep=3pt];
	\tikzstyle{flavour} = [regular polygon,regular polygon sides=4,inner 
sep=3pt, draw];
	\node (g0) [gauge,label=below:{$1$}] {};
	\node (g1) [gauge,right of=g0,label=below:{$2$}] {};
	\node (g2) [right of=g1] {$\ldots$};
	\node (g3) [gauge,right of=g2,label=below:{$k{-}1$}] {};
	\node (g4) [gauge,right of=g3,label=below:{$k$}] {};
	\node (g5) [gauge,right of=g4,label={[rotate=-45]below 
right:{$g_1{+}(p{+}n)$}}] {};
	\node (g6) [gauge,right of=g5,label={[rotate=-45]below 
right:{$g_2{+}2(p{+}n)$}}] {};
	\node (g7) [gauge,right of=g6,label={[rotate=-45]below 
right:{$g_3{+}3(p{+}n)$}}] {};
	\node (g8) [gauge,right of=g7,label={[rotate=-45]below 
right:{$g_4{+}4(p{+}n)$}}] {};
	\node (g9) [gauge,right of=g8,label={[rotate=-45]below 
right:{$g_5{+}5(p{+}n)$}}] {};
	\node (g10) [gauge,right of=g9,label={[rotate=-45]below 
right:{$g_6{+}6(p{+}n)$}}] {};
	\node (g11) [gauge,right of=g10,label={[rotate=-45]below 
right:{$g_7{+}4(p{+}n)$}}] {};
	\node (g12) [gauge,right of=g11,label={[rotate=-45]below 
right:{$2r{+}2(p{+}n)$}}] {};
	\node (g13) [gauge,above of=g10,label=above:{$g_8{+}3(p{+}n)$}] {};
	\draw (g0)--(g1) (g1)--(g2) (g2)--(g3) (g3)--(g4) (g4)--(g5) (g5)--(g6) 
(g6)--(g7) (g7)--(g8) (g8)--(g9) (g9)--(g10) (g10)--(g11) (g11)--(g12) 
(g10)--(g13);
	\end{tikzpicture}
	}
& 
$\substack{\{g_1, \ldots, g_8, r,p\} \\\text{determined by} \\ \text{ fluxes 
$(m_i,m'_j)$} \\ \text{cf. } \eqref{eq:ranks_case1} , \\  
\eqref{eq:ranks_case2}, \\ \text{or } \eqref{eq:ranks_case_4}  } $
\\ \bottomrule
 \end{tabular}
\caption{Contrasting the electric and magnetic description of the Higgs branch 
on a generic point and the origin of the tensor branch. The diagrams are 
schematics for the general cases discussed in Section \ref{sec:general_case}. A 
given embedding $\Z_k\hookrightarrow E_8$, labeled by fluxed $(m_i,m'_j)$ 
determines the brane configurations and the resulting magnetic quivers.}
\label{tab:contrast_finite_vs_infinite}
\end{table}

In Table \ref{tab:contrast_finite_vs_infinite}, the two phases are summarised 
with their respective electric and magnetic description.
For the weakly coupled phase, the Type IIA / Type I${}^\prime$ brane 
configuration provides a conventional low-energy effective description and the 
Higgs branch is a classical hyper-K\"ahler quotient. The corresponding magnetic 
quiver is composed of three characteristic parts:
\begin{compactenum}[(i)]
 \item A tail of length $k$ from the $\C^2\slash \Z_k$ ALE space.
 \item An affine $E_8$-type Dynkin part, where the ranks are determined by the 
fluxes of the chosen embedding.
\item A $\uo$-bouquet of size $(n+p)$, where $n$ denotes the number of \Mf s 
and $p$ is determined by the fluxes too.
\end{compactenum}
In contrast, at the UV-fixed point one has the 6d SCFT with non-local 
contributions from tensionless strings and a Higgs branch description from 
this is unknown. However, the magnetic phase of the brane configuration allows 
to derive an magnetic quiver, whose Coulomb branch readily describes the Higgs 
branch at infinite coupling. In particular, the changes to the magnetic quiver 
are simple \emph{additions} of $(n+p)$ affine $E_8$ Dynkin quivers, due to the 
nature of the $E_8$ instanton transition discussed in Section 
\ref{sec:small_E8}.
In \cite[Sec.\ 4.3]{Mekareeya:2017jgc}, the magnetic quivers for the 6d SCFTs
have been conjectured and the Coulomb branches have been argued to be 
related to the $E_8$ instanton moduli space on $\C^2\slash \Z_k$. In this 
paper, the magnetic quivers are \emph{derived} quantities and the Higgs 
branch over every singular locus of the tensor branch can be described in 
this fashion.

  \section{Conclusion}
\label{sec:Conclusion}
The physics of multiple \Mf\ branes near an \Mn\ plane on an $A$-type 
ALE singularity $\C^2 \slash \Z_k$ is very rich and it is encapsulated in a large 
family of $6$-dimensional $\Ncal=(1,0)$ supersymmetric gauge theories. 
For a given embedding $\Z_k \hookrightarrow E_8$, a system of $n$ \Mf\ 
branes exhibits a multitude of phases which are connected via three principal 
transitions:
\begin{compactenum}[(i)]
 \item An \Mf\ outside the \Mn\ can either be on the singularity or away 
from it.
\item Two \Mf s outside the \Mn\ can be separated or coincident, which 
leads to discrete gauging in the $6$d theory.
\item An \Mf\ can move into the \Mn, which is the small $E_8$ instanton 
transition.
\end{compactenum}
Among the phases $\Pcal_i$ where all \Mf s are on the singularity, there is 
only one 
phase which admits a weakly coupled electric quiver description. This leads to 
a conventional $6$-dimensional $\Ncal=(1,0)$ quiver gauge theory whose 
classical Higgs 
branch is a 
hyper-Kähler quotient. However, as put forward in Conjecture \ref{conj:Higgs}, 
there are many more Higgs branches at singular points of the tensor branch.

As demonstrated in this paper, any phase of the $6$-dimensional 
$\Ncal=(1,0)$ 
theory can be systematically captured by an associated magnetic 
quiver. Each 
quiver is 
derived from the phase of the Type IIA or Type I${}^\prime$ in which all \Ds\ 
branes are suspended between \De\ branes. This can be understood in analogy to 
the magnetic quiver of the $3$-dimensional $\Ncal=4$ quiver gauge theories. 
Consequently, it is the suspension pattern of \Df\ branes in the 
\Ds-\De-\NS\ configuration that dictates the form of the magnetic quiver. 
The derivations rules for the magnetic quiver can be summarised as in 
Conjecture 
\ref{conj:magnetic_quiver}.
The significance of the magnetic quiver $\magQuiv (\Pcal_i)$ for 
a phase $\Pcal_i$ is 
\begin{align}
 \Higgs^{6d}\left(\text{phase }\Pcal_i\right) =  \Coulomb^{3d} 
\left(\substack{\text{magnetic} \\ \text{quiver}} \ \magQuiv(\Pcal_{i}) \right)
\end{align}
where the 6d Higgs branch for $\Pcal_i$ does not admit any electric quiver 
description in any of the strongly coupled phases. However, every Higgs 
branch $\Higgs^{6d}(\Pcal_i)$ admits a description as space of dressed monopole 
operators. 

The magnetic quivers and their associated brane configurations provide  
physical explanations for effects in $6$d 
$\Ncal=(1,0)$ theories. On the one hand, the \emph{discrete gauging} effects 
and their manifestation as discrete $S_n$-quotients on the magnetic quiver are 
understood by the physics of $n$ indistinguishable $\tfrac{1}{2}$ BPS objects. 
On the other hand, the \emph{small $E_8$ instanton transition} implies that an 
$E_8$ 
global symmetry has to arise. Given that it is notoriously difficult to 
generate exceptional symmetries in brane systems, it is remarkable that 
magnetic quivers easily accommodate for this. In particular, it provides a Type 
IIA brane realisation for the closure of the minimal nilpotent orbit of $E_8$.

The multitude of Higgs branch phases has a structure reminiscent to Hasse 
diagrams of nilpotent orbit closures, due to inclusion relations as for instance 
shown in Section \ref{sec:geometry}. The structure of the Hasse-type phase 
diagram can be analysed by transverse slices, obtained for instance by quiver 
subtraction \cite{Cabrera:2018ann}.

Moreover, the formalism presented in this paper facilitates the understanding 
of Higgs branches of theories with 8 supercharges at finite and infinite gauge 
coupling as spaces of dressed monopole operators. The standard lore of Higgs 
branches with 8 supercharges being hyper-Kähler quotients only applies for 
finite gauge coupling. At infinite coupling the Higgs branches are no 
longer hyper-Kähler quotients, but can be described as hyper-Kähler spaces via 
Coulomb branches of $3$d $\Ncal=4$ gauge theories. This provides a uniform and 
systematic approach to all Higgs 
branch phases.

\paragraph{Outlook.}
An open questions remains regarding the two brane configurations 
\eqref{eq:1M5_in_M9} and \eqref{eq:1M5_in_M9_alternative} describing the Higgs 
branch after a single $E_8$ instanton transition. The analysis presented in 
here is 
inconclusive whether this observation hints on a new geometric feature of the 
small instanton transition. It might very well be that it is a simple 
rearrangement of the \Ds\ branes. This should be addressed in future work.

The arguments presented allow to speculate about $4$ and $5$-dimensional gauge 
theories with $8$ supercharges. In fact, starting from the \Dthree-\Dfive-\NS\ 
configuration in Type IIB one may consider either of the following two settings:
\begin{compactenum}[(i)]
 \item One T-duality to obtain a \Df-\Ds-\NS\ configuration in Type IIA, which 
yields $4$-dimensional $\Ncal=2$ world-volume theory. Turning to the magnetic 
phase where the \Df\ branes are suspended between the \Ds\ branes would induce 
a magnetic quiver derived from the way \Dtwo\ branes are suspended.
\item Two T-dualities to arrive at a \Dfive-\Dseven-\NS\ configuration of 
Type IIB with an $5$-dimensional $\Ncal=1$ world-volume theory. Here, the 
magnetic phase is reached when all $5$-branes are suspended between 
$7$-branes and the magnetic quiver encodes the suspension patter of 
\Dthree\ branes. This viewpoint has recently been employed in 
\cite{Cabrera:2018jxt} for the description of $5$d $\Ncal=1$ SQCD.
\end{compactenum}
Again, these magnetic descriptions have multiple advantages: firstly, they are 
applicable even when some (or all) gauge couplings of the electric quiver are 
tuned to infinity. Secondly, the $4$d or $5$d Higgs branches at infinite 
gauge coupling are described as spaces of dressed monopole operators.
Thirdly, the magnetic quivers allow to derive 
dimensions and symmetries of the Higgs branches at infinite coupling via 
the understanding of $3$d $\Ncal=4$ Coulomb branches.
\paragraph{Acknowledgements.}
We like to express our gratitude to  
Costas Bachas, 
Antoine Bourget, 
Andr\'es Collinucci,
Julius Grimminger,
Babak Haghighat,
Chris Hull,  
Rudolph Kalveks, 
Noppadol Mekareeya, 
Satoshi Nawata,
Tom Rudelius, and 
Zhenghao Zhong
for useful discussions.
We thank the Galileo Galilei
Institute for Theoretical Physics for the hospitality and the INFN for partial 
support during the
initial stage of this work at the workshop “Supersymmetric Quantum Field 
Theories in the Non-perturbative Regime” in 2018.
We  thank  the  Simons  Center  for  Geometry  and  Physics,  Stony  Brook  
University  for  the  hospitality  and  the  partial  support  during  the
intermediate  stage  of  this  work  at  the  Simons  Summer  workshop  2018.
S.C. is supported by an EPSRC DTP studentship EP/M507878/1.  
A.H. is supported by STFC Consolidated Grant ST/J0003533/1, and EPSRC Programme 
Grant EP/K034456/1.
M.S. had been supported by Austrian Science Fund (FWF) grant P28590.  M.S. 
thanks the Faculty of Physics of the University
of Vienna for travel support via the “Jungwissenschaftsf\"orderung”.
The work of M.S. was supported by the National Thousand-Young-Talents Program 
of China.  M.S. thanks the Theoretical Physics Group of Imperial College London 
for hospitality.

\appendix
%
\section{Symmetries}
\label{app:global_sym}
A crucial consistence check is provided by checking that the symmetries 
and dimensions of the 6d Higgs branches and 3d Coulomb branches match. Here, 
the symmetries of the hyper-K\"ahler moduli spaces are recalled.
%
%
\subsection{6d Higgs branches}
The hyper multiplets transform as $\oplus_I n_I \Rcal_I$ under the gauge group 
$\otimes_I G_I$, 
where $n_I$ denotes multiplicities. The resulting flavour symmetry $G_F$ 
depends on 
the representation $\Rcal_I$ as follows:
\begin{compactenum}[(i)]
 \item If $\Rcal_I$ is a complex representation, then the flavour symmetry 
contains a $\urm(n_I)$ factor.
\item If $\Rcal_I$ is a real representation, then the flavour symmetry is 
enhanced to $\usprm(2n_I)$.
\item If $\Rcal_I$ is a pseudoreal representation, then the flavour symmetry is 
enhanced to $\sorm(2n_I)$.
\end{compactenum}
As already remarked in \cite{Hanany:2018vph}, most of the appearing $\uo$ 
factors are not global symmetries in the $6$d field theory as they are 
anomalous. Nevertheless, these $\uo$ factors are isometries or, more generally, 
symmetries of the Higgs branch moduli space. Therefore, it is legitimate to 
use the $\uo$ gradings along the Higgs branch. 
%
%
\subsection{3d Coulomb branches}
If the gauge group $G$ of a 3d $\Ncal=4$ gauge theory contains abelian 
factors then Coulomb branches exhibit a topological symmetry 
$G_J^{\mathrm{UV}}=\uo^{\#\{\uo \text{ in }G\}}$. This UV symmetry may be 
enhanced in the IR to $G_J^{\mathrm{IR}}$ such that $G_J^{\mathrm{UV}}$ is a 
maximal torus of $G_J^{\mathrm{IR}}$.
For quiver gauge theories, the Coulomb branch symmetry can be read off from the 
quiver as follows
\begin{compactenum}[(i)]
 \item The subset of balanced gauge nodes forms the Dynkin diagram of the 
non-abelian part of $G_J^{\mathrm{IR}}$.
\item The number of unbalanced gauge nodes minus one provides the number of 
$\uo$ factors inside $G_J^{\mathrm{IR}}$.
\end{compactenum}
Recall that a unitary gauge node is \emph{balanced} if the number of flavours 
equals twice the rank. Note that in some cases the Coulomb branch symmetry read
following the previous procedure might be enhanced to a bigger group $G^{IR}_{enh}$, such that
$G_J^{\mathrm{IR}}\subset G^{IR}_{enh}$. Hence, this procedure provides a 
\emph{minimum amount of symmetry} for the Coulomb branch.

%
%
 \bibliographystyle{JHEP}     
 {\footnotesize{\bibliography{references}}}

\end{document}